\documentclass[10pt, a4paper]{article}
\usepackage[a4paper, total={6in, 10in}]{geometry}
\usepackage{graphicx} 
\usepackage{amsmath}
\usepackage[
  backend=biber,
  style=numeric,
  sorting=none,
  maxnames=10,   
  minnames=1,    
  giveninits=true,
  doi=true,
  url=false,
  isbn=false,
  eprint=false
]{biblatex}
\usepackage{xurl} 
\renewbibmacro{in:}{}
\usepackage{authblk}
\usepackage[colorlinks=true, urlcolor=blue, linkcolor=blue, citecolor=blue]{hyperref}

\DeclareNameAlias{author}{given-family}                
\DeclareFieldFormat[article]{title}{\mkbibquote{#1}}   
\DeclareFieldFormat[article]{volume}{\mkbibbold{#1}}   
\DeclareFieldFormat{pages}{#1}                         
\DeclareFieldFormat{doi}{DOI:\ \href{https://doi.org/#1}{\nolinkurl{#1}}}
\DefineBibliographyStrings{english}{andothers = {et\ al\adddot}}

\AtEveryBibitem{%
  \ifentrytype{article}{%
    \clearfield{number}
  }{}%
}

\DeclareBibliographyDriver{article}{%
  \printnames{author}%
  \setunit{\addcomma\space}%
  \printfield[title]{title}%
  \setunit{\addcomma\space}%
  \printfield{journaltitle}%
  \setunit{\space}%
  \printfield[volume]{volume}%
  \setunit{\addcomma\space}%
  \printfield{pages}%
  \setunit{\addspace}%
  \printfield{year}%
  \iffieldundef{doi}{}{\setunit{\addcomma\space}\printfield{doi}}%
  \finentry
}
\DeclareFieldFormat{eprinttype}{\MakeLowercase{#1}} 
\DeclareFieldFormat{eprint:arXiv}{%
  \href{https://arxiv.org/abs/#1}{#1}}               
\DeclareFieldFormat{url}{\href{#1}{\nolinkurl{#1}}}
\DeclareBibliographyDriver{misc}{%
  \usebibmacro{begentry}%
  \printnames{author}%
  \setunit{\addcomma\space}%
  \printfield[title]{title}%
  \setunit{\adddot\space}%
  \printfield{year}%
  \iffieldundef{eprint}{}{%
    \setunit{\adddot\space}%
    \printtext{%
      \printfield{eprinttype}\addcolon\space
      \printfield[eprint:arXiv]{eprint}%
      \iffieldundef{eprintclass}{}{ \addspace[\printfield{eprintclass}] }%
    }%
  }%
  \iffieldundef{url}{}{\setunit{\adddot\space}url:\space\printfield{url}}%
  \setunit{\adddot}%
  \usebibmacro{finentry}%
}

\DeclareBibliographyDriver{online}{\usebibmacro{begentry}%
  \printnames{author}\setunit{\addcomma\space}%
  \printfield[title]{title}\setunit{\adddot\space}%
  \printfield{year}%
  \iffieldundef{eprint}{}{%
    \setunit{\adddot\space}%
    \printtext{\printfield{eprinttype}\addcolon\space
      \printfield[eprint:arXiv]{eprint}%
      \iffieldundef{eprintclass}{}{ \addspace[\printfield{eprintclass}] }}%
  }%
  \iffieldundef{url}{}{\setunit{\adddot\space}url:\space\printfield{url}}%
  \setunit{\adddot}\usebibmacro{finentry}}

\graphicspath{ {./images/} } 

\title{Data--Driven Reduced--Order Modeling of Phase Mixing Dynamics from Particle Kinetic Simulation
}

\author[1]{Darian FIGUERA-MICHAL}
\author[1]{Sungpil YUM}
\author[2]{Jae-Min KWON}
\author[1]{Eisung YOON\thanks{Corresponding author: esyoon@unist.ac.kr}}

\affil[1]{Department of Nuclear Engineering, UNIST, Ulsan 44919, Republic of Korea}
\affil[2]{Korea Institute of Fusion Energy, Daejon 34133, Republic of Korea}
\date{}

\begin{document}

\maketitle

\begin{abstract}
    Phase mixing is a fundamental kinetic process that governs dissipation and stability in collisionless plasmas, but its inherent filamentation in velocity space creates major challenges for both high-fidelity simulations and reduced-order modeling. This work presents the first exploratory evaluation of a joint Proper Orthogonal Decomposition and Sparse Identification of Nonlinear Dynamics (POD--SINDy) framework applied to particle-in-cell simulations of phase mixing. Simulation datasets were generated under progressively complex conditions, starting from a passive kinetic case without self-consistent electric fields, extending to self-consistent simulations with nonlinear electric field feedback, and finally to a noisy dataset with reduced particle resolution. In the passive kinetic regime, POD–SINDy achieved near-optimal reconstructions with only five modes, reproducing filamentation with errors below four percent. In self-consistent electrostatic cases, variance spread across more modes due to nonlinear interactions and noise, slowing singular value decay and making strict low-rank embeddings more demanding. Nevertheless, retaining ten modes was sufficient to recover the dominant structures, yielding reconstruction errors of about seven percent for the low-noise case and thirteen percent for the noisy dataset. Across all scenarios, SINDy provided sparse and interpretable equations for modal amplitudes that remained predominantly linear despite the underlying nonlinear data, while POD truncation effectively filtered particle noise and preserved coherent dynamics. These findings demonstrate that POD–SINDy constitutes a compact and interpretable approach to reduced-order modeling of phase mixing, capable of retaining essential physics across regimes of increasing complexity while achieving data compression from three to five orders of magnitude depending on dataset complexity.
\end{abstract}

\section{Introduction}

Phase mixing is an essential mechanism in plasma kinetics, governing how perturbations in the distribution function of the particles evolve toward finer filamentary structures in velocity space and ultimately propitiate macroscopic dissipation. As first emphasized by Hammett and Perkins \cite{Hammet}, reduced fluid models that fail to incorporate this kinetic effect produce qualitatively incorrect predictions, motivating the development of moment closures that explicitly account for phase mixing. While Mouhot and Villani \cite{Mouhot} confirmed that phase mixing is an intrinsic property of the Vlasov equation itself, phase mixing also dictates the effective dissipation and stability of collisionless plasmas, shapes the saturation of instabilities \cite{Santos}, and governs the transfer of energy across scales \cite{Parker}. Nevertheless, capturing filamentation structures poses formidable challenges for theoretical modeling and simulations as it requires kinetic solvers with high computational cost.

This tension between the centrality of phase mixing and the complexity of its modeling provides a natural motivation to turn toward reduced--order modeling \textbf{(ROM)} frameworks, which have the potential to extract and compress the dominant dynamics into tractable forms without losing the physics of the phase mixing. The central idea behind ROMs is that many nonlinear systems evolve on attractors or manifolds of considerably lower dimensions in the full state space. Identifying these subspaces can enable efficient simulation and deeper physical insight, while discarding negligible fluctuations.

Among the most established techniques for ROM generation, we can find the Proper Orthogonal Decomposition \textbf{(POD)}, which was first formalized for turbulence by Berkooz, Holmes, and Lumley \cite{Berkooz}. POD extracts an optimal set of orthogonal modes that maximize the capture of the system variance, providing in this way a compact basis for Galerkin projection of dynamical equations. POD has shown strong capabilities on capturing coherent structures \cite{Holmes}, which have had direct impact on broad applications across engineering and physical science \cite{Rowley, Lu}. POD is then recognized as a systematic method to balance compression with fidelity, yielding modes that retain dominant dynamics. 

Complementing modal decomposition approaches, sparse regression techniques have emerged to discover governing equations directly from data. The Sparse Identification of Nonlinear Dynamics \textbf{(SINDy)} algorithm, introduced by Brunton, Proctor, and Kutz \cite{Brunton}, assumes that only a small subset of equational terms from a candidate library are active in describing a given system. By enforcing sparsity, SINDy recovers parsimonious and interpretable models in the form of dynamical equations. SINDy serves then as a powerful tool for equation discovery \cite{Rudy, Champion19}, complementing modal approaches by providing explicit dynamical laws.

Together, POD and SINDy represent two complementary strategies. POD allows to potentially compress data into energetic modes, while SINDy seeks to discover symbolic equations governing their evolution. Their joint use has already shown promise in fluid dynamics, where POD provided a reduced representation and SINDy extracts governing equations for modal amplitudes \cite{oishi23}. In this regard, the natural question motivating this study is whether such a joint POD--SINDy framework can be successfully extended to the plasma kinetics of phase mixing. Although separately, both POD and SINDy have been introduced into the plasma community demonstrating significant promise individually but also revealing limitations that suggest the need for a combined framework.

On the POD side, dimensionality reduction has been applied directly to particle-in-cell \textbf{(PIC)} simulations showing that POD basis can drastically reduce computational cost while capturing essential dynamics \cite{Nicolini}. Later works have sought to improve robustness by addressing invariance \cite{Tyranowski}, representation of multistage kinematic dynamics \cite{tsai}, and conservation \cite{Issan}. These studies confirm POD's ability to compress kinetic simulations effectively but also reveal recurring limitations related to the fact that conservation, stability, and interpretability are not inherent to the method and must be addressed separately. In parallel, SINDy has been explored as a mean of discovering reduced plasma models directly from data. SINDy approach has been shown to recover reduced fluid-like equations that retain essential kinetic physics such as Landau damping, though sensitivity to particle noise remains as a persistent challenge in plasma applications \cite{Alves}. This strategy was further extended to derive symbolic closures for ten-moment fluid models from PIC data, providing explicit terms where traditional closure theories rely on assumptions \cite{Donaghy}. Thus, SINDy's strengths lies in producing interpretable dynamical equations, but its effectiveness depends heavily on the quality of the reduced basis provided to it, as well as on its robustness against noise and multiscale coupling.

Analyzed together, these findings suggest a natural complementarity between the two approaches. POD excels at compressing kinetic simulation data into coherent modes, but the dynamics of these modes are not guaranteed to be physically interpretable. SINDy, on the other hand, excels at discovering sparse governing equations, but it requires a reduced representation of phase space data that is robust to noise. When combined, POD provides an efficient modal basis capturing dominant particle kinematics, while SINDy can identify parsimonious dynamical equations for the corresponding amplitudes, offering both compression and interpretability. This synergy has already been demonstrated in fluid dynamics context \cite{oishi23}, but has not yet been systematically tested in plasma phase mixing. Addressing this potential methodology is the central motivation for the present study, which aims to provide the first exploratory assessment of POD--SINDy framework in this context.

This study does not aim to deliver a fully optimized reduced-order model but rather to investigate the performance of the POD--SINDy framework when applied to plasma data of increasing dynamical complexity. A controlled sequence of test cases was considered, starting from a passive kinematic simulation without nonlinear feedback, then extending to simulations with nonlinear effects induced by self-consistent electric fields, and finally to a self-consistent electrostatic dataset with elevated noise to test robustness under degraded statistical quality. The analysis is deliberately exploratory and restricted to the initial stage of phase mixing process, providing a first assessment of how POD–SINDy balances compactness, interpretability, and dynamical fidelity across progressively complex regimes. In doing so, it identifies the conditions under which the framework succeeds or struggles, offering insights that can guide future developments and extensions.

Through this study, we therefore establish a unique contribution by providing a first systematic exploration of POD--SINDy in early stages of plasma phase mixing, conducted across datasets of increased difficulty. The proposed work sets a benchmark for assessing data-driven reduced-order models in kinetic plasma physics, clarifies current limitations, and opens the path toward more robust hybrid strategies.

The remainder of this paper is organized as follows. Section 2 presents the methodology, including details of the kinetic simulation setup, the reduced-order modeling framework, and the evaluation metric. Section 3 reports the results and discussion, structured in two parts: first, the assessment of ROM performance in the passive kinematic simulation case which will serve as benchmark, and second, the evaluation in self-consistent electrostatic kinetic simulations with reduced and increased noise level. Finally section 4 summarizes the conclusions, highlighting the main findings, limitations, and perspectives for future research.

\section{Methodology}
\subsection{Data generation via PIC simulations}

This study employs a one-dimensional, one-velocity electrostatic particle-in-cell (PIC) solver to generate the numerical data for the early stages of phase mixing process. A dimensionless Vlasov-Poisson system is implemented in the code with fixed neutralizing background ions and fully kinetic electrons, normalized such that time is expressed in units of the inverse plasma frequency ($\omega^{-1}_{pe}$), space in Debye lengths ($\lambda_{De}$), and velocity in the electron thermal speed ($v_{th}=\omega_{pe}\lambda_{De}$). 

Particle positions are initialized from a spatial distribution function perturbed by two sinusoidal modes to seed the plasma oscillation:
\[n_{e}(x,t=0) = n_{0}[1+A_{1}\cos({k_{1}x+\phi_{1}})+A_{2}\cos({k_{2}x+\phi_{2}})],\]
where $n_0$ is the background electron density, $A_1$, $A_2$ are the perturbation amplitudes, $k_1$, $k_2$ are the mode wave numbers, and $\phi_1$, $\phi_2$ are phase offsets. The algorithm deposits charge density on a uniform spatial grid of 256 cells representing a periodic domain tied to the fundamental wave number such that domain length is $L=\frac{2\pi}{k_1}$.

Parameters such as the wave numbers ($k_1$, $k_2$) and phase offsets ($\phi_1$, $\phi_2$) were adopted from the reference setup in \cite{QIN}, although in our implementation different amplitude values were explored, as shown in table \ref{table:data1}. The simulation time step was held constant ($\Delta t = 10^{-3}\omega_{pe}^{-1}$) and the final time was fixed at $t_f = 5\omega_{pe}^{-1}$ for all cases.

\begin{table}[h!]
    \centering
    \begin{tabular}{||c c c c c c ||} 
         \hline
         $k_1$ & $k_2$ & $\phi_{1}$ & $\phi_{2}$ & $A_1$ & $A_2$ \\ [0.5ex] 
         \hline\hline
         0.6 & 1.2 & 0 &  0.38716 & 0.1 & 0.8 \\ [1ex] 
         \hline
    \end{tabular}
    \caption{Particle density perturbation parameters.}
    \label{table:data1}
\end{table}

The code was developed in Python and accelerated with Numba for high-performance execution on GPU architectures. It follows the general structure and initialization strategy presented in \cite{QIN}, while introducing modifications tailored to the present objectives in this study. In particular, velocities are drawn from a Maxwellian with prescribed thermal speed $v_{th}=0.5$, representing the standard deviation of the distribution.

During the execution of the PIC solver, the self-consistent electric field is computed via spectral Poisson solver and particles are pushed using a standard leapfrog integrator. The code also presents a passive simulation option in which the electric field is suppressed ($E=0$) in order to isolate free-streaming phase mixing for its initial evaluation.

Using this framework, three distinct datasets were produced for subsequent analysis: 

\begin{enumerate}
    \item \textbf{40K-NoE} dataset, corresponding to 40,000 particles per cell with passive kinematic phase mixing dynamics ($E=0$).
    \item \textbf{40K-E} dataset, with the same particle resolution but including the self-consistent electric field.
    \item \textbf{5K-E} dataset, with 5,000 particles per cell and the enabled self-consistent electric field.
\end{enumerate}

\begin{figure}[h]
    \centering
    \includegraphics[width=\linewidth]{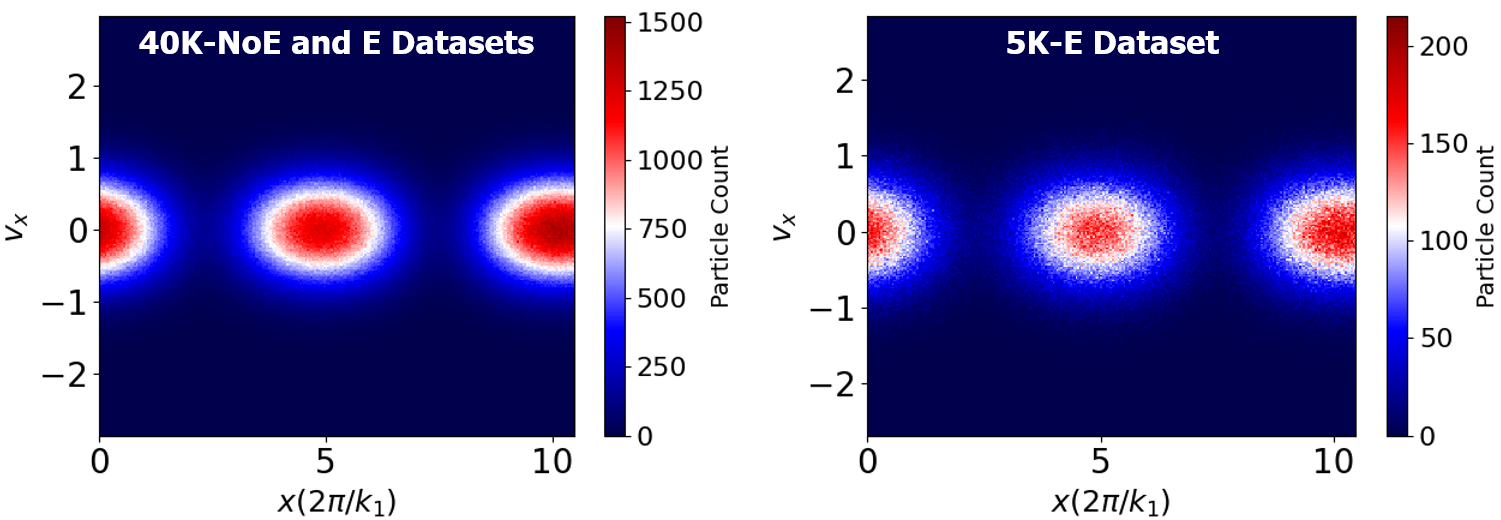}
    \caption{Initial snapshots for generated datasets. (Left) 40K dataset for both passive simulation and self-consistent electric field simulation cases, (right) 5K dataset with self-consistent electric field. Effect of initial sinusoidal density perturbation as well as effect of noise can be appreciated.}
    \label{fig:fig1}
\end{figure}

In each case, the primary output consists of particle phase space snapshots $(x,v)$ written at regular time intervals every 10 time steps generating a total of 500 snapshots. The first snapshot of each dataset can be observed in figure \ref{fig:fig1}, in which initial density perturbation and noise effect can be noticed. These datasets provide complementary views of Landau damping and passive kinematic phase mixing across different noise levels and particle statistics, and they form the basis for evaluation of the reduced-order modeling and data-driven analysis presented in this work.

\subsection{Proper orthogonal decomposition by method of snapshots}

One of the main objectives of this work is to achieve a reduced--order model whose computation can be done efficiently while retaining the physically important meaning of the original data itself. Although novel in its application on phase space data, POD has been widely used in the fluid dynamics field as a methodology for dimensionality reduction and on a variety of cases showing its reliability. For this reason, we seek to evaluate this technique for dimensionality reduction of the phase space evolution data obtained from 1D kinematic particle simulation.

The application of the POD by the method of snapshots \cite{oishi23, taira20} is based on the following process:

\begin{enumerate}

    \item Firstly, a flattened matrix $\mathbf{X}$ must be generated by converting each simulation snapshot of the phase space into a column vector. This is done by arranging the snapshot data row by row into a single column, where the x- and v-axes maintain their grid resolution of 256 uniform cells, resulting in column vectors with $256 \times256$ elements. The resulting matrix $\mathbf{X}$ consists of these column vectors, each corresponding to a snapshot in the order of the simulation data, and from which the mean matrix is subtracted.

    \item We then solve the eigendecomposition to obtain the eigenvector and eigenvalue matrix as follows:

    \begin{equation}
    \mathbf{X}^T \mathbf{X} \boldsymbol{\Psi} = \boldsymbol{\Psi} \boldsymbol{\lambda},
    \end{equation}
    
    where $\mathbf{X}$ is the original flattened matrix in which the mean field was previously subtracted, $\mathbf{X}^T$ is its transpose, $\boldsymbol{\Psi}$ is the eigenvector matrix and $\boldsymbol{\lambda}$ is the diagonal matrix of eigenvalues.

    \item With the obtained eigenvector and eigenvalue matrices, we can calculate the POD modes as follows:

    \begin{equation}
        \boldsymbol{\Phi} = \mathbf{X} \boldsymbol{\Psi} \boldsymbol{\lambda}^{-1/2},
        \label{pod1}
    \end{equation}

    \item The reconstruction of the original matrix is performed by:
    
    \begin{equation}
        \mathbf{X_n} = \boldsymbol{\Phi_n} \boldsymbol{\lambda}^{1/2} \boldsymbol{\Psi_n}^{T}, 
        \label{pod2}
    \end{equation}
    \begin{equation}
        \mathbf{X_n} = \boldsymbol{\Phi_n} \mathbf{A_n},
        \label{pod3}
    \end{equation}
    
    where \textit{n} represent the selected number of modes to reconstruct and $\mathbf{A_n}$ is the amplitude matrix for the selected modes, after which the previously subtracted mean field must be added. 
    
\end{enumerate}

\subsection{SINDy Algorithm for Mode's Amplitude Identification}

As we can see from equations (\ref{pod1}) to (\ref{pod3}), if we are able to identify the expressions for the amplitudes of the modes, then we can successfully evaluate their dynamics at any time within the phase space domain of the 1D particle simulation. SINDy algorithm is particularly important for this task, as it is a proven methodology for system identification \cite{Brunton, oishi23}.

\begin{figure}[h]
    \centering
    \includegraphics[width=0.55\linewidth]{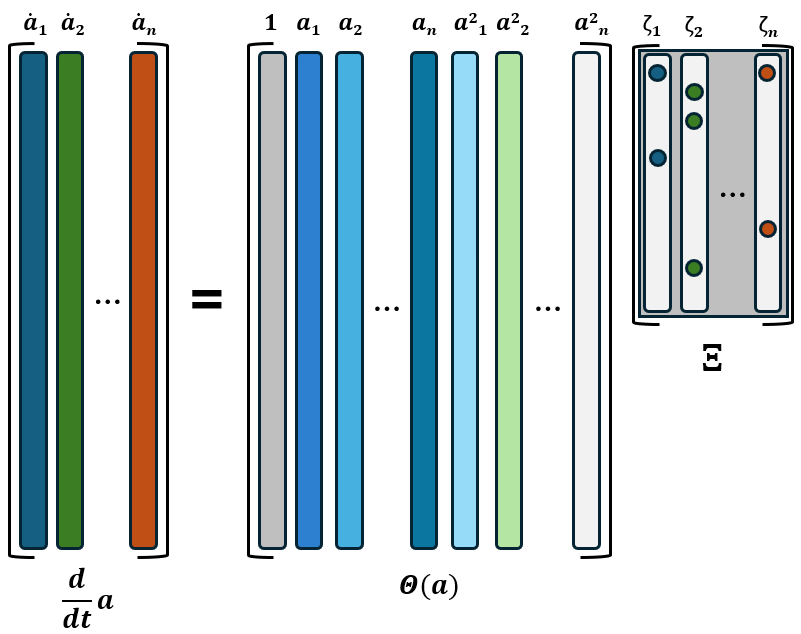}
    \caption{Schematic view of SINDy for obtaining a symbolic expression of mode amplitudes evolution.}
    \label{fig:fig2}
\end{figure}

As mentioned above, through the application of POD we can find the temporal amplitudes of the ROM. Based on the methodology shown by \cite{oishi23}, the evolution of a mode amplitude $a(t)$ can then be described by a differential equation of the form:
\[\dot{a} = f(a),\]
then, the characterization of a model $f$ that describes the amplitudes dynamics is attempted by SINDy by facing a sparse optimization problem indicated as follow:

\begin{enumerate}

    \item We firstly estimate that an acceptable approximation of the model $f$ is obtained by $\dot{a} \approx \Theta(a) \Xi$, where  $\Theta(a)$ is a library of candidate functions and $\Xi$ is a sparse matrix of coefficients that allows to include active terms in the dynamics from the library, as shown in figure \ref{fig:fig2} . The library of candidate functions is built based on the requirements of each case, but in this work, and considering the shape of the selected modes amplitude, we choose polynomial function up to second order to be considered to construct the final expression of the amplitude’s evolution.

    \item The matrix of coefficients $\Xi$ is calculated using the \textit{sparse relaxed regularized regression} (SR3) algorithm which provides a more robust handling of noisy data \cite{Champion20}, allowing in this way to solve the following sparse optimization problem:

    \begin{equation}
    \underset{\Xi}{\arg\min} \| \dot{a} - \Theta(a)\Xi \|_2^2 + \gamma R(\Xi) ,
    \end{equation}
    \(\text{where }\) $\mathbf{R(\Xi)}$ is the regularization term and $\mathbf{\gamma}$ a hyperparameter for the regularization.

\end{enumerate}

\subsection{General Workflow for POD--SINDy Methodology}

\begin{figure}[h]
    \centering
    \includegraphics[width=\linewidth]{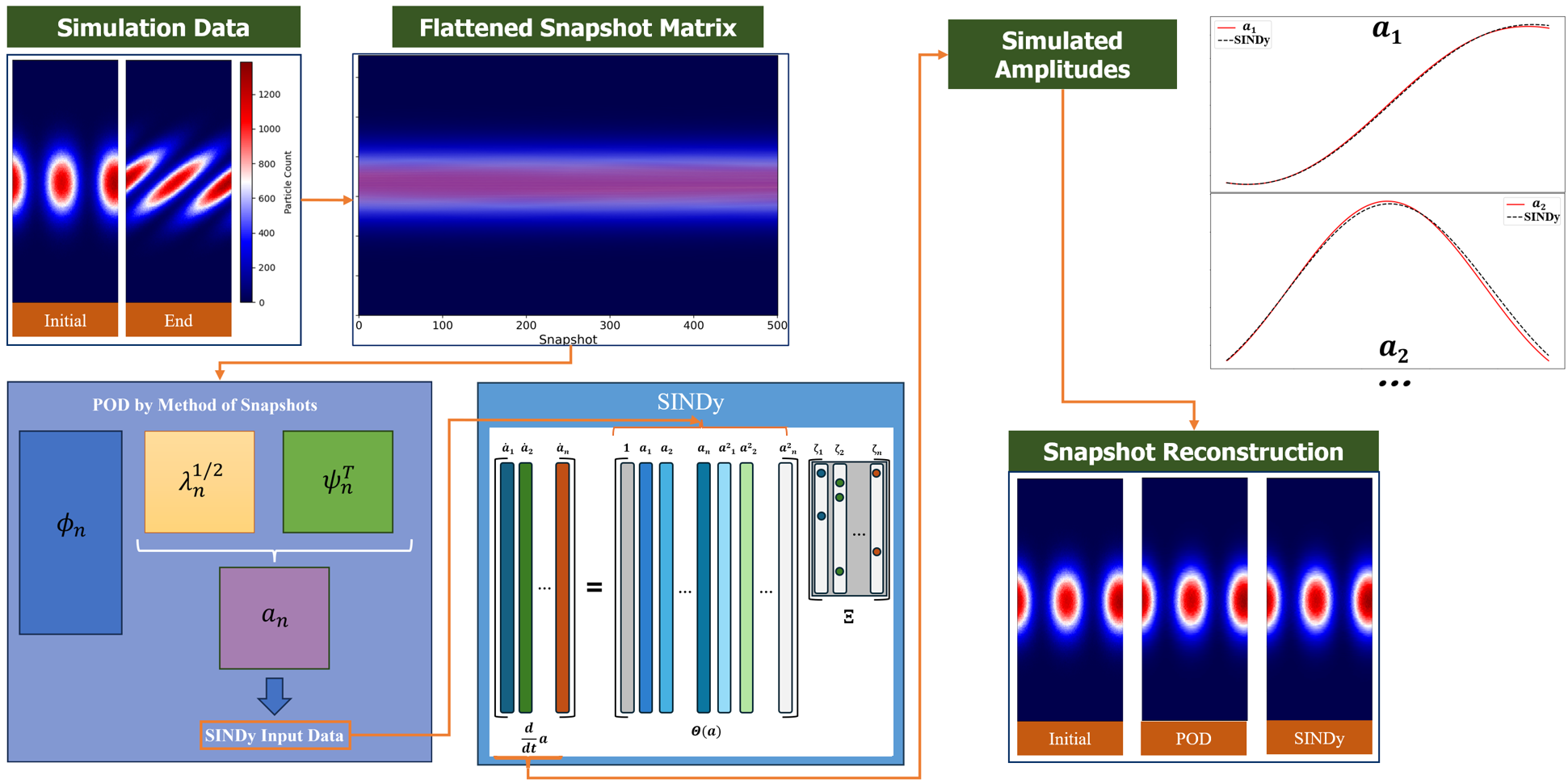}
    \caption{Schematic view of POD--SINDy framework for ROM of particle kinetic simulation.}
    \label{fig:fig3}
\end{figure}

Figure \ref{fig:fig3} presents a schematic of the workflow for the proposed joint POD--SINDy methodology. The goal of this framework is to obtain a ROM that symbolically describes the dynamics of the POD modes while efficiently capturing the phase mixing process with a substantial reduction in data size. For each dataset, with varying noise levels, the simulation snapshots of size \(m \times m\) are first reshaped into a flattened matrix \(\mathbf{X}\) of size \(m^2 \times n\), where \(n\) is the number of snapshots, and from which the mean matrix is subtracted. POD is then applied to the result of this mean-subtracted matrix. From the resulting modal decomposition, we extract a chosen number of modes together with their amplitudes, selected according to their contribution to the system's energy. These amplitudes form the input to SINDy, which identifies a sparse set of equations governing their evolution. Once SINDy characterizes the mode dynamics, the equations are solved to obtain simulated amplitudes, which replace the original POD amplitudes. Finally, the simulated amplitudes and selected modes are combined to reconstruct the flattened matrix \(\mathbf{X}\) and the mean matrix is added. This framework allows for the POD--SINDy model to produce the reconstructed snapshots and evaluate them in comparison with original data and POD-only reconstruction.

The quality of reconstructed snapshots will be evaluated by calculating their relative root mean square error (RRMSE) as follows:

\begin{equation}
\text{RRMSE}(\%) = \sqrt{\frac{\sum_{i=1}^{N} (y_i - \hat{y}_i)^2}{\sum_{i=1}^{N} (\hat{y}_i)^2}} \times 100 ,
\end{equation}
\(\text{where }\) $\hat{y}_i$ represent the ground truth simulation value and $y_i$ the value from reconstructed data. This will help us to numerically appreciate the level of error between the original data and the reconstructed one.

\section{Results and Discussion}
\subsection{ROM performance on passive kinetic simulation}

In the passive kinetic simulation, as in the case of the 40K-NoE dataset, the phase mixing evolution is governed by a purely advective phenomena which is free from the self-consistent electric field feedback. This provides a simplified and clean baseline to assess the performance of the POD--SINDy framework as a viable reduced--order modeling strategy for kinematic simulations. In this sense, our central question lies on whether a compact set of POD modes, including the corresponding SINDy dynamics on those modal amplitudes, can recover the essential early filamentation structure with a minimal dimensionality. The inquiry is initiated by analyzing the energy capture and the associated modal hierarchy, thereby quantifying how the variance is distributed across the ordered modes. This step provides the basis for examining the spatial structures of both leading low-order and higher-order modes, which in turn motivates an evaluation of their dynamical behavior as identified through the SINDy framework.

\begin{figure}[h]
    \centering
    \includegraphics[width=0.85\linewidth]{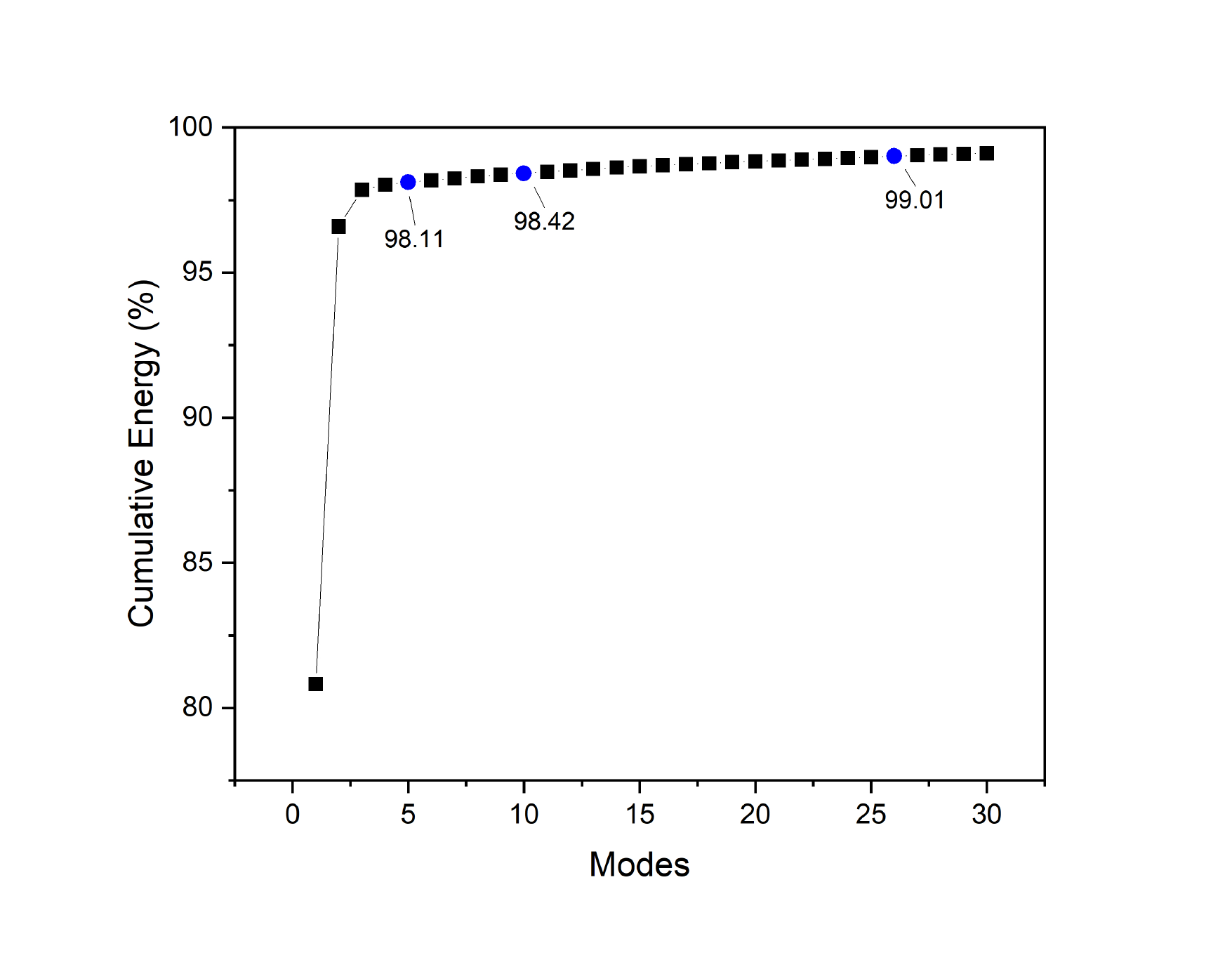}
    \caption{Evolution of cumulative energy per mode in the passive kinetic simulation 40K-NoE dataset.}
    \label{fig:CDF_40kNoE}
\end{figure}

\subsubsection{Energy capture and modal hierarchy}

Figure \ref{fig:CDF_40kNoE} shows us the evolution of the cumulative energy of the modes generated by performing POD to the passive kinetic simulation 40K dataset. This curve tells us how much of the system's variance, or dynamics, can be captured as more modes are selected for the truncation of the ROM. We can clearly appreciate how the first few modes are able to capture almost all of the variance present in the modal system. In fact, with just 5 modes we already recover approximately 98\% of the total energy. Doubling to 10 modes we can observe that the percentile gain is very small (less than 0.3\%). This shows that in the case of the passive kinetic dataset, the mode variance is concentrated in very few coherent structures, which implies that the dimensionality of the system is extremely low with the selection of the first 5 modes. 

\begin{figure}[p] 
    \centering 
    
    \begin{tabular}{ c  c }

        \includegraphics[width=0.4\textwidth]{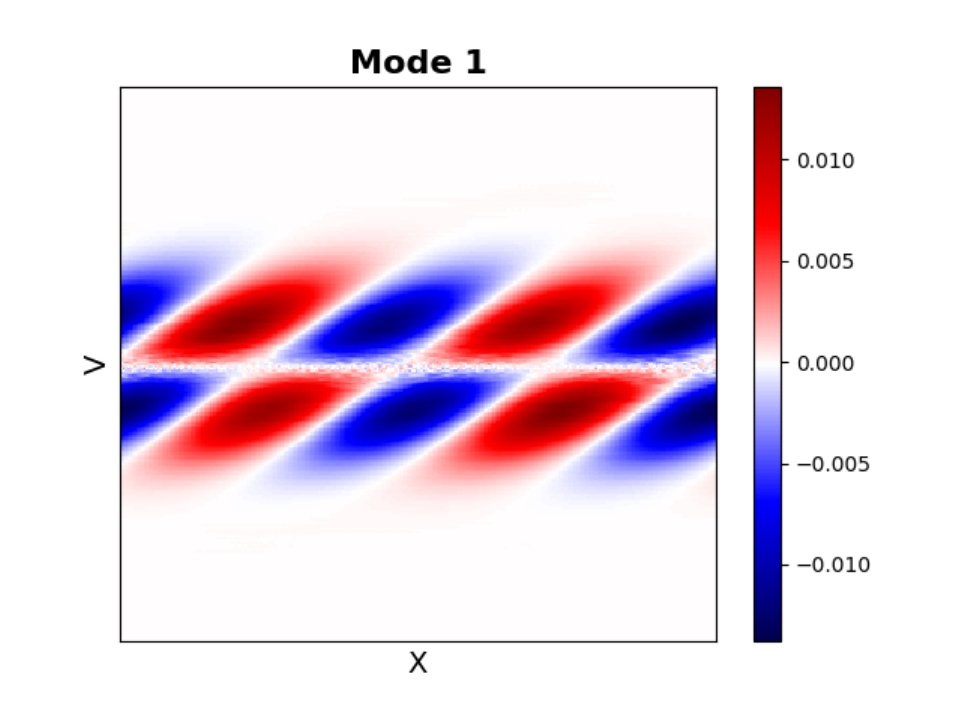} & 
        \includegraphics[width=0.4\textwidth]{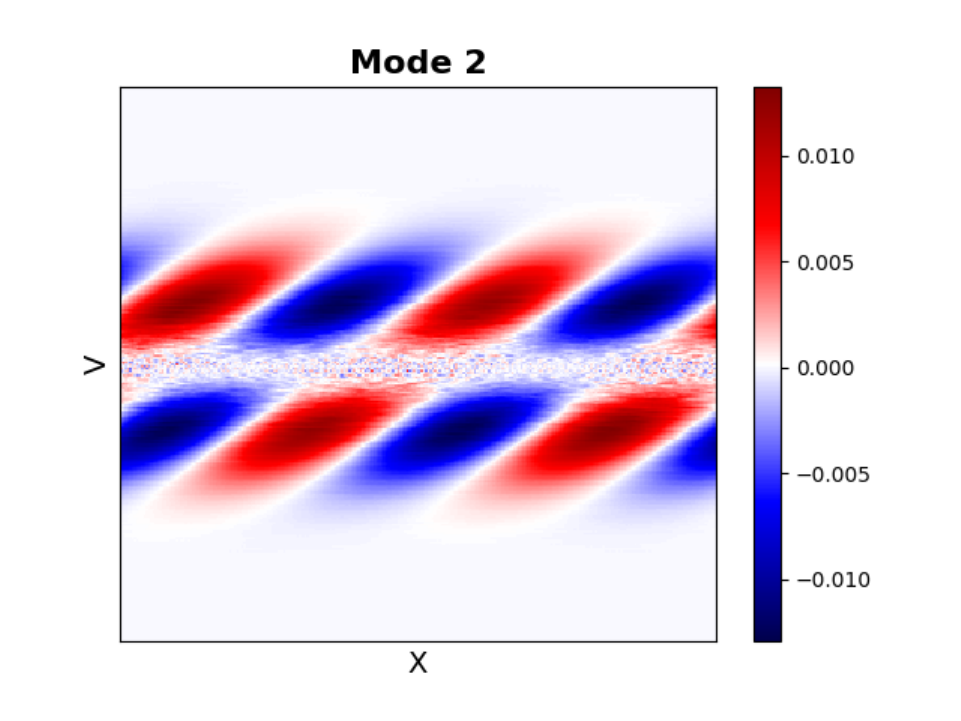} \\
        
        \includegraphics[width=0.4\textwidth]{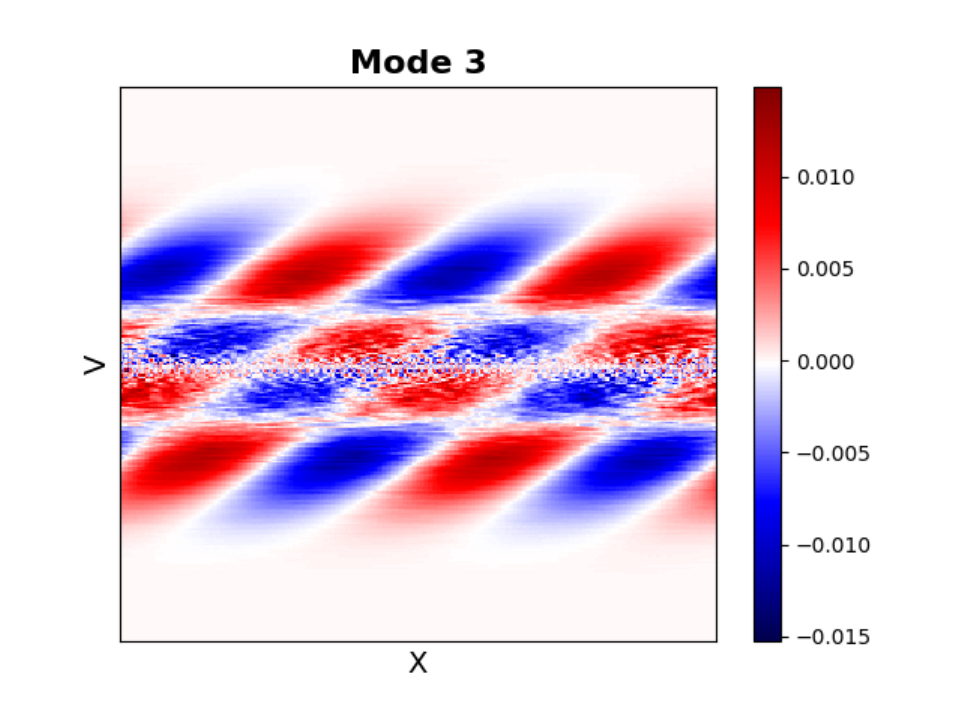} & 
        \includegraphics[width=0.4\textwidth]{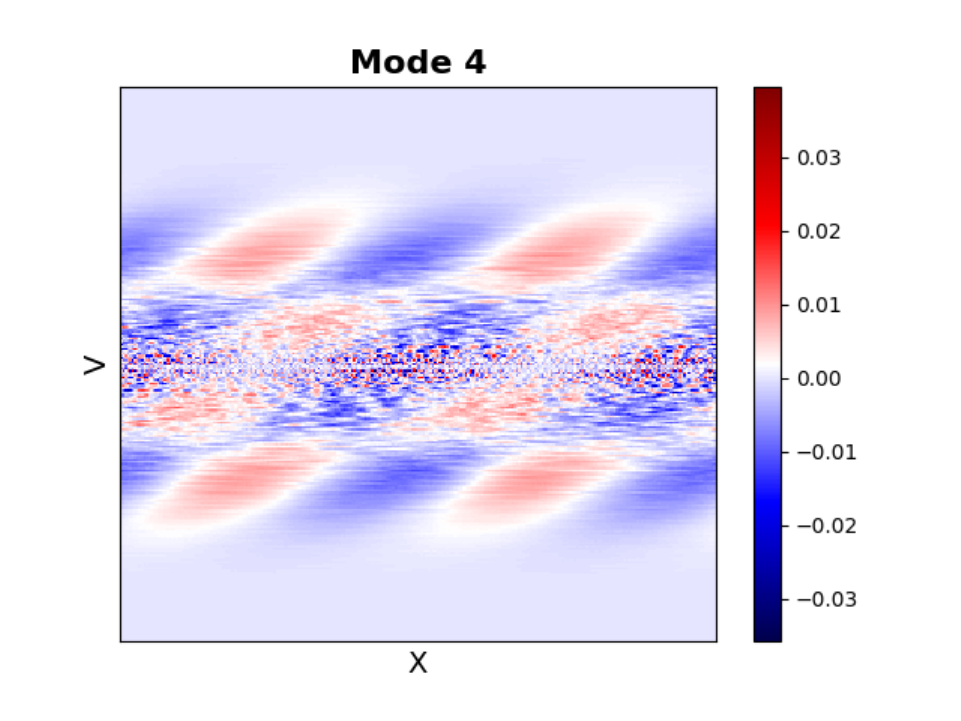} \\

        \includegraphics[width=0.4\textwidth]{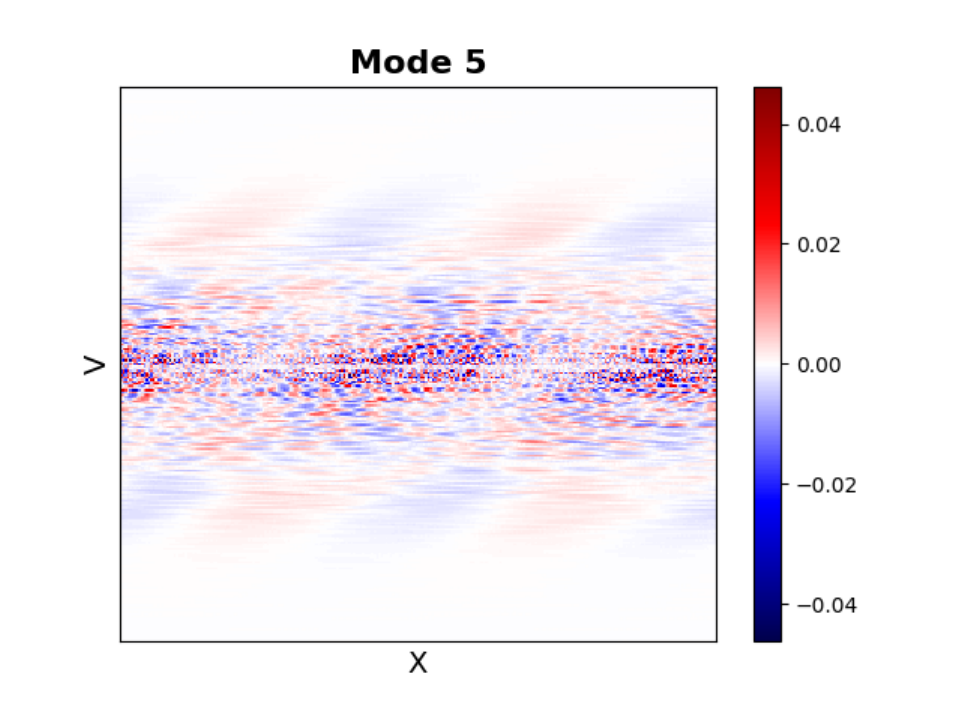} & 
        \includegraphics[width=0.4\textwidth]{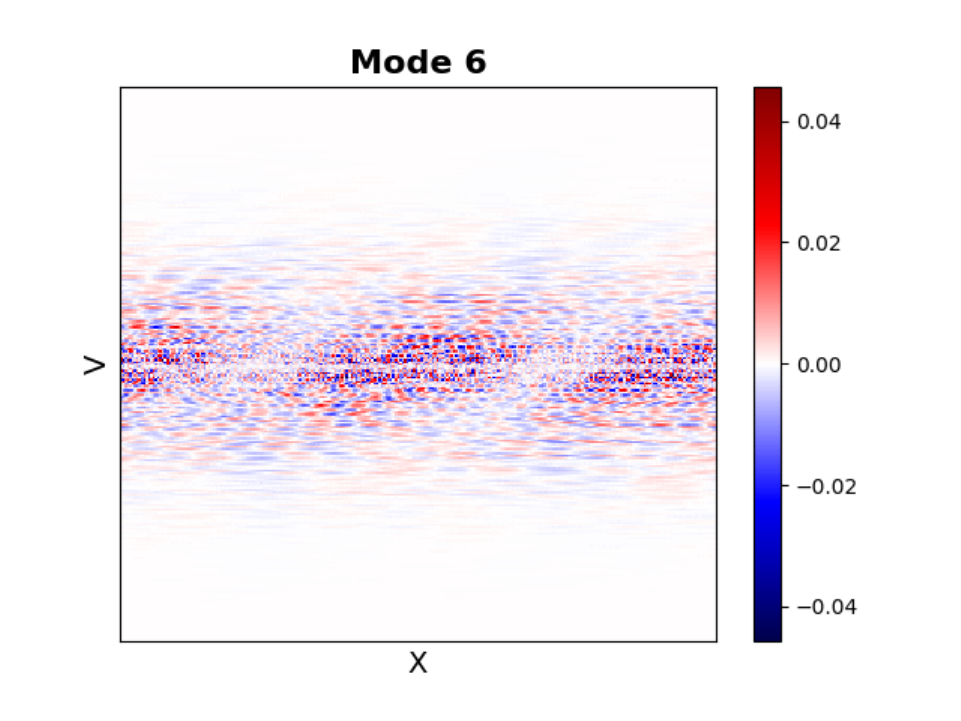} \\

        \includegraphics[width=0.4\textwidth]{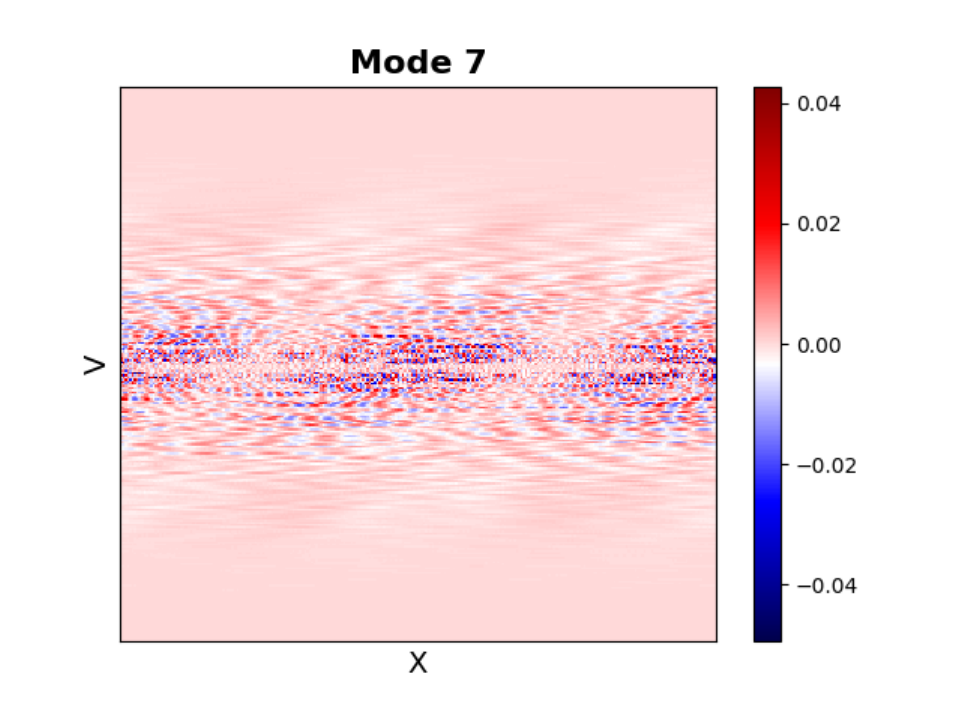} & 
        \includegraphics[width=0.4\textwidth]{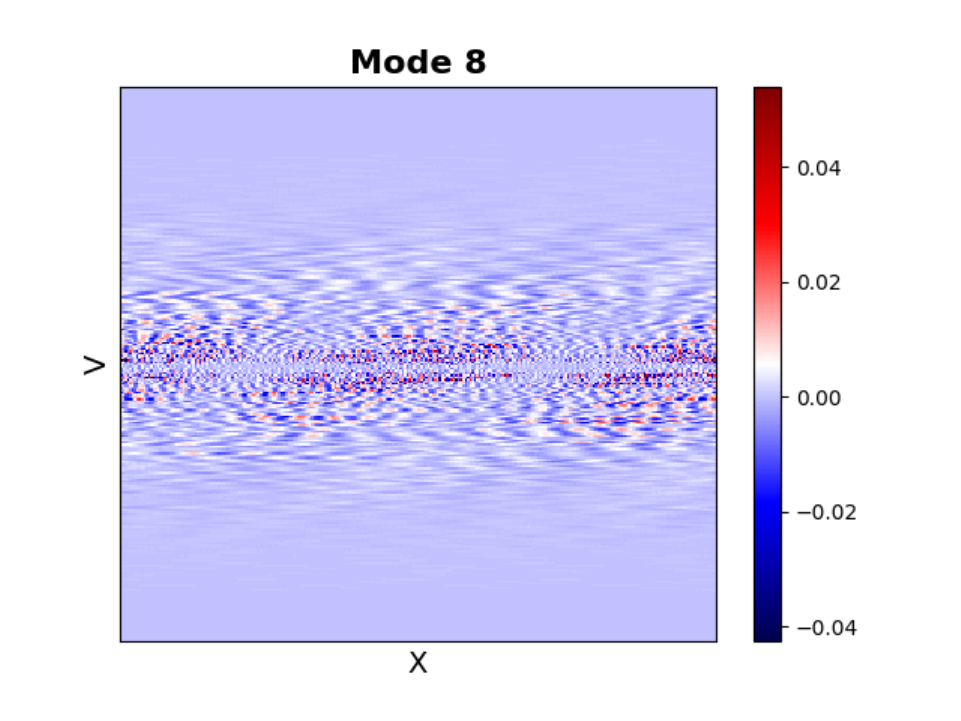} \\

        \includegraphics[width=0.4\textwidth]{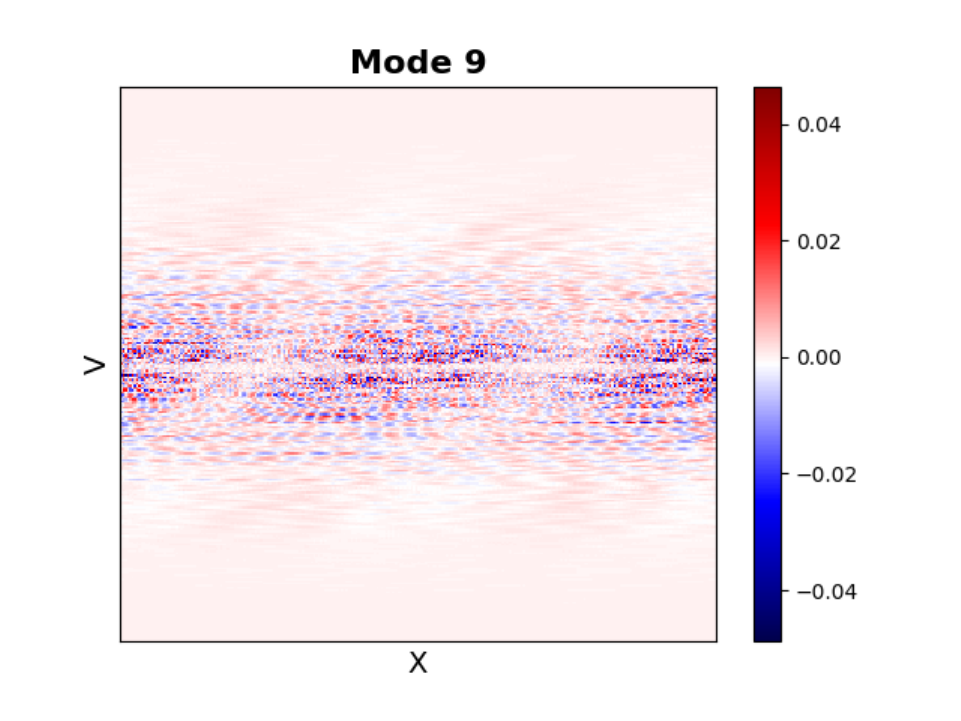} & 
        \includegraphics[width=0.4\textwidth]{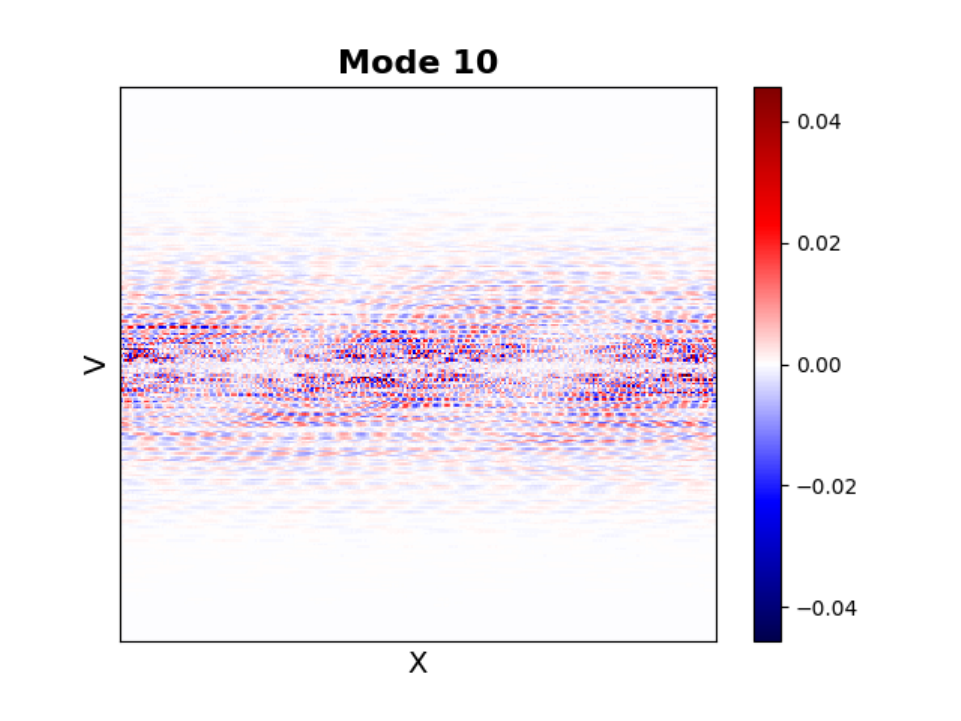} \\
  
    \end{tabular}
    
    \caption{Spatial distribution of first ten modes from POD in the passive kinetic simulation 40K dataset.}
    \label{fig:Modes_40kNoE}  

\end{figure}

This reduced variance has a direct impact on the spatial structures of the obtained modes. Figure \ref{fig:Modes_40kNoE} shows the spatial distribution of the first ten modes in the passive kinetic dataset. Each panel exhibits the presence of coherent structures that start to diminish as the mode order increases. The first two modes, shown in the first row of the figure, present very clean and well-defined coherent structures as they dominate the ROM dynamics by concentrating a considerable part of the variance of the system. They capture the large-scale linear structures associated with the early stages of phase mixing process without presence of a self-consistent electric field. Third mode, shown at the second row the figure, still exhibits coherent structures, although fine added oscillations in the spatial pattern start to manifest, indicating that this mode contributes to finer-scale structures in the overall system. Fourth and fifth modes, at the second and third row of the figure, show that their spatial patterns become noisier and less structured, which is a clear effect of the low variance that these modes possess in the system individually. These modes mostly capture smaller fluctuations rather than the dominant dynamics. As we move to higher-order modes, as shown from the sixth to tenth mode, the effect of the low variance that these modes carry becomes more pronounced as the coherent structures fully distort and disappear. This analysis showcases how the chosen low-rank truncation of five modes has the potential to achieve an effective ROM, as the leading low-order modes successfully represent the main and organized behavior of the system, while the higher-order modes act more like corrections of small-scale dynamics in the system.

\subsubsection{POD--SINDy framework performance}

Preliminary analysis of the cumulative energy and spatial characteristics of the modes allowed us to truncate the ROM with 5 modes. This sets the initial step to test whether the integration of SINDy methodology can allow us to obtain a symbolic model that could effectively model the mode amplitudes' dynamics consistently. 

\begin{figure}[h]
    \centering
    \includegraphics[width=0.4\linewidth]{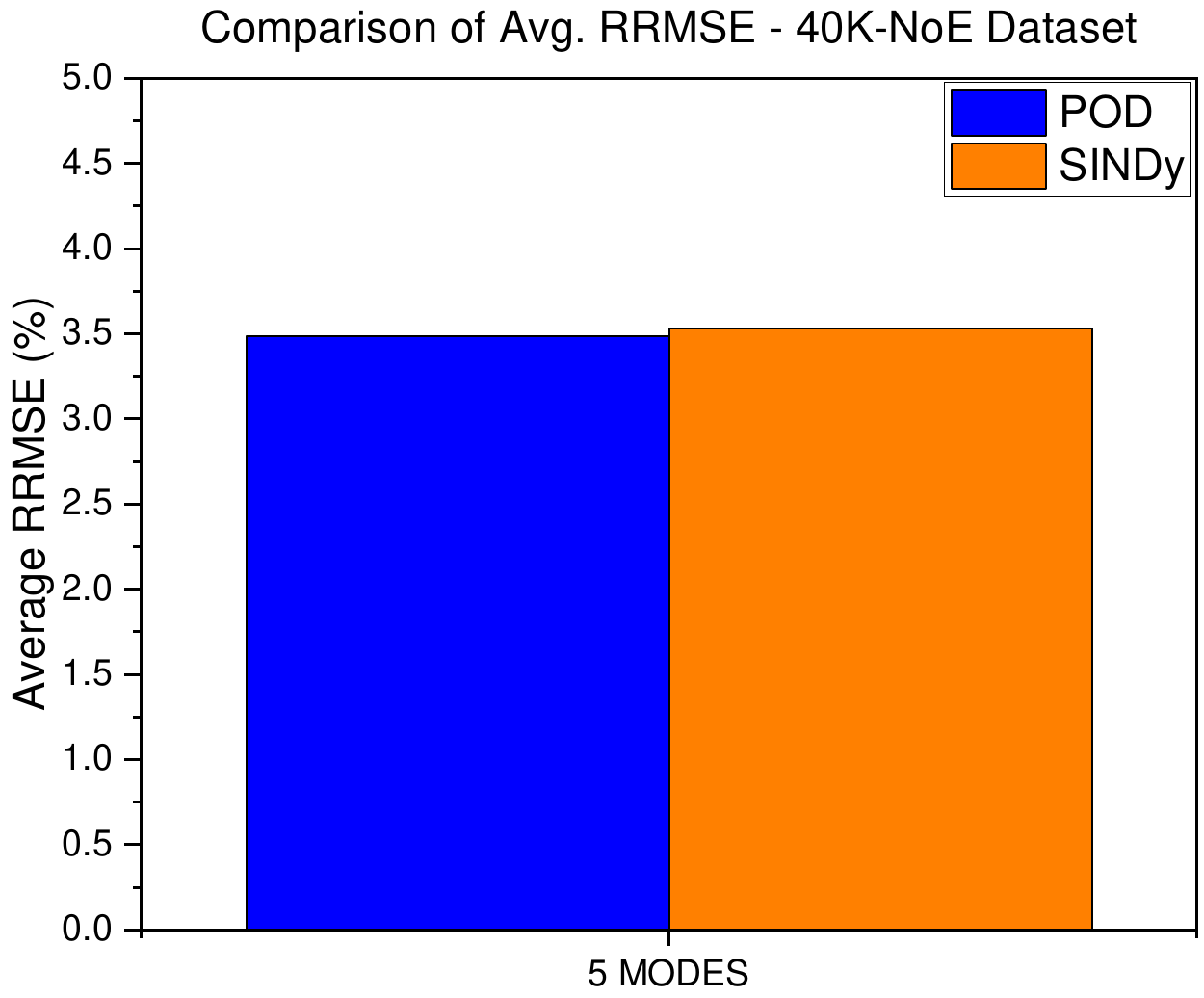}
    \caption{Comparison of average RRMSE from reconstructed snapshots by POD and SINDy-simulated amplitudes with 5 modes in the passive kinetic simulation 40K dataset.}
    \label{fig:AvgRRMSE_40kNoE}
\end{figure}

Figure \ref{fig:AvgRRMSE_40kNoE} shows the comparison of the average RRMSE calculated from the reconstructed snapshots employing POD data and SINDy-simulated amplitude data, relative to the original snapshots from the passive kinetic simulation  40K dataset. This allows us to evaluate and compare the reconstruction performance of the POD-derived ROM and the POD--SINDy framework.

We can appreciate that with the selected truncation of just five modes, both approaches are able to achieve very low average errors of approximately 3.5\%. It is important to notice that not only the reconstruction errors are small but also that both POD and POD--SINDy approach deliver nearly identical results. The reason for this strong agreement lies in the nature of the passive kinetic simulation case, in which the dynamics are governed by linear advection without nonlinear electric field coupling. This makes the system highly suitable for low-rank approximations and sparse regression models.

\begin{figure}[h]
    \centering
    \includegraphics[width=\linewidth]{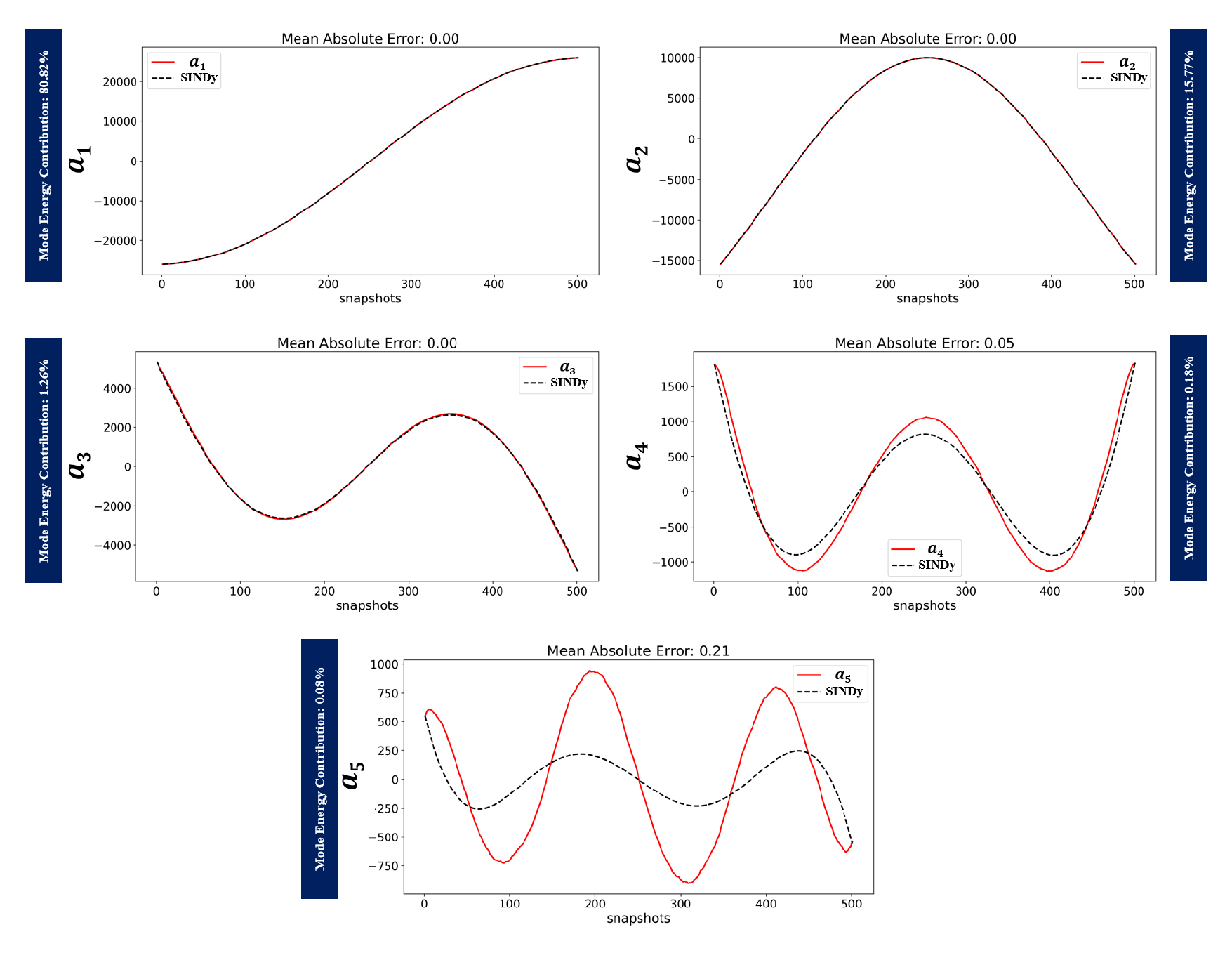}
    \caption{Comparison between original and SINDy-simulated amplitudes of the selected modes.}
    \label{fig:SINDY_40kNoE}
\end{figure}

These results suggest that the compact reduced-modal representation is already sufficient to describe closely the passive kinematic dataset, as SINDy can reproduce the same dynamics without needing to evolve the full particle system. This becomes evident when we observe the results from the comparison between original POD amplitudes and SINDy simulated amplitudes, as shown in figure \ref{fig:SINDY_40kNoE}. The mean absolute error is employed as evaluation metric for comparison between POD original and SINDy simulated amplitudes:

\[MAE = \frac{1}{N}\sum^{N}_{i=1}|y_i-\hat{y}_i|.\]
In the plots we can appreciate how the obtained models for the selected modes amplitudes closely match original POD data. In fact, the simulated amplitudes for the first three low-rank dominating modes match perfectly up the two decimal places their respective POD amplitude data. As the mode rank increases, we can observe how the error between the simulated SINDy amplitude and original POD amplitude starts to increase. Amplitude of mode four has an increased error of 0.05 while closely following the original amplitude dynamics. For the fifth mode the error becomes more pronounced as it rises to 0.21 and the simulated SINDy amplitude fails to match the original amplitude data. However, it is important to notice that this increased error has little influence on the overall reconstruction of the original kinematic data as the fifth mode only contributes with 0.08\% of the total energy of the ROM system. These results showcase the effectiveness of SINDy on capturing the dynamics of the original amplitude from the dominant modes, allowing to closely match POD performance in reconstruction capability for the passive kinematic dataset.

Nevertheless, the ability to closely match the original amplitude data is not the only quality that SINDy is able to provide to the joint POD--SINDy framework. SINDy is able to lift the veil from the black-box of the unknown ROM dynamics by providing a set of ODE that give light to a symbolic representation to the mode's amplitudes, described as follow: 

\[\dot{a}_1 = 103.68 + 0.001(6a_2-a_4)\]
\[\dot{a}_2 = 0.001(13-6a_1-6a_3)\]
\[\dot{a}_3 = 24.41 + 0.001(6a_2+9a_4)\]
\[\dot{a}_4 = 0.001(42+a_1-9a_3+9a_5)\]
\[\dot{a}_5 = -2.34 + 0.001(a_2-10a_4)\]

Above equations provide a compact system of ODEs which are identified for the first five modes in the passive kinetic simulation dataset. These discovered equations provide insight of the coupled relationship of the interaction between mode amplitudes while remaining linear and sparse, meaning that only few linear terms are necessary to describe the underlying dynamics where only the essential coupling between modes remain. 

In the passive kinematic dataset, the dominant interactions appear as linear terms with only minor cross-mode contributions, a result consistent with the underlying Vlasov advection, which is linear in nature and free of nonlinear feedback. It is important to notice that SINDy does not force nonlinearity as it simply identifies the simplest governing structure present in the amplitude data, which for this system is largely linear and sparse. This highlights the ability of SINDy to not only capturing the dynamics of dominant modes but also provide interpretability of such dynamics through a set of sparse ODEs. 

\begin{figure}[p] 
    \centering 
    
    \begin{tabular}{ | c | c | c |} 
        
        \hline
        \textbf{Original Data} & \textbf{POD Reconstructed} & \textbf{POD--SINDy Reconstructed} \\
        \hline 
        \rule{0pt}{15pt} 
        
        \includegraphics[width=0.3\textwidth]{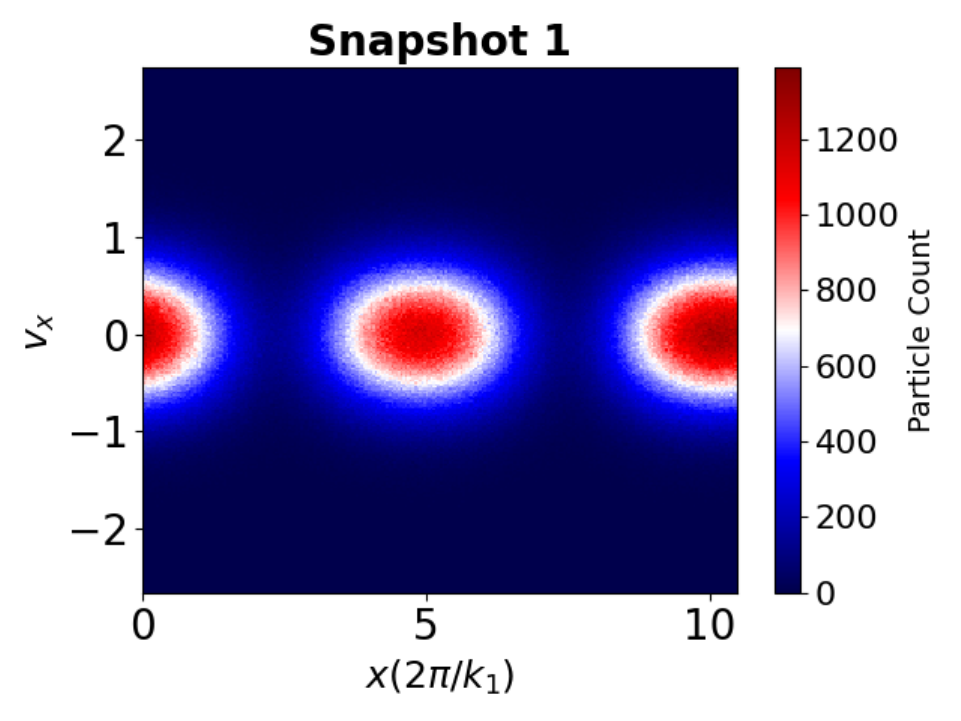} & 
        \includegraphics[width=0.3\textwidth]{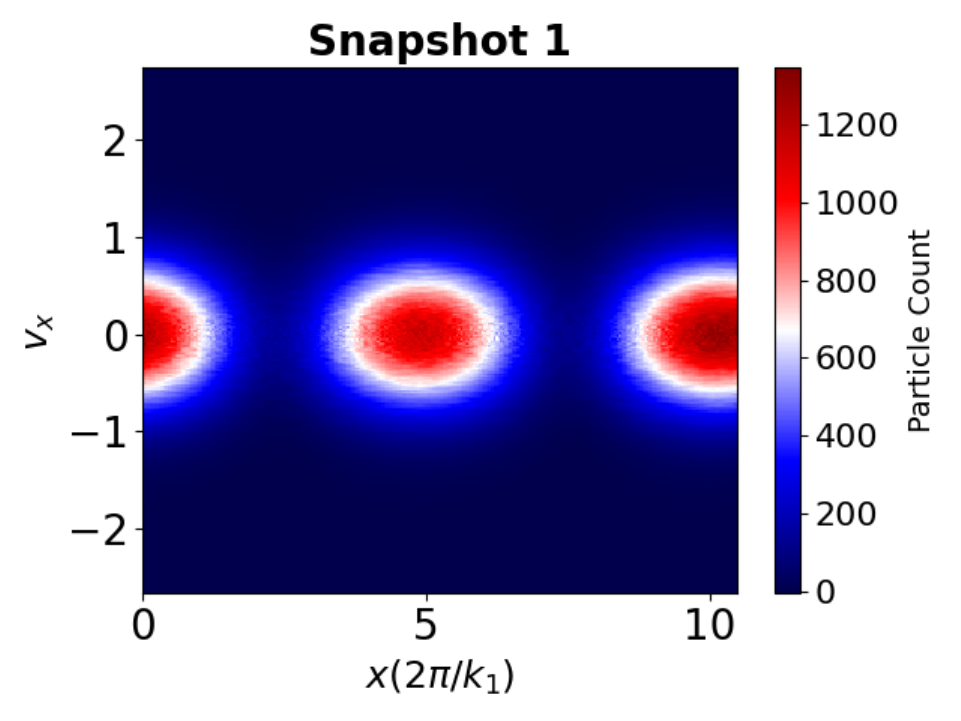} & 
        \includegraphics[width=0.3\textwidth]{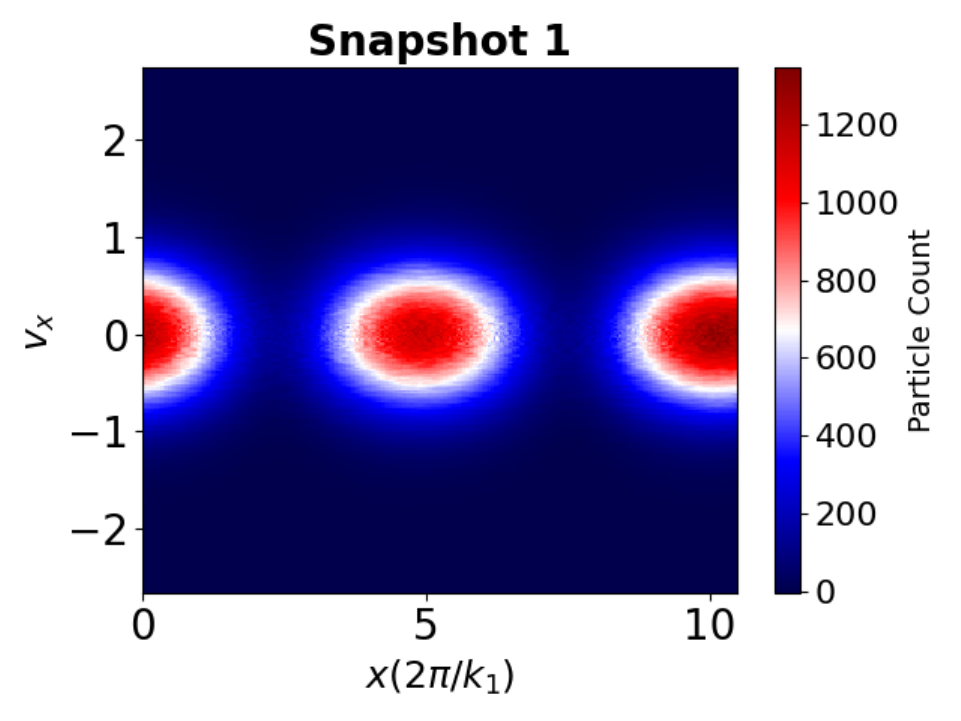} \\
        
        \includegraphics[width=0.3\textwidth]{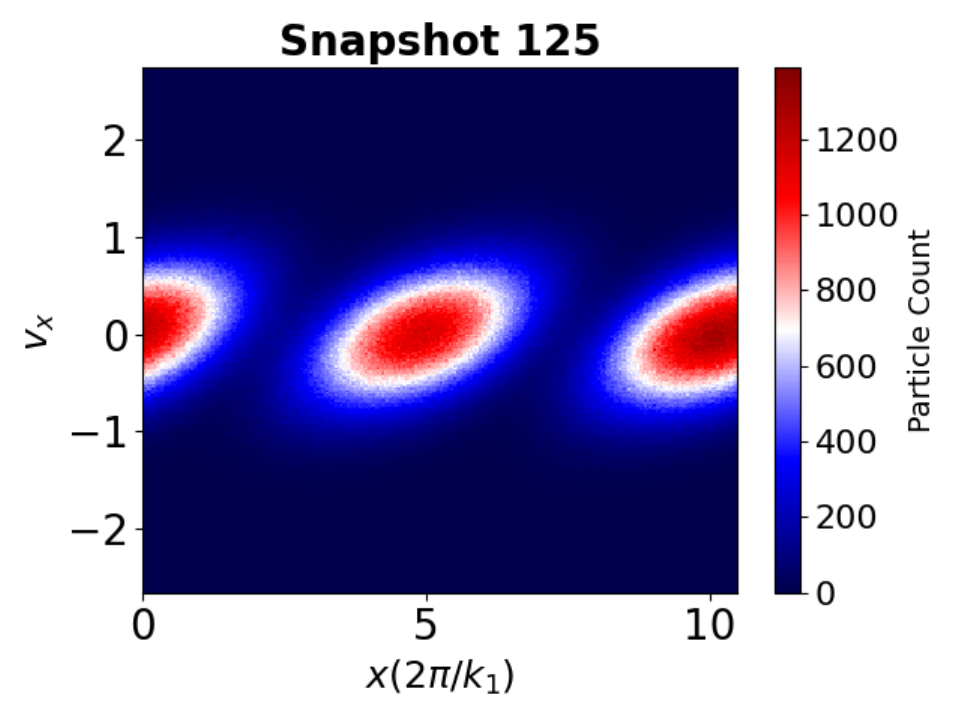} & 
        \includegraphics[width=0.3\textwidth]{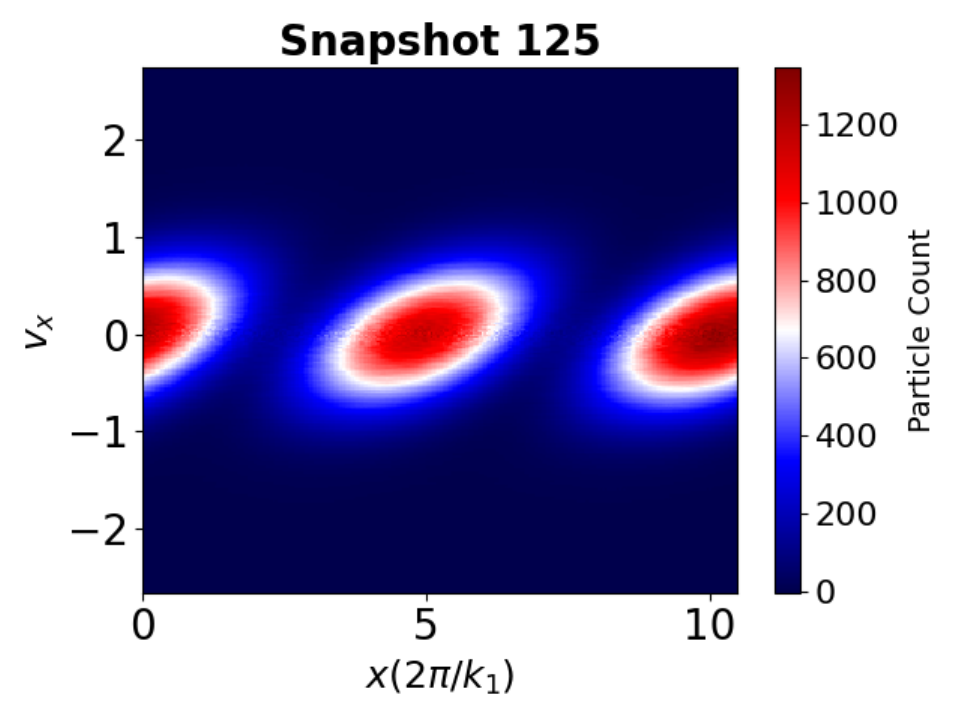} & 
        \includegraphics[width=0.3\textwidth]{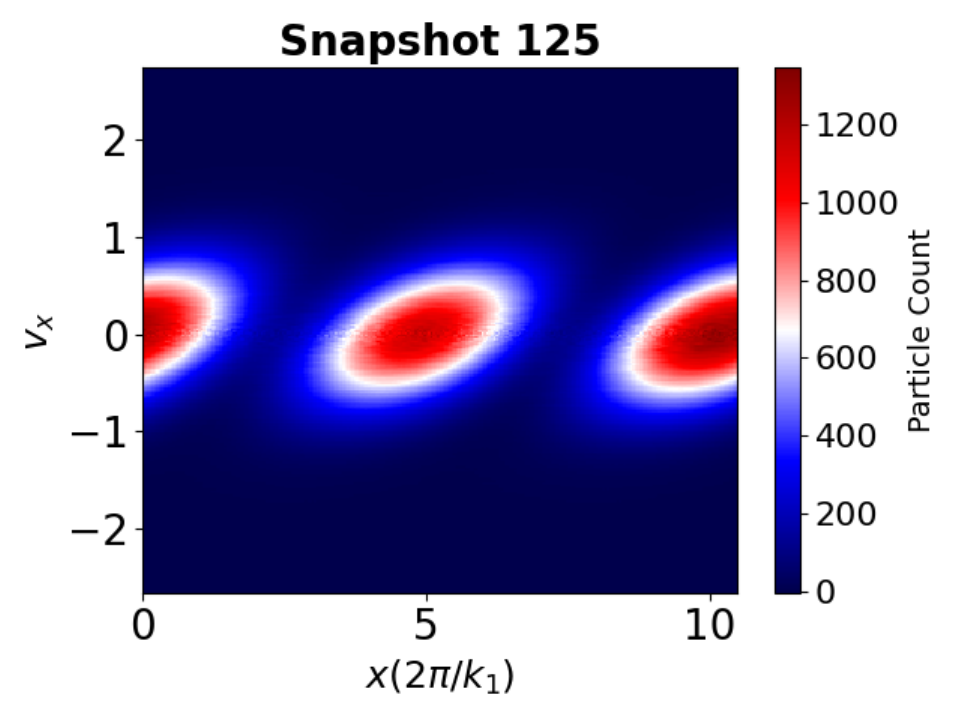} \\

        \includegraphics[width=0.3\textwidth]{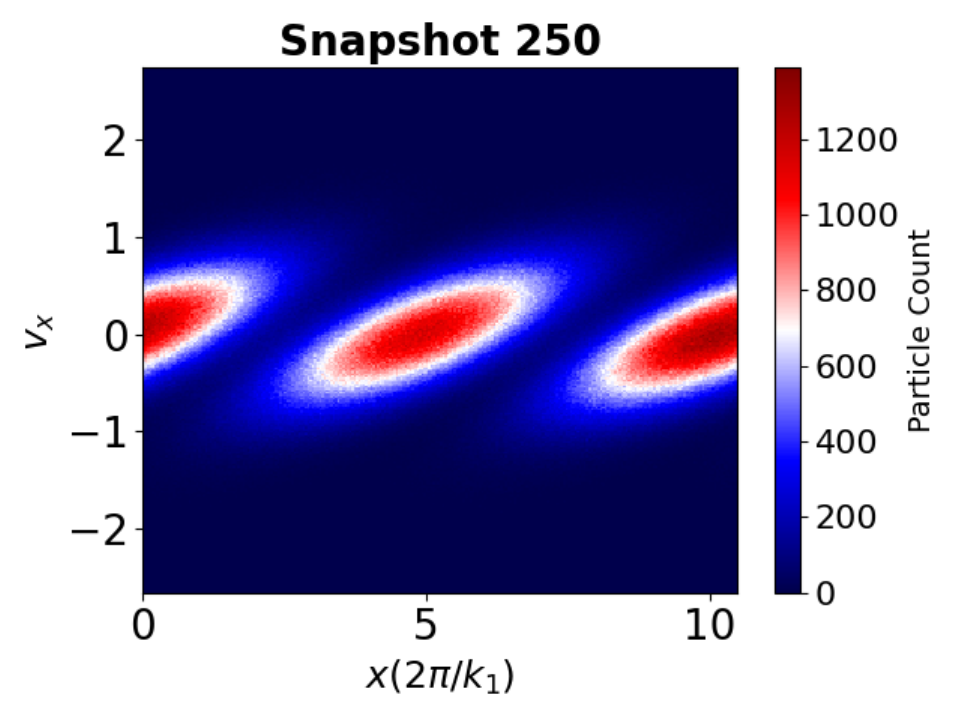} & 
        \includegraphics[width=0.3\textwidth]{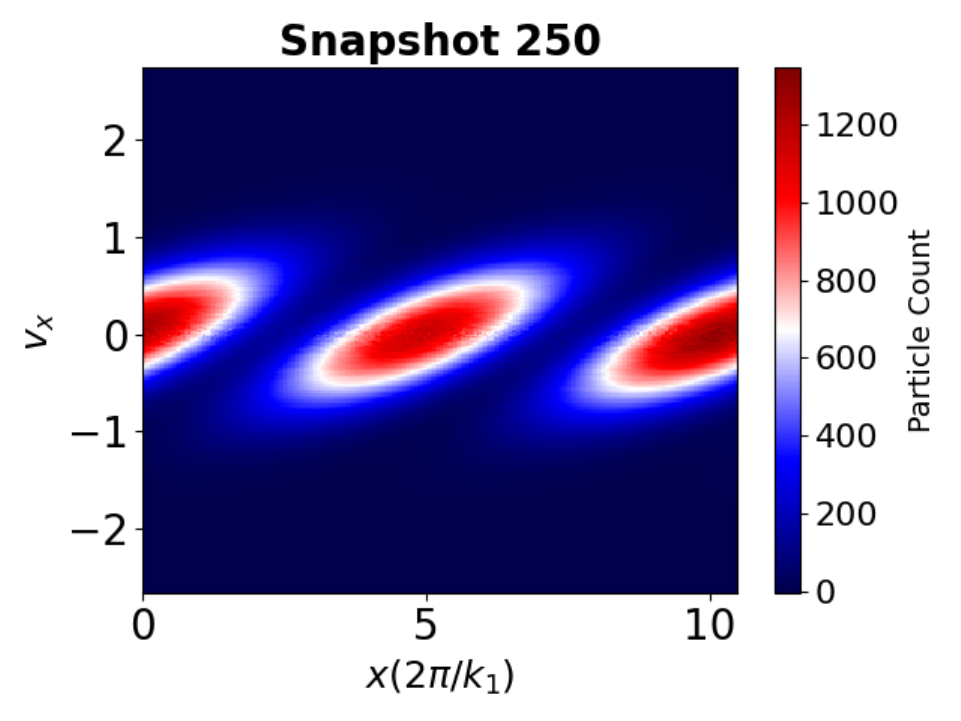} & 
        \includegraphics[width=0.3\textwidth]{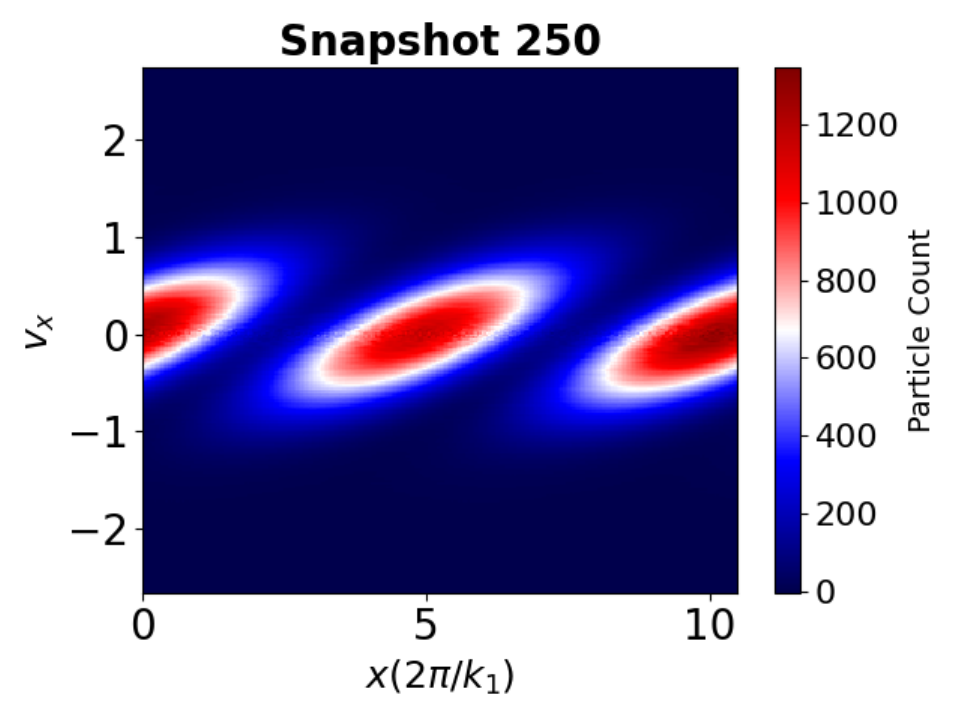} \\

        \includegraphics[width=0.3\textwidth]{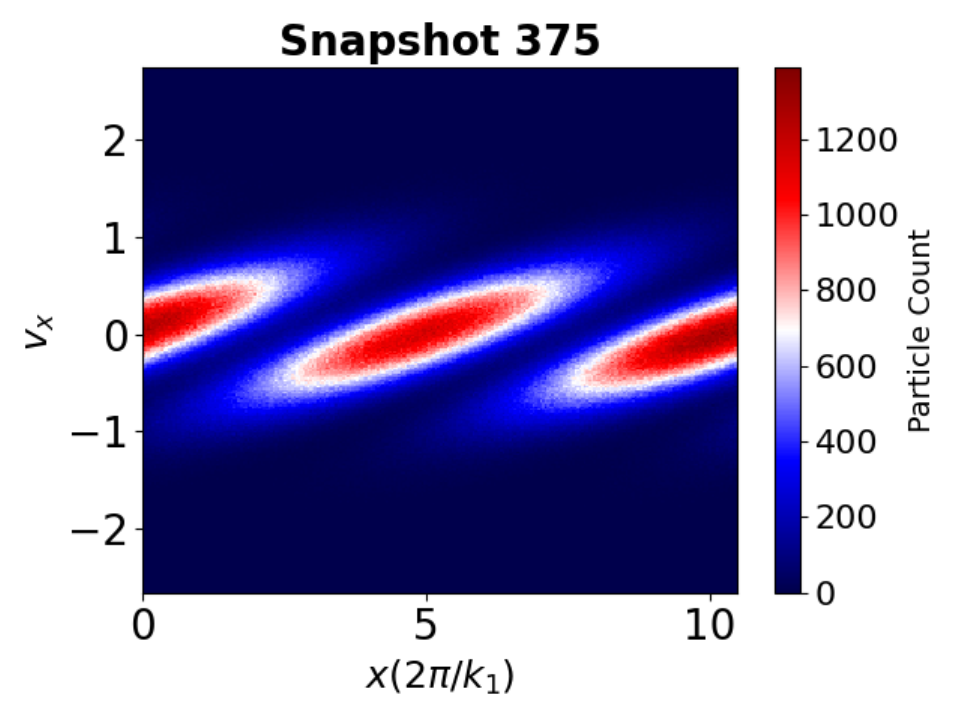} & 
        \includegraphics[width=0.3\textwidth]{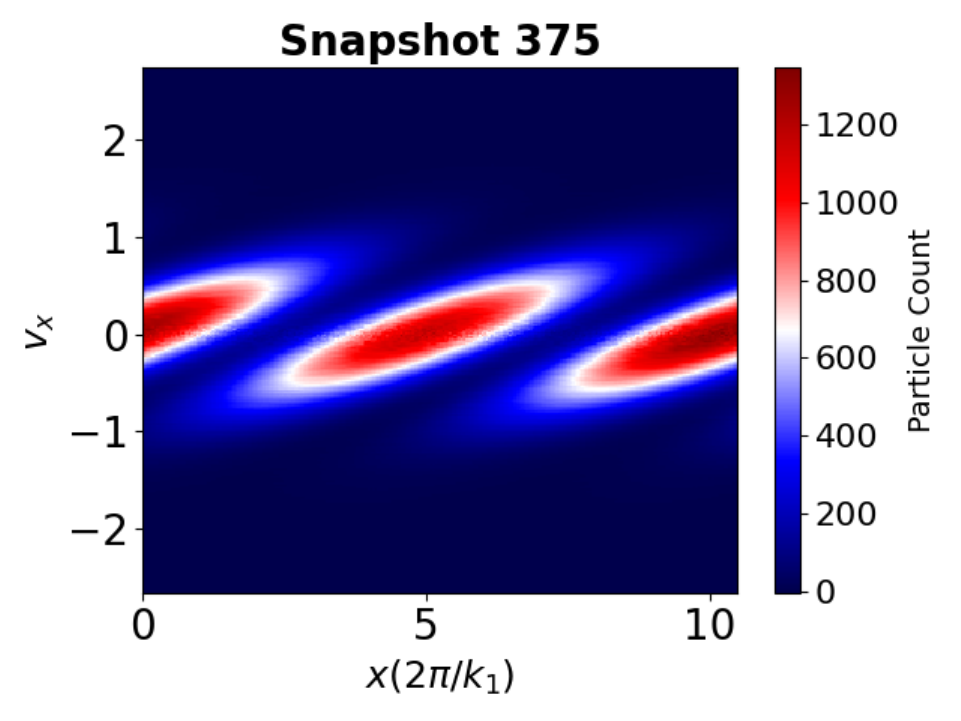} & 
        \includegraphics[width=0.3\textwidth]{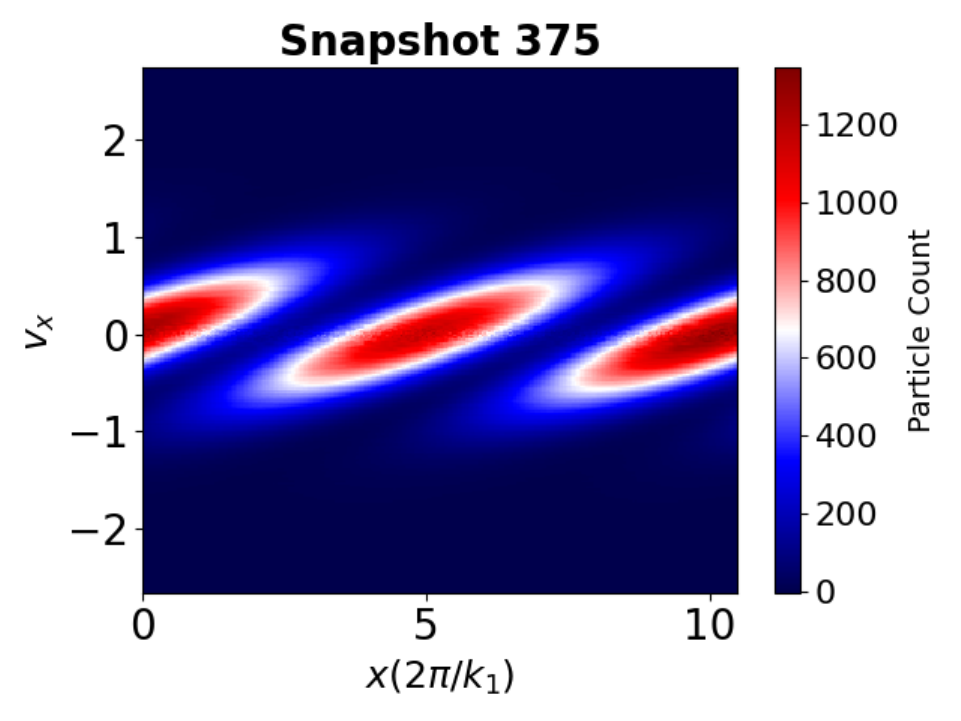} \\

        \includegraphics[width=0.3\textwidth]{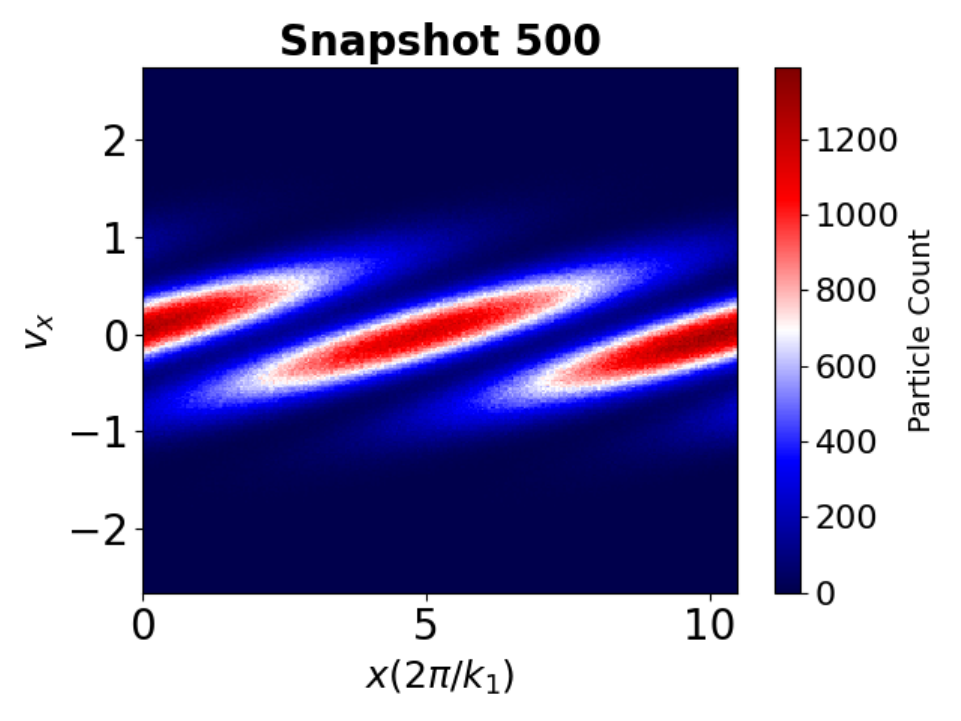} & 
        \includegraphics[width=0.3\textwidth]{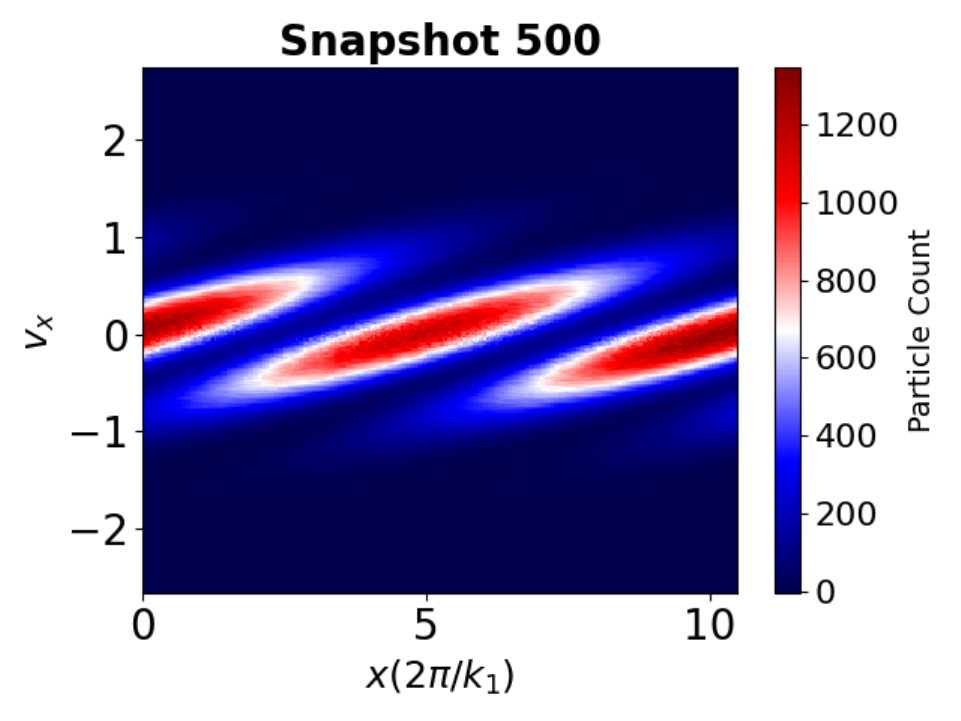} & 
        \includegraphics[width=0.3\textwidth]{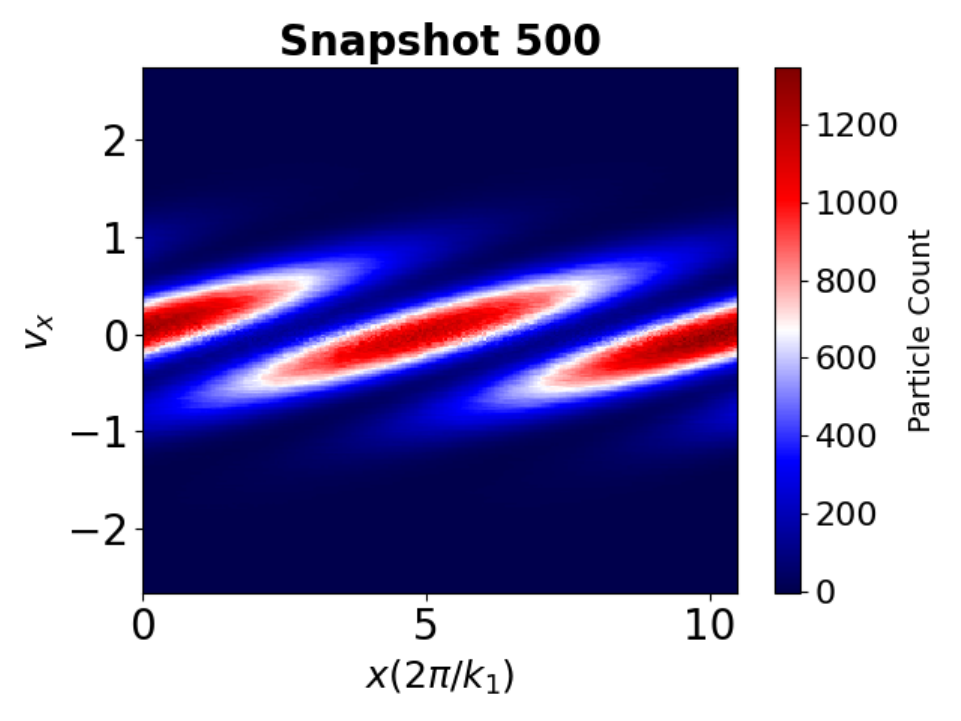} \\
    \hline
    \end{tabular}
    
    \caption{Comparison of phase mixing process with original data (left), POD reconstructed data (center) and POD--SINDy reconstructed data (right) for passive kinetic simulation 40K dataset.}
    \label{fig:Snapshots_40kNoE}  

\end{figure}

Finally, we can appreciate the performance of the POD--SINDy framework in the case of the simple passive kinetic simulation 40K dataset, by evaluating the reconstruction of phase space snapshots and comparing them with POD reconstructed data with five modes, and original phase space data. As shown in figure \ref{fig:Snapshots_40kNoE}, we can observe how both POD and POD--SINDy reconstructed phase space data closely follow the original evolution of the phase mixing process, capturing efficiently the early stages of filamentation and spreading in phase space with remarkable accuracy in the snapshot reconstruction in which errors are below 4\%. This confirms the capability of the POD--SINDy framework of providing a reduced representation that can preserve the essential physics without needing the full particle dataset in the case of the linear passive kinetic simulation dataset, which established a benchmark where the obtained ROM achieves a near-optimal efficiency.

\subsection{ROM performance on self-consistent electrostatic kinetic simulations}

Previously we analyzed the performance of the POD--SINDy framework in the case of the simplified passive kinetic simulation in which the phase mixing dynamics are advective and linear. This served as benchmark or starting point for further evaluation of the methodology. In this section, we will expose the ROM framework to more challenging datasets in which self-consistent electric field and higher particle noise are included. The inclusion of such self-consistent electric field fundamentally enriches the dynamics of the phase-mixing process by introducing nonlinear interactions among the particles, which translates into an increasing level of complexity to the task of reduced--order modeling. 

In this section we will observe how a paradox emerges at this new regime: even though SINDy performance appears to reproduce modal amplitude dynamics with notable accuracy, the overall reconstruction error of phase space data is significantly higher than in the passive kinetic simulation benchmark. To understand this apparent contradiction, we will begin the analysis by examining the reconstruction error themselves and then trace their origin through analyses of POD energy capture, modal structures, and the dynamical equations uncovered by SINDy.

\subsubsection{POD--SINDy reconstruction errors}

\begin{figure}[h]
    
    \begin{tabular}{| c c |} 
        
        \hline
        \textbf{40K-E Dataset} & \textbf{5K-E Dataset} \\
        \hline
        \rule{0pt}{15pt} 
        
        \includegraphics[width=0.47\textwidth]{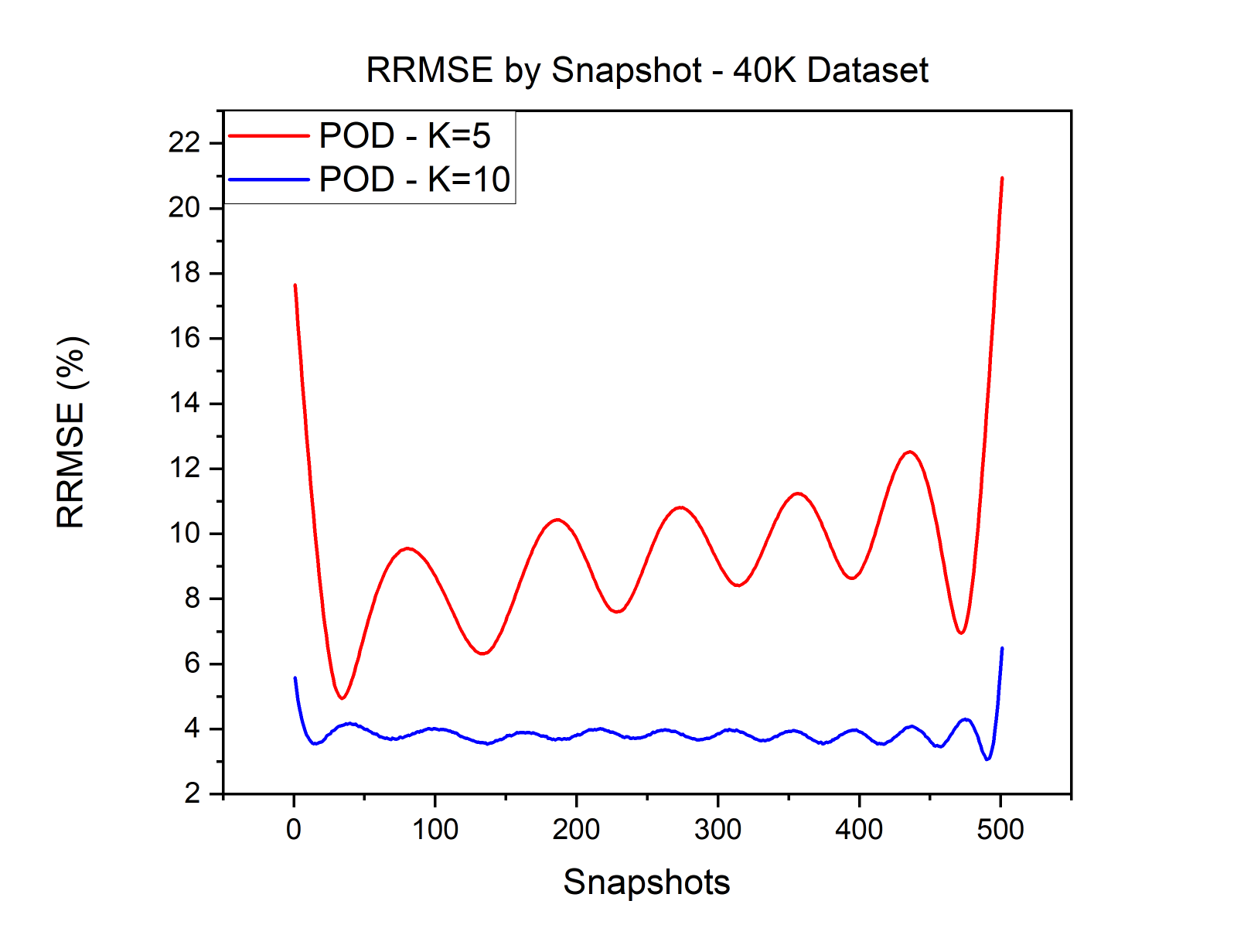} & 
        \includegraphics[width=0.47\textwidth]{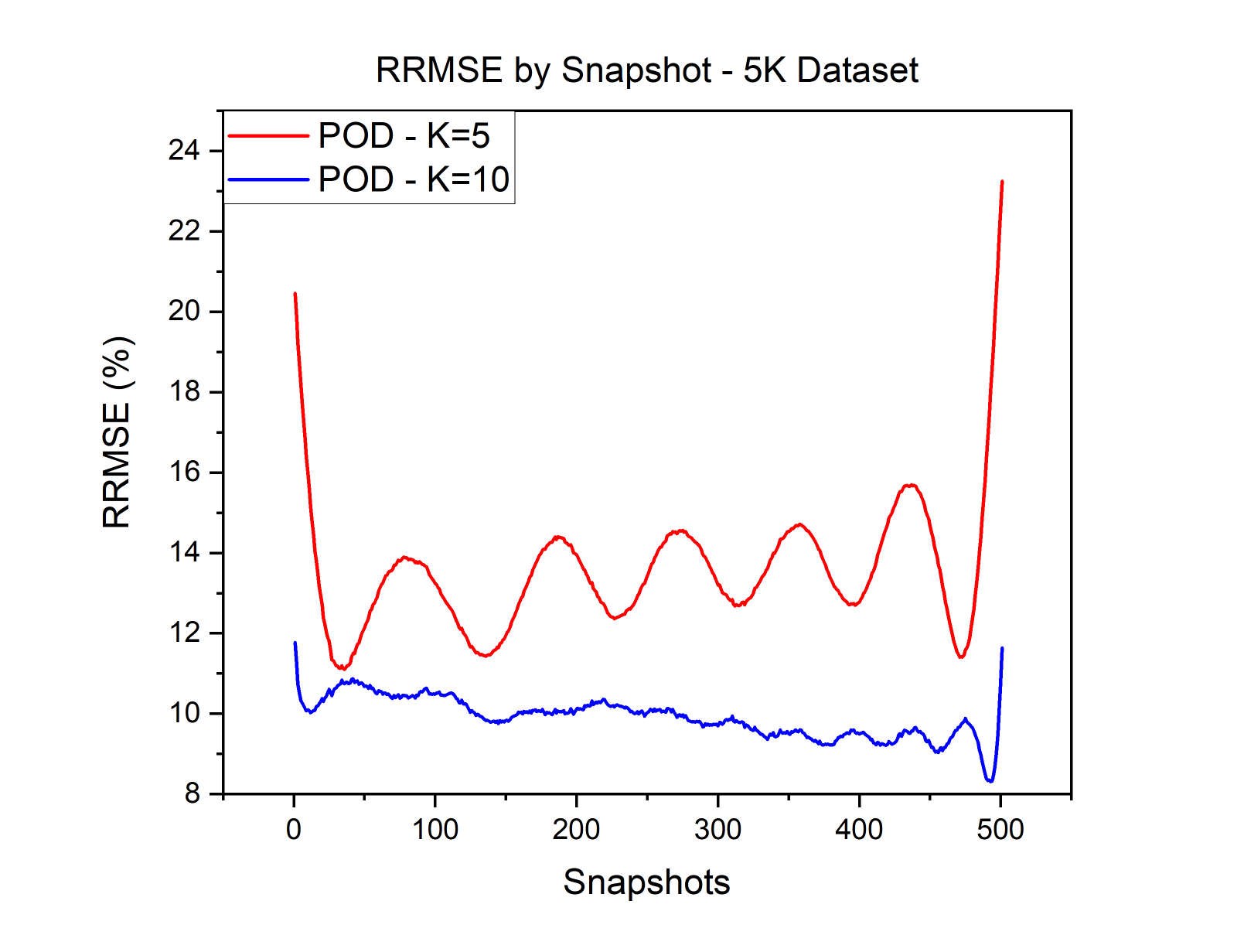} \\
        \hline        
    \end{tabular}
    
    \caption{Comparison of the evolution of the RRMSE by snapshot for reconstruction with five modes (red) and ten modes (blue) in self-consistent electrostatic kinetic simulation: (left) 40K-E dataset, (right) 5K-E dataset.}
    \label{fig:RRMSE_Snapshot}  

\end{figure}

Figure \ref{fig:RRMSE_Snapshot} shows the evolution of the RRMSE per snapshot with the selection of five modes (similar to benchmark dataset) and ten modes for truncation of the generated POD ROM. In the left figure we can appreciate the results for the low noise 40K-E dataset, while the right side exhibit the behavior of the RRMSE evolution for the high-noise 5K-E dataset. 

We can observe how with a truncation of five modes, similar to the benchmark scenario of passive kinetic  simulation dataset, the reconstruction error shows a dramatic fluctuation across the snapshots with increased baseline for the noisy 5K-E dataset compared to the 40K-E dataset. Once the mode truncation is doubled to ten modes, we can appreciate a dramatic reduction of the RRMSE fluctuation as well as a notorious reduction of the general baseline of the errors in both 40K-E and 5K-E dataset which drop significantly around 4\% and 10\% respectively. This gives a preliminary insight of the capability of POD technique to mitigate the dramatic effects of noise. Moreover, it is demonstrated that using ten modes as the reconstruction for the 40K-E dataset yields error levels comparable to those of the 40K-NoE dataset. Consequently, we adopt this truncation as the basis for evaluating the performance on the 5K-E dataset as well.

However, the RRMSE evolution exhibits clear peaks at the edges of the datasets in both truncation cases. We can notice how the amplitude of these peaks decrease as higher order modes are included in the truncation. This is a natural effect of POD as it tends to represent more accurately the average dynamics of the system but might struggle at the edges of the datasets, for which it requires the inclusion of higher order modes that will act as localized correction terms at those time steps. 

\begin{figure}[h]
    \centering 

    \begin{tabular}{ | c  c |} 
        
        \hline
        \textbf{40K-E Dataset} & \textbf{5K-E Dataset} \\
        \hline
        \rule{0pt}{15pt} 
        
        \includegraphics[width=0.48\textwidth]{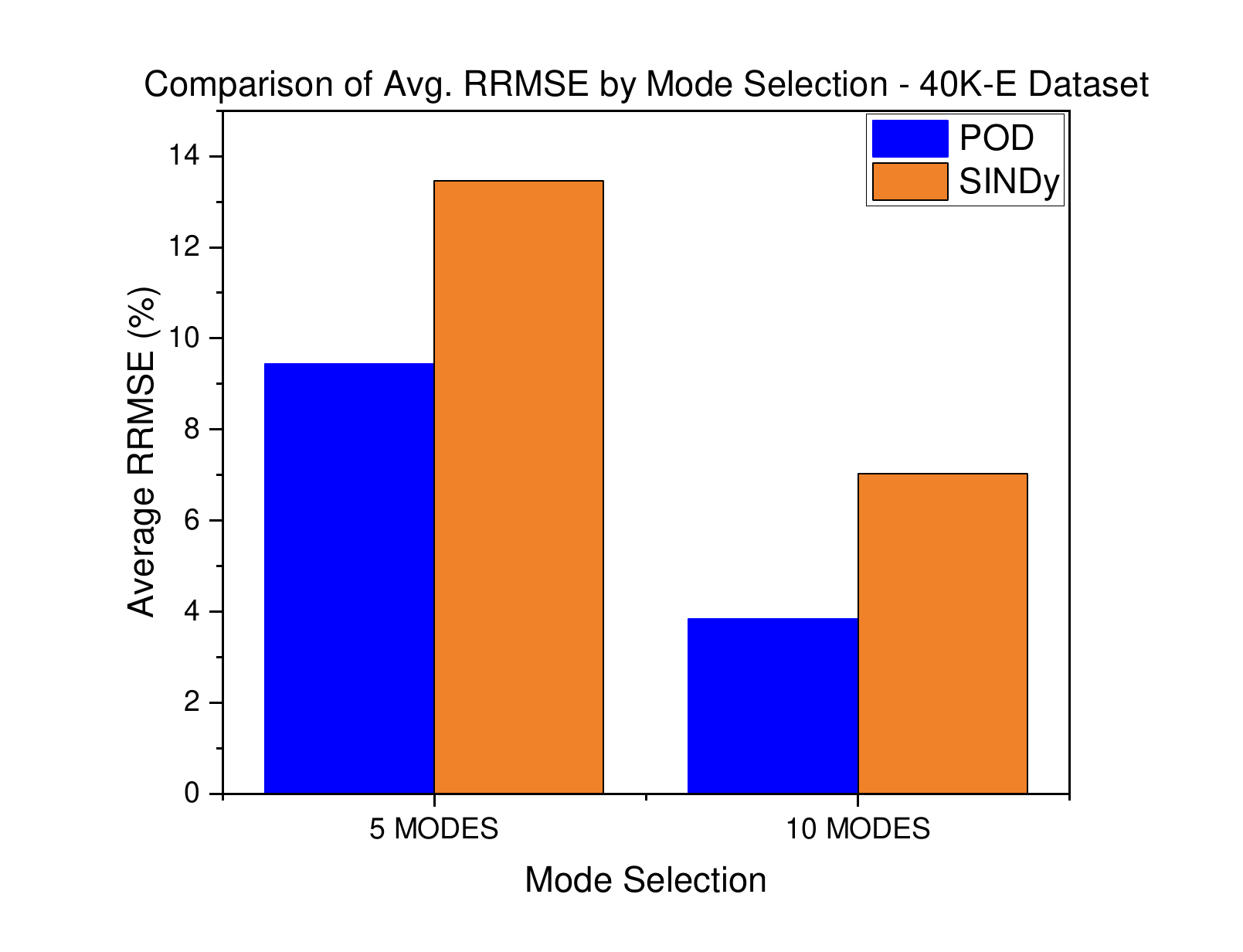} & 
        \includegraphics[width=0.48\textwidth]{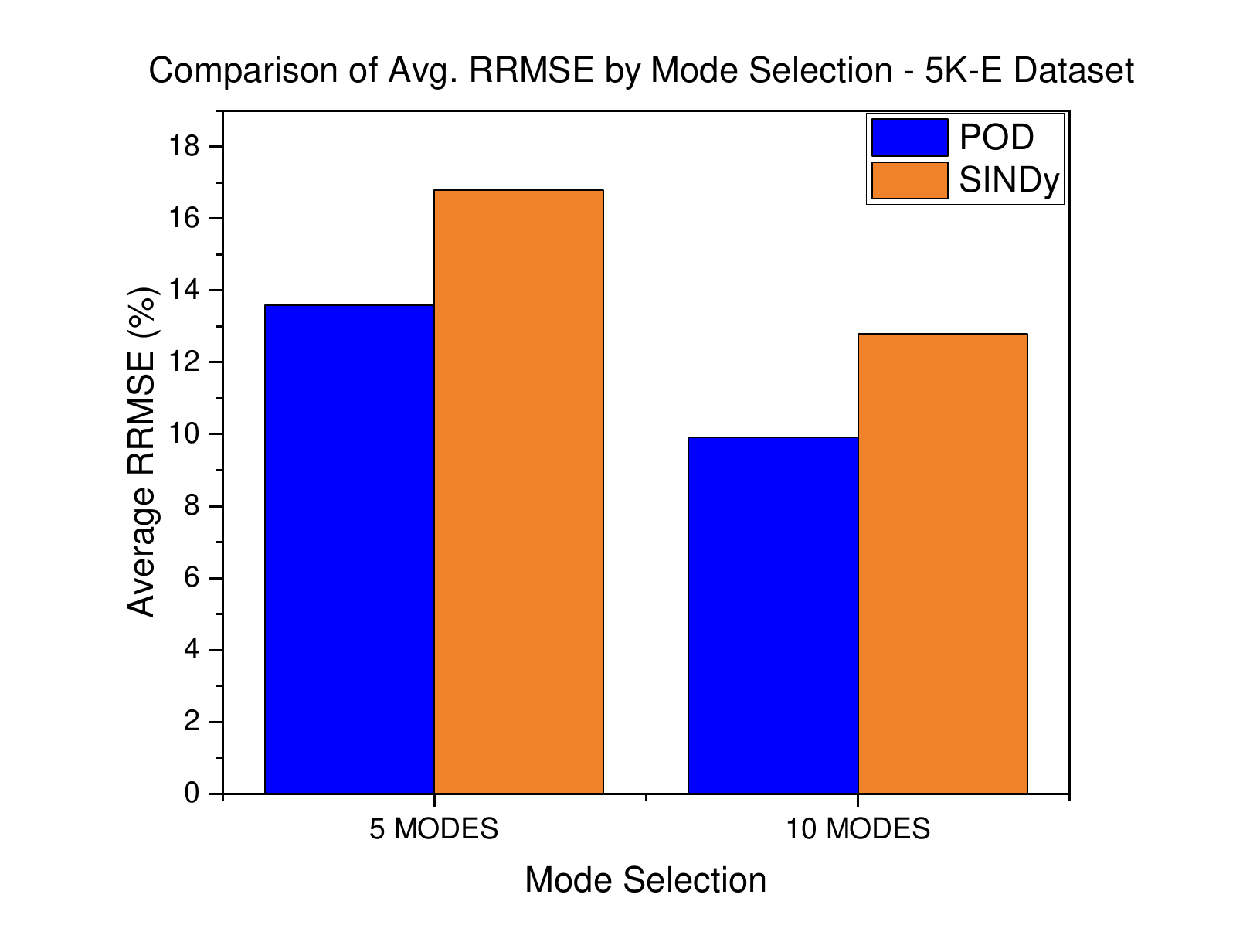} \\
        \hline         
    \end{tabular}
    
    \caption{Comparison of average RRMSE from reconstructed snapshots by POD (blue) and SINDy simulated data (orange) with five and ten modes in self-consistent electrostatic kinetic simulation: (left) 40K-E dataset, (right) 5K-E dataset.}
    \label{fig:AvgRRMSE_E}  

\end{figure}

Once the overall performance of the POD technique has been evaluated, we can proceed with the assessment of the average RRMSE between POD and SINDy simulated data relative to the reconstruction of the original phase space snapshots. In figure \ref{fig:AvgRRMSE_E} we can appreciate the comparison of the average RRMSE for POD--reconstructed data and SINDy--simulated reconstructed data in the 40K-E dataset (left) and the noisy 5K-E dataset (right). 

For the 40K-E dataset, we can observe how with only five modes the POD already gives us an average reconstruction error below 10\%, while SINDy is noticeably higher at around 13\%. However, as soon as ten modes are considered for truncation, both methods improve significantly allowing POD to drop the average reconstruction error closer to the benchmark dataset value below 4\%, while SINDy follows same tendency reducing to about 7\%. Similarly, for the noisy 5K-E dataset, the inclusion of higher order modes with a truncation of ten modes reduces the average RRMSE in both POD and SINDy cases, indicating that the achieved ROM is still able to capture a significant part of the dynamics despite the noise present in the dataset. It is important to notice at this point that reconstruction from SINDy simulated data cannot show better performance than POD reconstruction case, as SINDy results are obtained based on POD itself.

This overall increment of the average reconstruction error is then expected due to the introduction of more complex dynamics in the phase mixing process as well as the consideration of increased statistical noise by reduction of the number of particles per cell. Nevertheless, these results suggests that although SINDy struggles more with the nonlinear E-field dynamics compared to the simpler passive kinetic simulation case, it is still able to follow same trend as POD considering that the accuracy improves with the inclusion of higher order modes in the set of equations. 

\subsubsection{SINDy amplitude dynamics and equations}

Based on previous evaluation of the close behavior of SINDy reconstructed data with its homologous POD counterpart, this might reflect a potential success of SINDy on capturing the dominant dynamics of the low-order modes in the considered ten-modes truncation.

\begin{figure}[p]
    \centering
    \includegraphics[width=\linewidth]{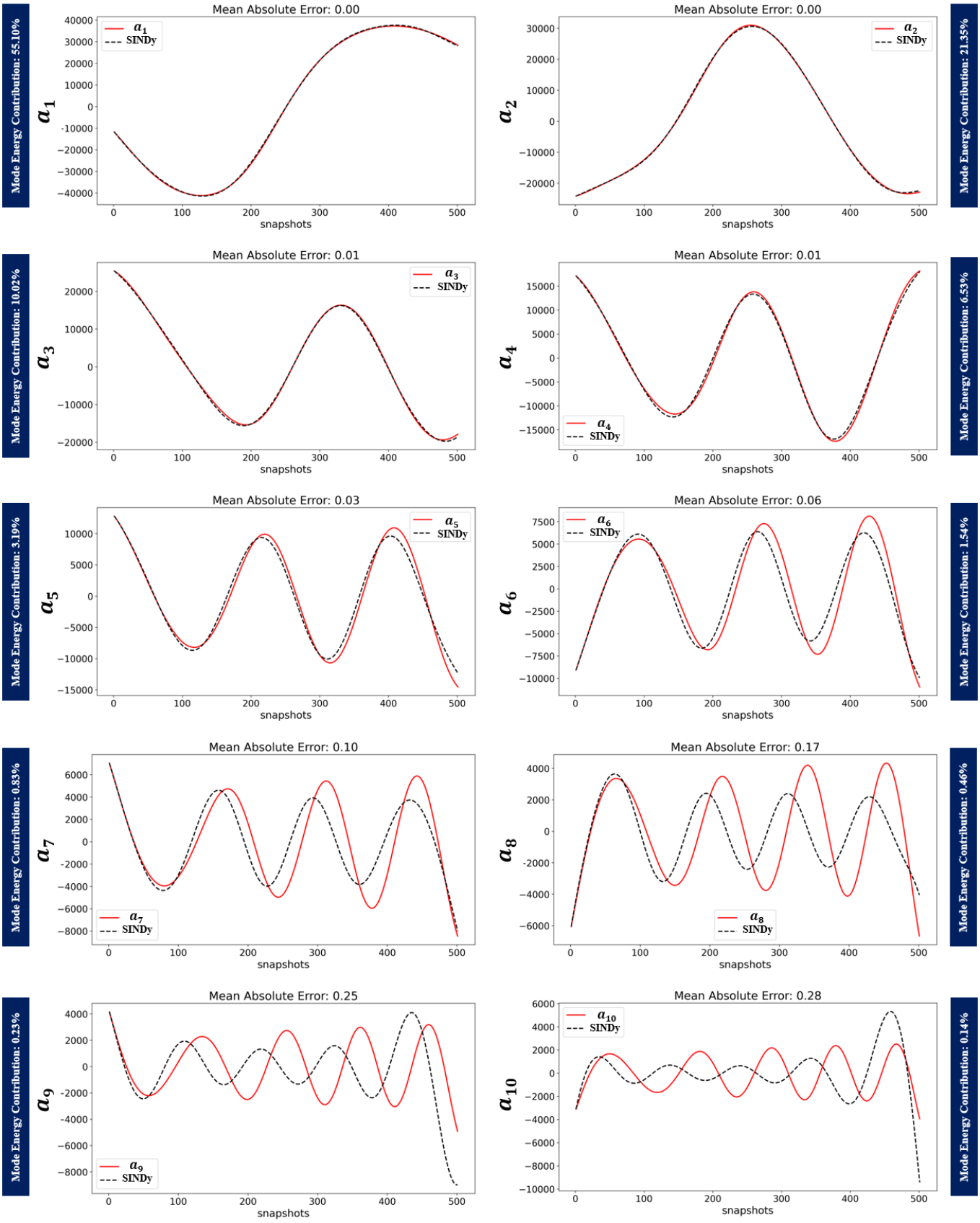}
    \caption{Comparison between original and SINDy simulated amplitudes of the selected modes in self-consistent electrostatic kinetic simulation 40K dataset.}
    \label{fig:SINDY_40kE}
\end{figure}

Figure \ref{fig:SINDY_40kE} shows the comparison between original POD and SINDy simulated amplitudes for the achieved ten modes ROM system in the case of the low-noise 40K-E dataset. As we can see in the plots, the SINDy simulated amplitudes closely follow the overall behavior of the original amplitude data. At the low-order modes, more specifically modes one to six, the error remains below 0.03 following almost exact dynamics as original data. Evaluation of higher order modes indicates they still manage to retrieve similar dynamics from the original amplitude data, with a slow increment of the error from 0.06 in the sixth mode to 0.28 in the tenth mode.

\begin{figure}[p]
    \centering
    \includegraphics[width=\linewidth]{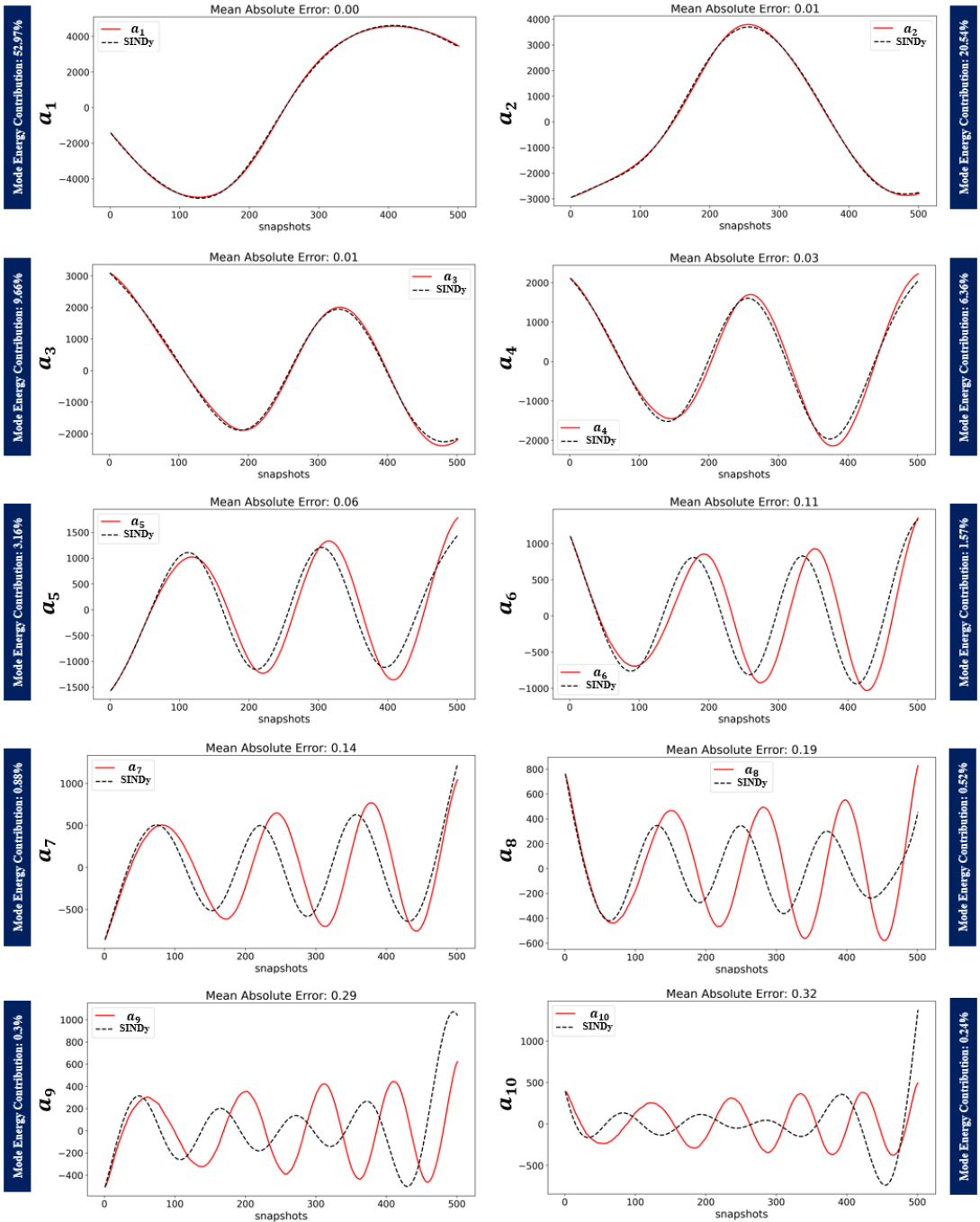}
    \caption{Comparison between original and SINDy simulated amplitudes of the selected modes in self-consistent electrostatic kinetic simulation 5K dataset.}
    \label{fig:SINDY_5kE}
\end{figure}

On the other hand, figure \ref{fig:SINDY_5kE} shows the same comparison between original POD and SINDy simulated amplitudes for the achieved ten modes ROM system in the case of the high-noise 5K-E dataset. It can be appreciated from the plots that they follow similar tendency as their low-noise 40K counterpart, since the SINDy simulated amplitudes for low-order modes successfully achieve low error (below 0.06) and follow similar dynamics with the original data. By analyzing high-order modes we can observe that the increment of error in the simulated amplitude becomes slightly steeper, an effect that is expected as noise is increased in the phase space data. However, SINDy is still able to capture the tendency of the amplitude evolution.

This evaluation of SINDy performance on capturing POD mode's amplitude dynamics in both the low- and high-noise self-consistent electrostatic dataset reflect strong similarities with the benchmark passive kinetic simulation dataset:

\begin{itemize}
    \item Low errors in dominant dynamics: SINDy was able to efficiently capture the amplitude dynamic of the low-order dominant modes, in both the low- and high-noise datasets, indicating a reliable performance in the POD--SINDy framework as ROM.

    \item Negligible influence of high order modes: increased error in high order modes is more notorious in both the 40K-E and 5K-E datasets. However, similarly to the benchmark 40K-NoE dataset, these higher order modes possess significant lower cumulative energy which diminish their influence in the major structures of the phase mixing dynamics.

    \item Accurate capture of amplitude dynamics: in both cases, the achieved SINDy amplitudes were able to follow the overall tendency of the original amplitude data, especially in low-order modes which slowly was diminished as higher order modes were considered.
    
\end{itemize}

Evaluation of achieved equations also show similar structure than the benchmark passive kinetic simulation dataset. The following equations represent the achieved ODEs of the first five mode amplitudes when ten modes are used for truncation:

\begin{itemize}
    
    \item 40K-E dataset:
    
    \[\dot{a}_1 = 79.41 + 0.001(a_1+14a_2-6a_3-a_4+2a_5+a_6-a_7+a_8+3a_9-3a_{10})\]
    \[\dot{a}_2 = 2.74 + 0.001(-7a_1-7a_3+4a_4+2a_5-7a_6+4a_7-3a_8-a_9)\]
    \[\dot{a}_3 = -86.45 + 0.001(a_1+9a_2-2a_3+13a_4-5a_5-2a_6-a_7-3a_8+2a_9-2a_{10})\]
    \[\dot{a}_4 = 0.001(2a_1-a_2-17a_3+17a_5-a_6+4a_7+a_8+2a_9-2a_{10})\]
    \[\dot{a}_5 = -54.90 + 0.001(-a_1+3a_2+a_3-17a_4+a_5-24a_6-a_7-7a_8-4a_9-a_{10})\]

    \item 5K-E dataset:

    \[\dot{a}_1 = 9.70 + 0.001(a_1+14a_2-6a_3-a_4-2a_5-a_6+a_7-a_8-a_9)\]
    \[\dot{a}_2 = 0.31 + 0.001(-7a_1-7a_3+4a_4-a_5+7a_6-4a_7+3a_8+a_9+a_{10})\]
    \[\dot{a}_3 = -10.55 + 0.001(a_1+9a_2-2a_3+13a_4+5a_5+2a_6+2a_7+3a_8)\]
    \[\dot{a}_4 = 0.22+ 0.001(2a_1-a_2-17a_3-17a_5+a_6-4a_7-a_8-2a_9+a_{10})\]
    \[\dot{a}_5 = 6.73 + 0.001(a_1-3a_2-a_3+17a_4+a_5-24a_6-7a_8-3a_9-a_{10})\]
    
\end{itemize}

From above set of ODEs we can appreciate how the general structure of the equations retain similarity with the benchmark passive kinetic simulation dataset. The increment of the number of terms in the achieved equations is a direct consequence of the inclusion of higher order modes in the evaluation process of the mode amplitudes by SINDy. Although the nonlinear polynomial library used for regression contains several hundred candidate terms, the resulting equations retain only about 8–12 active linear terms each, which corresponds to less than 5\% of the full candidate space. In this sense, the identified equations remain sparse: the dynamics of each mode are governed by a small subset of possible interactions. This sparsity is consistent with the structure of the benchmark passive kinematic dataset, where only limited mode-mode couplings dominate the evolution. Moreover, for both low- and high-noise datasets SINDy is able to generate almost similar equations to describe amplitudes dynamics. If we examine the obtained equations we can appreciate that their differences are minimal and they are mostly linked to changes in the constant terms, sign changes of high-order terms, and their inclusion in the obtained equations. This shows the potential of SINDy on capturing the mode dynamics even for datasets with increased noise levels.

It is important to highlight a notable characteristic of the identified equations. The coupling between modes remains linear, despite originating from a self-consistent electrostatic kinematic simulation dataset where nonlinearity is present. This outcome persists even though the employed SINDy algorithm is capable of identifying nonlinear terms. 

The fact that the equations identified by SINDy for the self-consistent electric field datasets remain predominantly linear, with a structure closely resembling the passive kinetic simulation benchmark, can be traced to the mathematical and physical properties of the POD decomposition. In its mathematical foundation, POD modes form an orthogonal basis that diagonalizes the covariance of the dataset, ensuring that most of the system’s variance is projected onto a small number of independent directions. This orthogonality allows to suppress fictitious cross-interactions in the reduced space, so that the dominant dynamics of modal amplitudes appear largely decoupled from nonlinear mode–mode couplings that exist in the full particle system. At the same time, the phase mixing process itself is fundamentally governed by linear Vlasov dynamics, with nonlinearity introduced primarily through field–particle interactions that affect the spatial distribution of the particles. Since POD concentrates coherent structures into leading modes and relegates fluctuations to higher-order components with low energy, SINDy tends to identify governing equations that are sparse and nearly linear, even in the presence of self-consistent fields and noise. Thus, the observed persistence of linear equation structure across both passive kinetic  and nonlinear cases is a natural consequence of the combined effect of the underlying linear physics of phase mixing and the orthogonality of the POD basis, while the mode's spatial distribution carry the nonlinear patterns present in the data.

Previous analysis shows SINDy proficiency on not only capturing compact set of ODEs, which are sparse and capable of describing a nonlinear system using simplified linear coupling of mode's dynamics, but it also proved efficient on reconstructing the original amplitude data, especially for the dominant low-order modes which retain most of the system variance. This give light to the question stated at the beginning of this section: increment of the average reconstruction error in the POD--SINDy methodology is rooted on the POD truncation rather than on SINDy performance, as will be shown below in section \ref{POD-ENERGY-SECTION}. 

\subsubsection{POD energy capture and spatial distribution}
\label{POD-ENERGY-SECTION}

\begin{figure}[h]
    \centering 
    
    \begin{tabular}{| c  c |} 
        
        \hline
        \textbf{40K-E Dataset} & \textbf{5K-E Dataset} \\
        \hline 
        \rule{0pt}{15pt} 
        
        \includegraphics[width=0.45\textwidth]{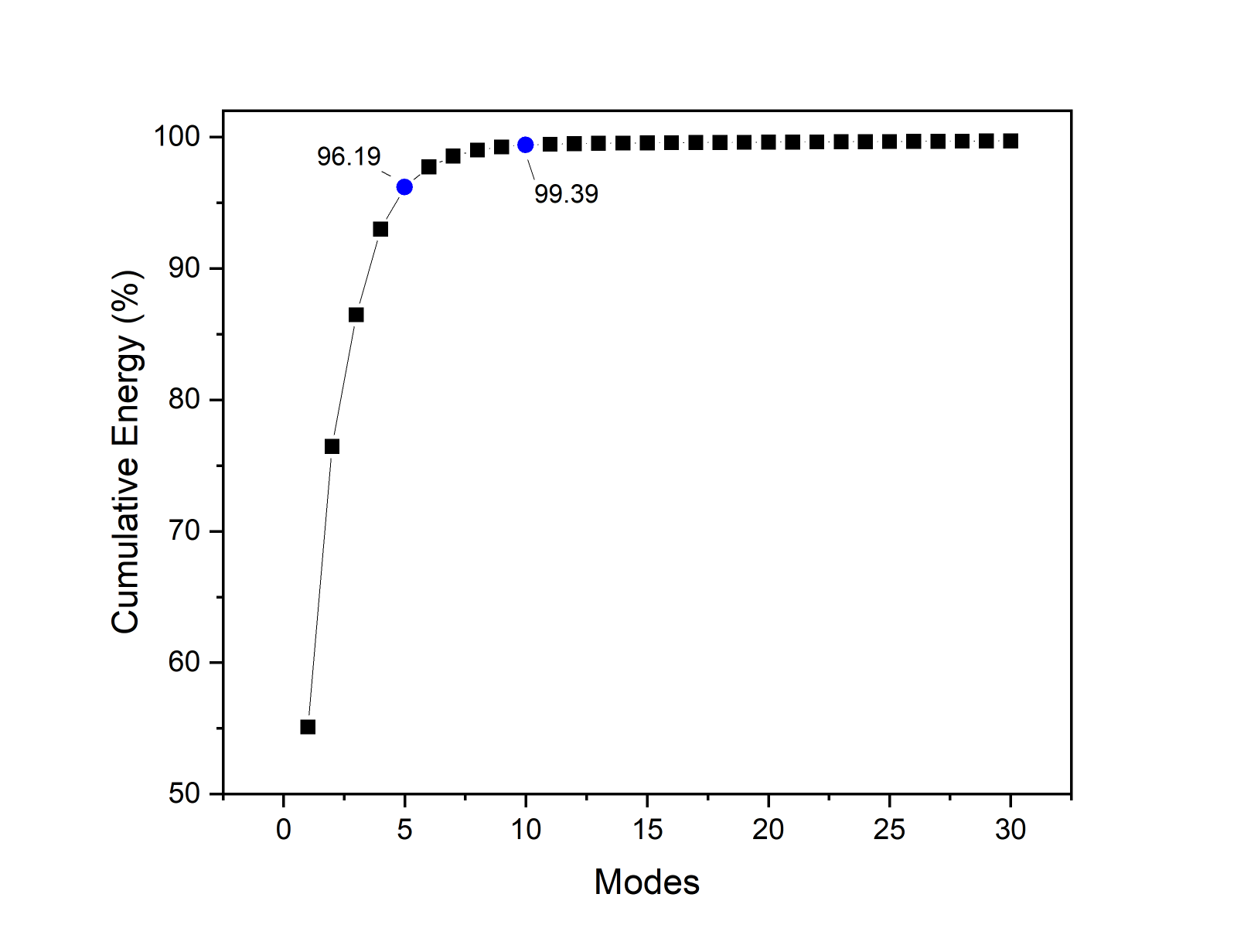} & 
        \includegraphics[width=0.45\textwidth]{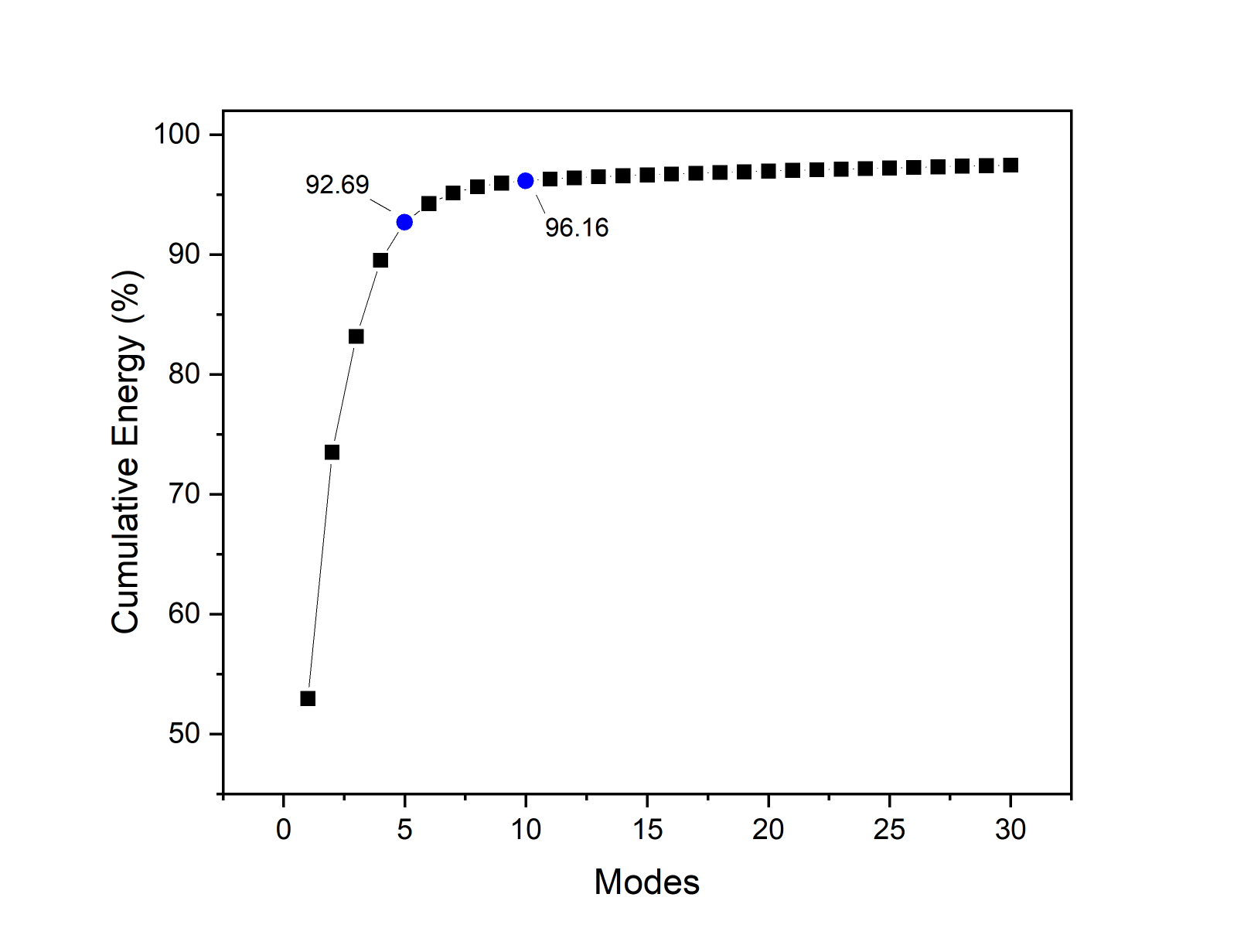} \\
                
    \hline
    \end{tabular}
    
    \caption{Comparison of evolution of cumulative energy per mode in self-consistent electrostatic kinetic simulation: (left) 40K-E dataset, (right) 5K-E dataset.}
    \label{fig:CDF_E}  

\end{figure}

In order to understand the effect of introducing self-consistent electric field and noise in the POD--SINDy framework performance related to increased reconstruction error it is necessary to first evaluate how the variance of the dynamics is influenced in the system. 

Figure \ref{fig:CDF_E} shows the cumulative energy for the self-consistent electrostatic datasets with low (left) and high (right) particle noise levels. Compared with the benchmark passive kinetic simulation case, in which only five modes were sufficient to capture over 98\% of the variance of the POD system, the number of modes necessary to reach similar levels are increased considerably. In the case of the 40K-E dataset, ten modes are necessary to achieve similar cumulative energy around 99\%. This slower energy capture tell us that the variance of the system dynamics is no longer concentrated in just few dominant structure, but rather it is spread across more modes that have influence in the major dynamics of the system. This can be seen as a direct consequence of the richer dynamics introduced by the self-consistent electric field and the nonlinear interactions it generates. Therefore, while the passive kinetic simulation case reflected simple linear advection, the application of POD on this dataset exhibits a slower decay of each mode energy, which is a clear sign of the increased dynamical complexity.

The evaluation of the 5K-E dataset indicates that the introduction of noise has a higher impact on the variance as with even ten modes the cumulative energy only reaches approximately 96\% of the total energy. With fewer particles per cell, statistical noise reduces the efficiency of each mode on capturing the major dynamics of the dataset, so more modes are required to reach the same cumulative energy levels. This behavior highlights an important point related to POD performance on noisy datasets. The compactness of the reduced-order model is directly influenced by particle resolution. With higher noise, POD compression becomes more challenging as the energy that was concentrated in a few dominant modes now gets distributed over more low-order modes. Effect of noise in the system is then captured by high-order modes.

\begin{figure}[h]
    \centering 
    
    \begin{tabular}{| c  c |} 
        
        \hline
        \textbf{40K-NoE Dataset} & \textbf{40K-E Dataset} \\
        \hline 
        \rule{0pt}{15pt} 
        
        \includegraphics[width=0.47\textwidth]{images/40k_NoE/POD_Modes/40k_NoE_mode_1.pdf} & 
        \includegraphics[width=0.47\textwidth]{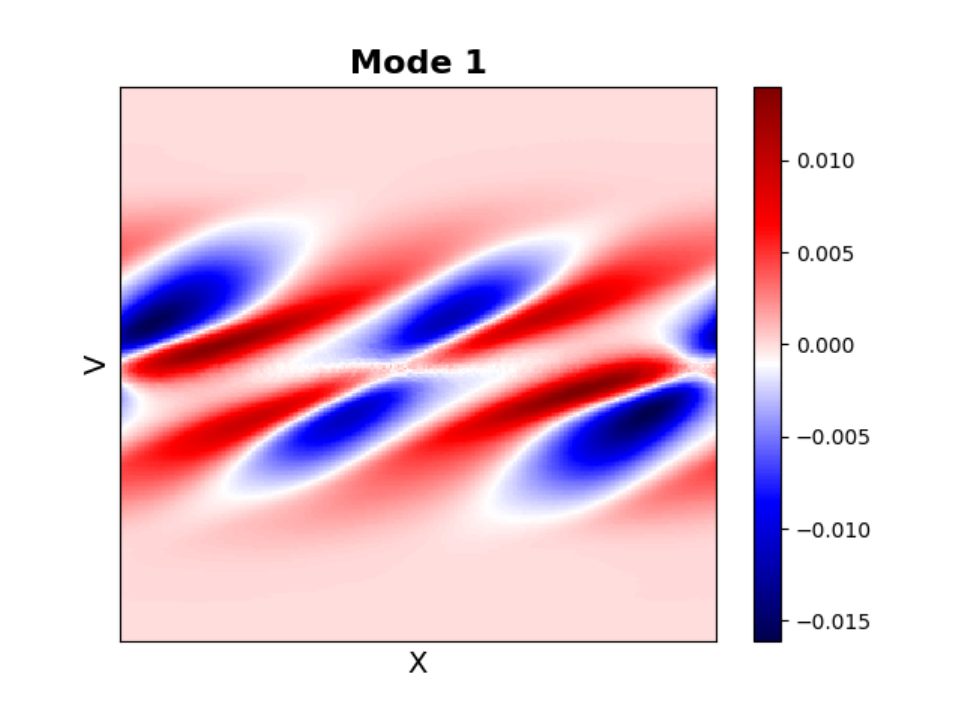} \\
                
    \hline
    \end{tabular}
    
    \caption{Comparison of spatial distribution of the first mode from POD in passive kinetic simulation (left) and self-consistent electrostatic kinetic simulation (right).}
    \label{fig:Modes_comparison}  

\end{figure}

Above analysis becomes clearer when we consider the spatial distribution of the obtained modes. Figure \ref{fig:Modes_comparison} shows the comparison of the spatial distribution of the first POD mode in the passive kinetic simulation case (left) and the self-consistent electrostatic simulation case (right). We can appreciate how in the passive kinetic simulation case the low-order dominant modes exhibit a relatively clear and smooth spatial distribution, with large sinusoidal structures that mostly reflect the linear advection which is dominant in the phase mixing process at the passive kinetic simulation datasets. Once the self-consistent electric field is included, the dominant modes adopt a much finer and filamented structure, capturing in this way the nonlinear structures present in the phase space evolution of the particles. This strongly supports previous analysis in which it was shown that the variance of the system is no longer concentrated in global sinusoidal features but are rather distributed into more localized, complex spatial structures. Therefore, the electric field transforms simple advective modes into richer, more intricate structures that reflects the onset of nonlinear and stronger phase-mixing dynamic introduced by particle-field interactions.

\begin{figure}[p] 
    \centering 
    
    \begin{tabular}{ c  c }

        \includegraphics[width=0.4\textwidth]{images/40k_E/POD_Modes/40k_mode_1.pdf} & 
        \includegraphics[width=0.4\textwidth]{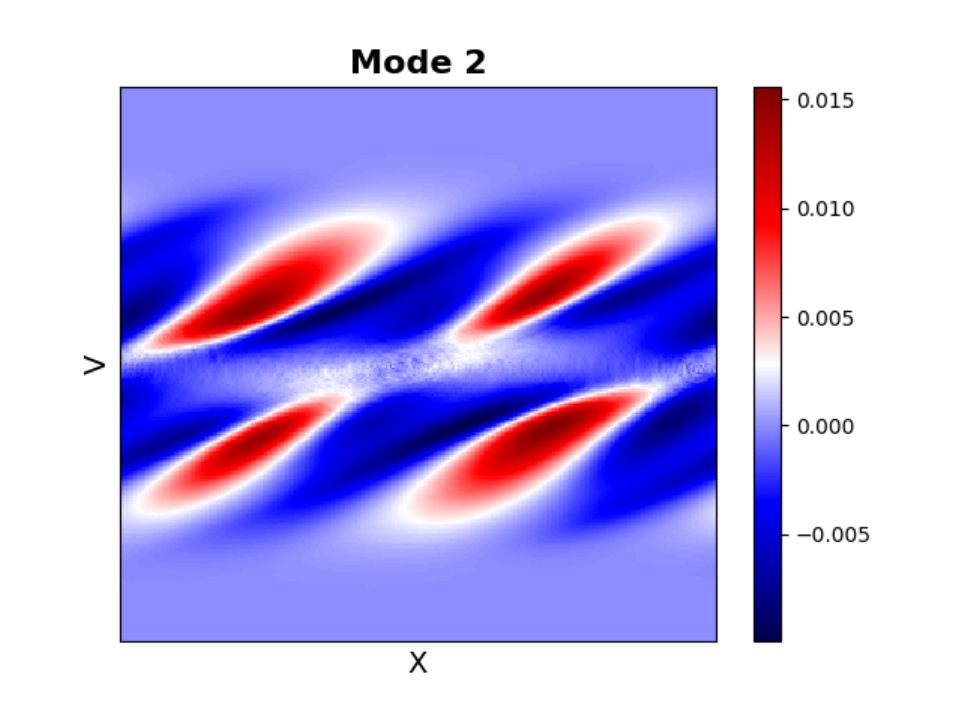} \\
        
        \includegraphics[width=0.4\textwidth]{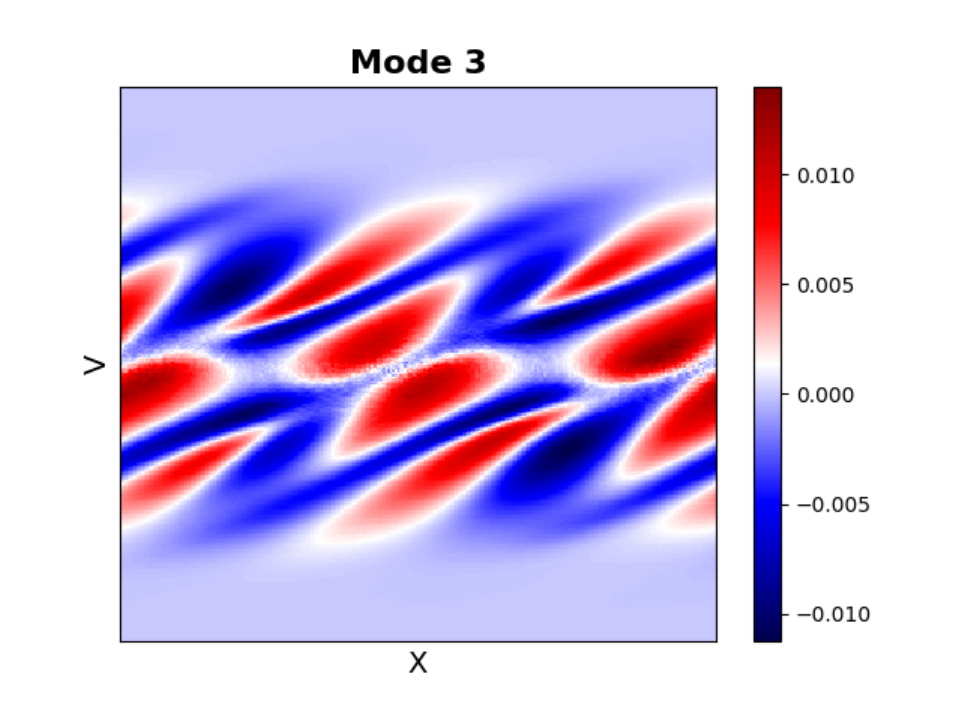} & 
        \includegraphics[width=0.4\textwidth]{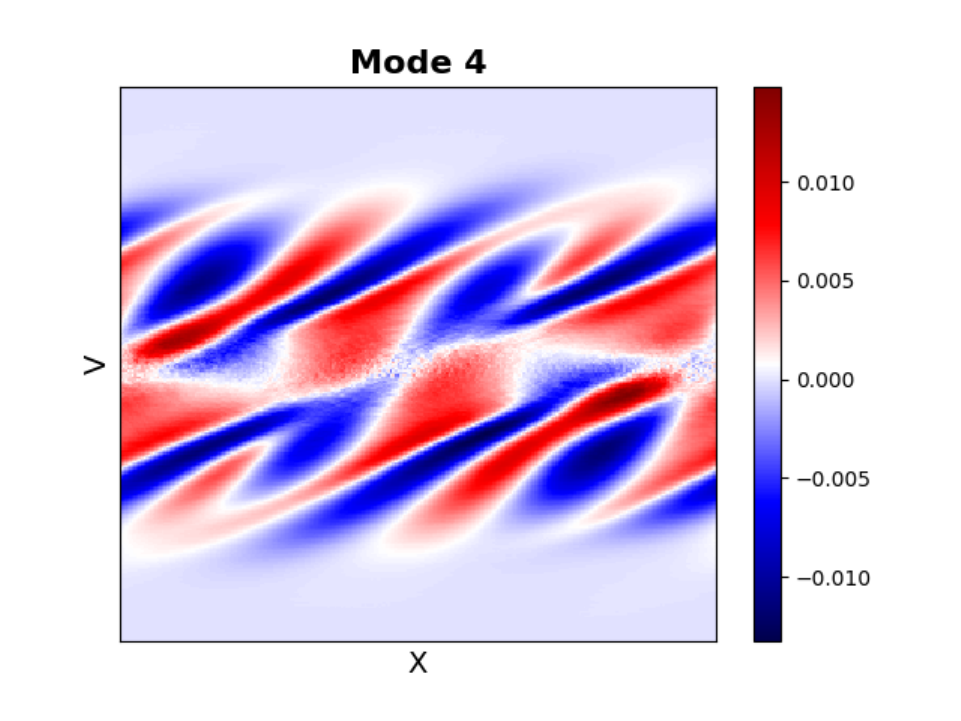} \\

        \includegraphics[width=0.4\textwidth]{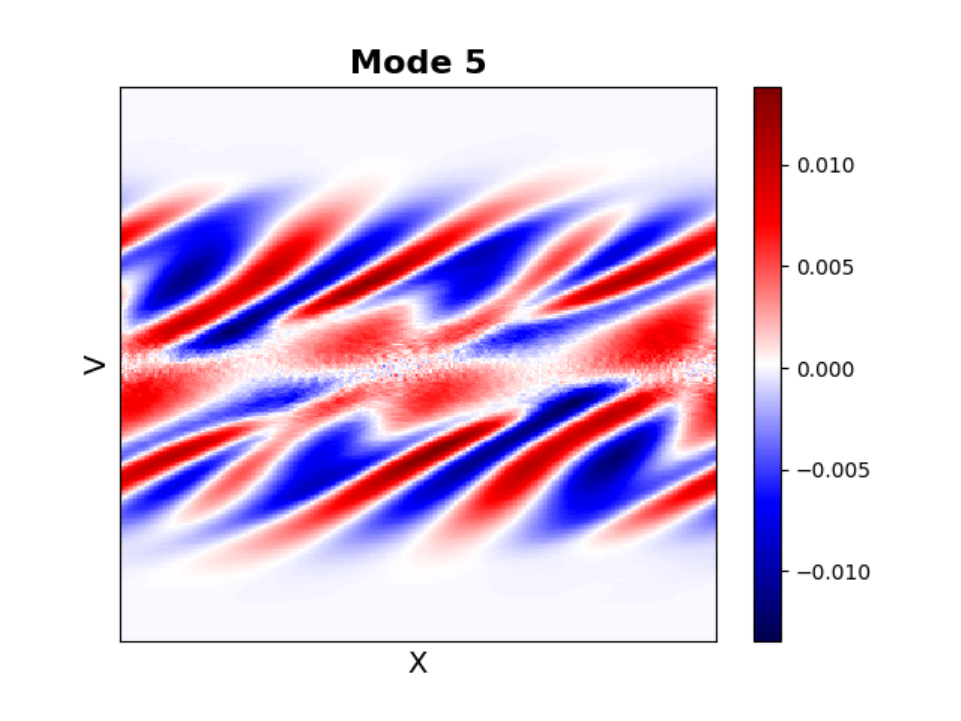} & 
        \includegraphics[width=0.4\textwidth]{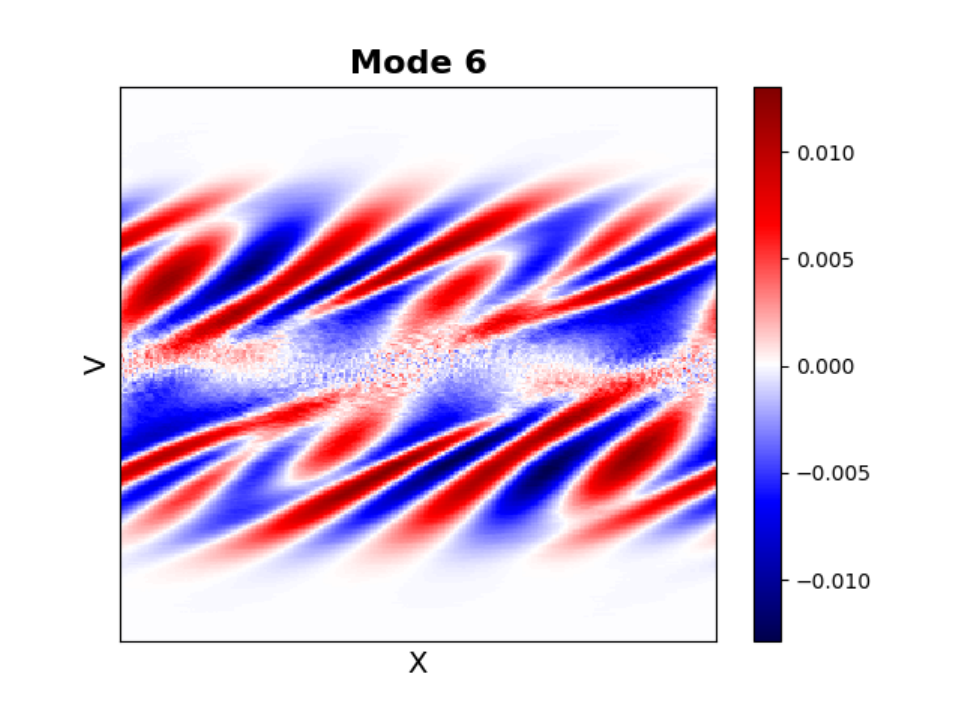} \\

        \includegraphics[width=0.4\textwidth]{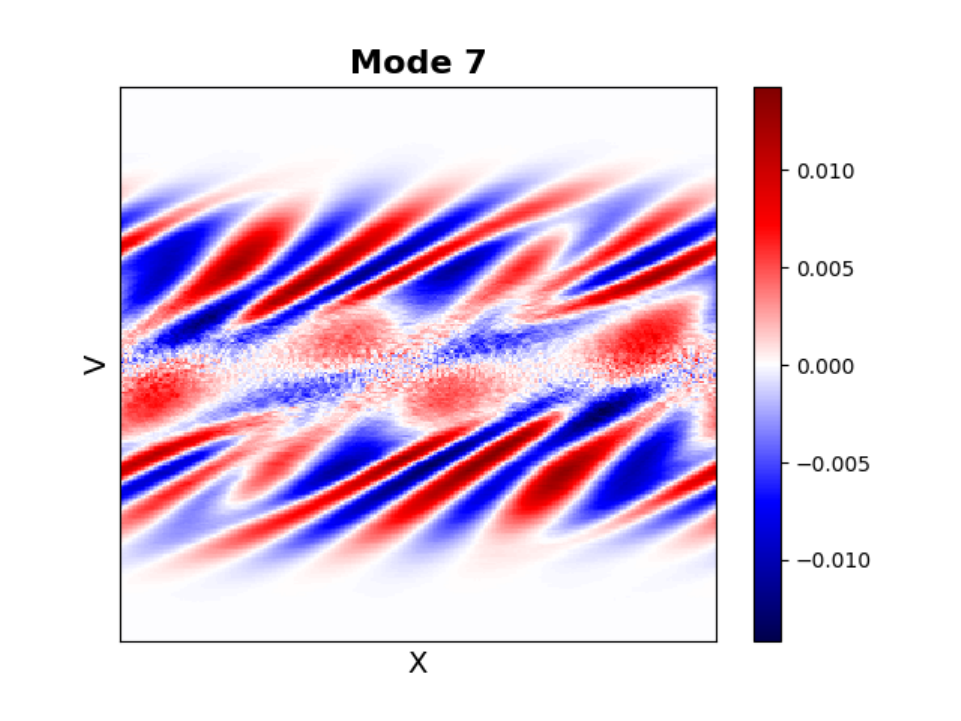} & 
        \includegraphics[width=0.4\textwidth]{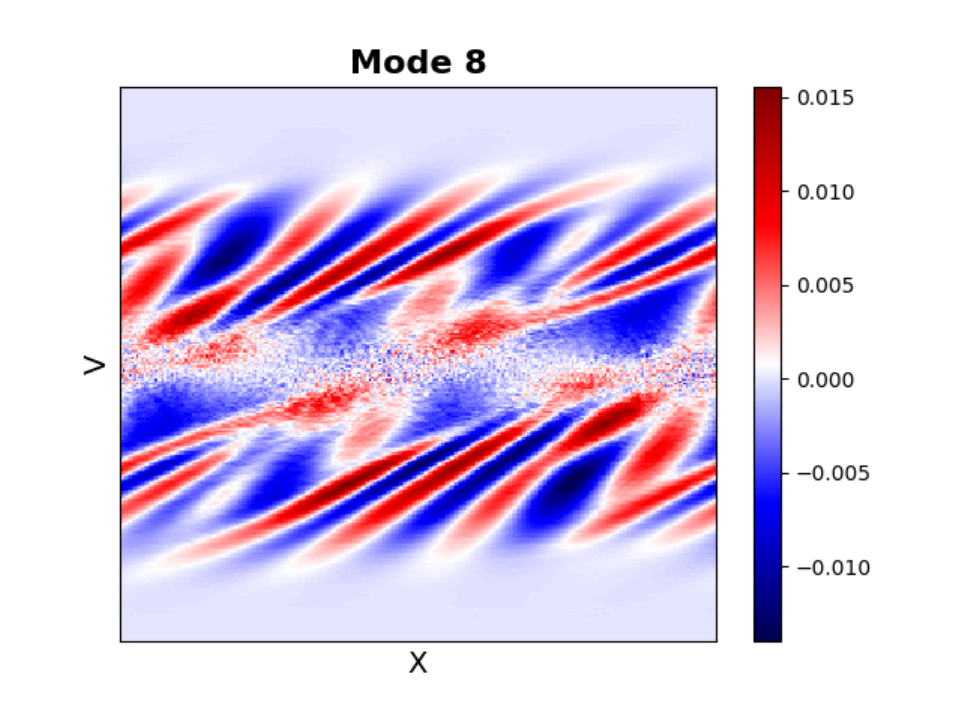} \\

        \includegraphics[width=0.4\textwidth]{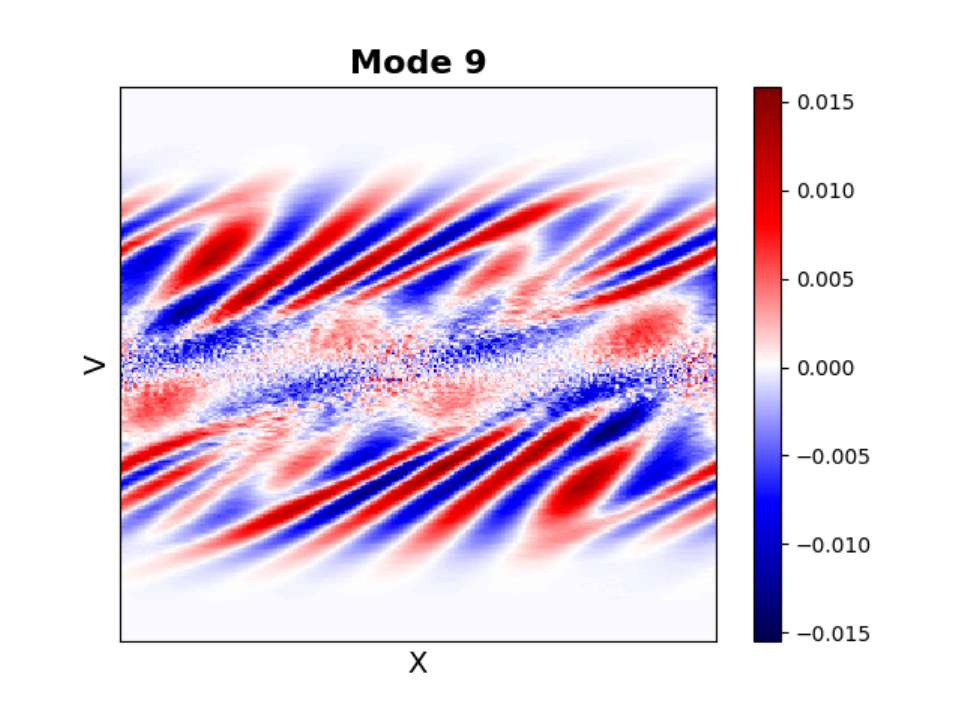} & 
        \includegraphics[width=0.4\textwidth]{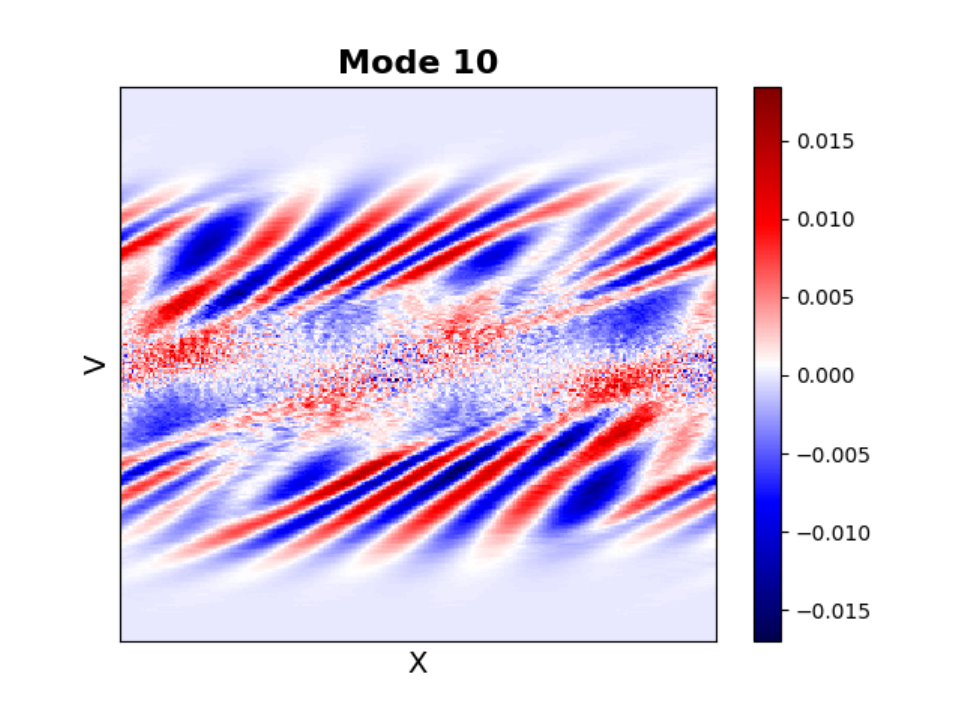} \\
  
    \end{tabular}
    
    \caption{Spatial distribution of first ten modes from POD in the self-consistent electrostatic kinetic simulation 40K dataset.}
    \label{fig:Modes_40kE}  

\end{figure}

\begin{figure}[p] 
    \centering 
    
    \begin{tabular}{ c  c }

        \includegraphics[width=0.4\textwidth]{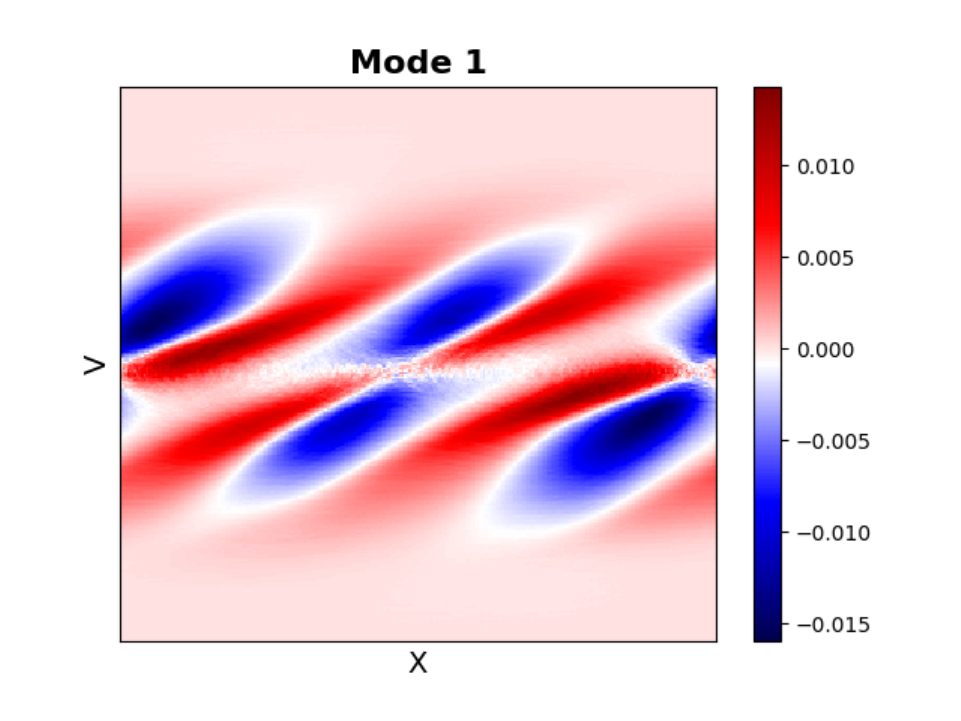} & 
        \includegraphics[width=0.4\textwidth]{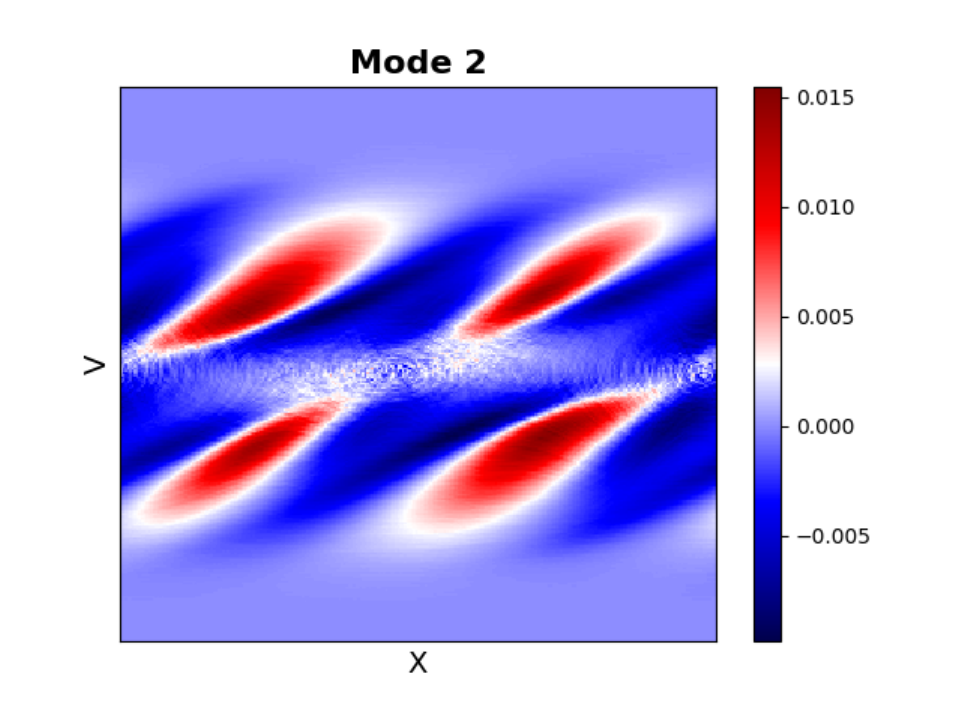} \\
        
        \includegraphics[width=0.4\textwidth]{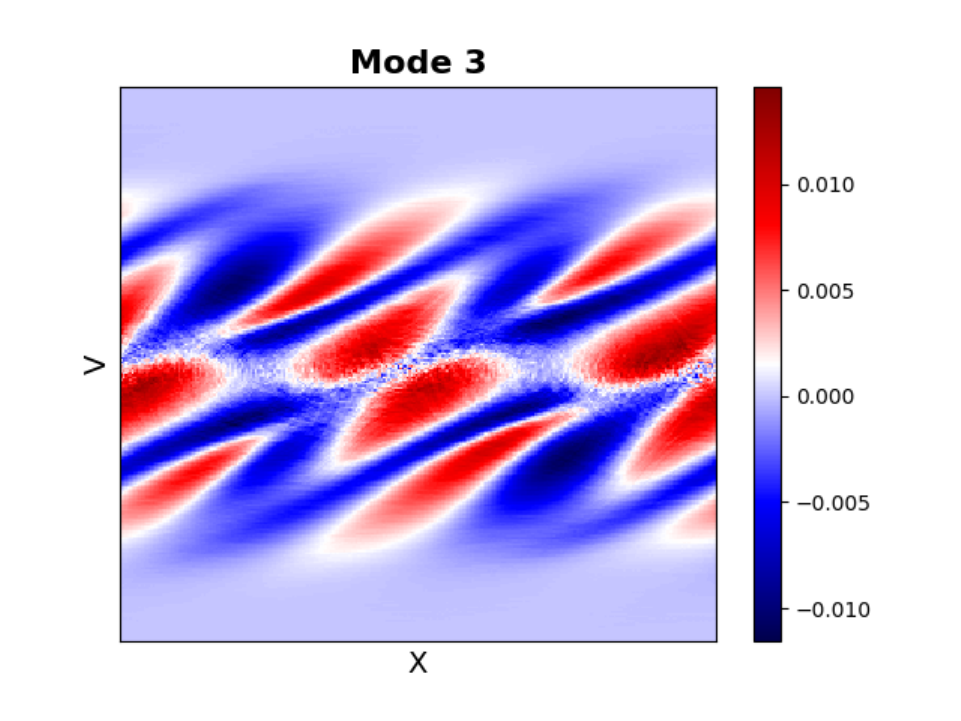} & 
        \includegraphics[width=0.4\textwidth]{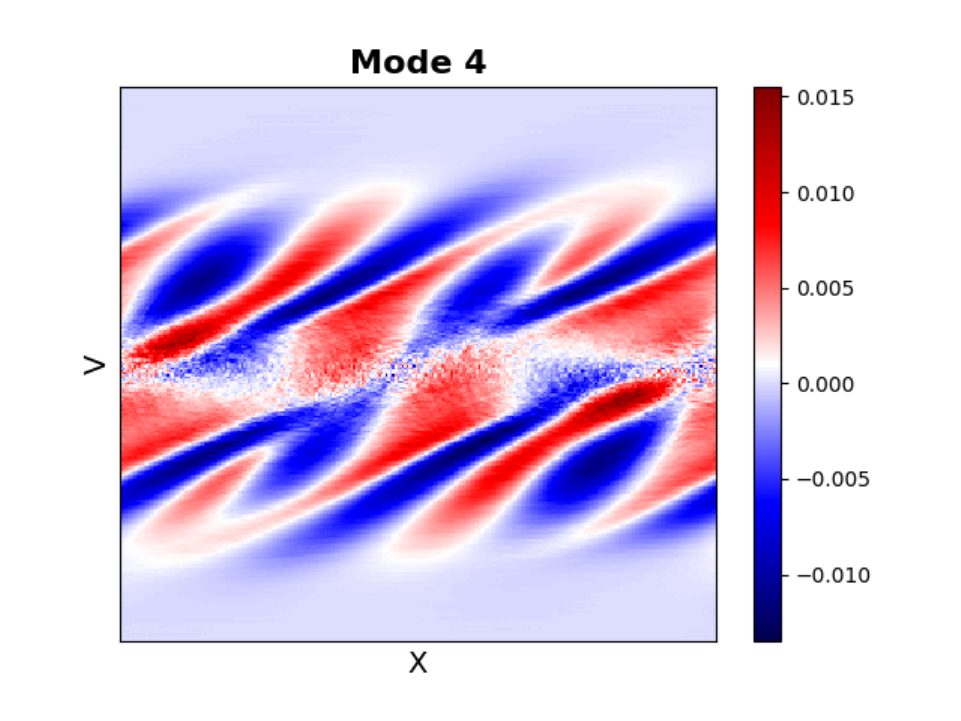} \\

        \includegraphics[width=0.4\textwidth]{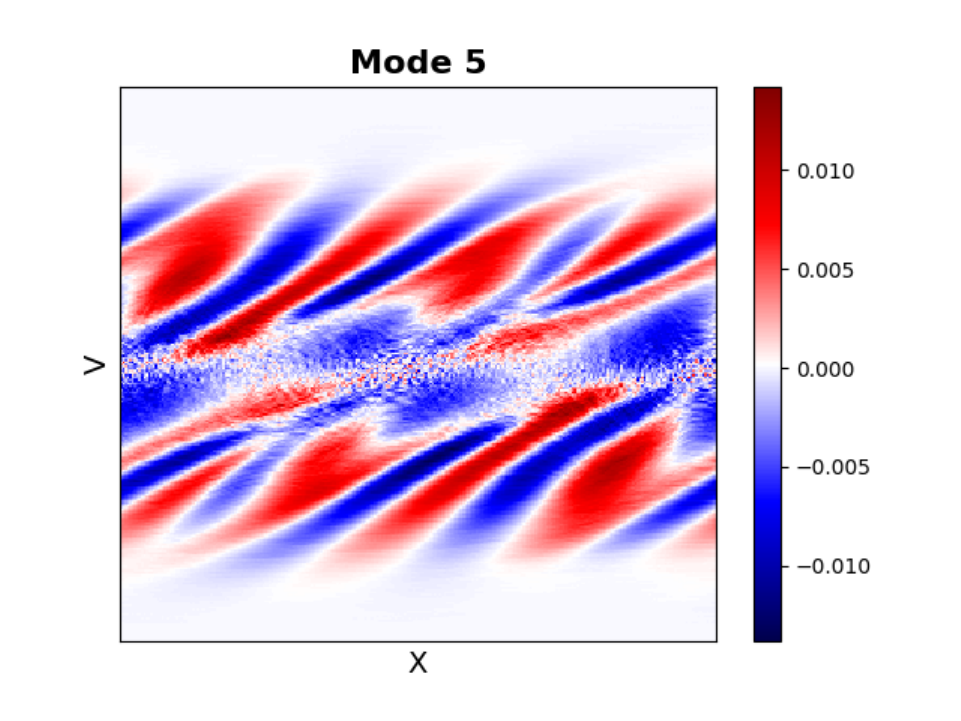} & 
        \includegraphics[width=0.4\textwidth]{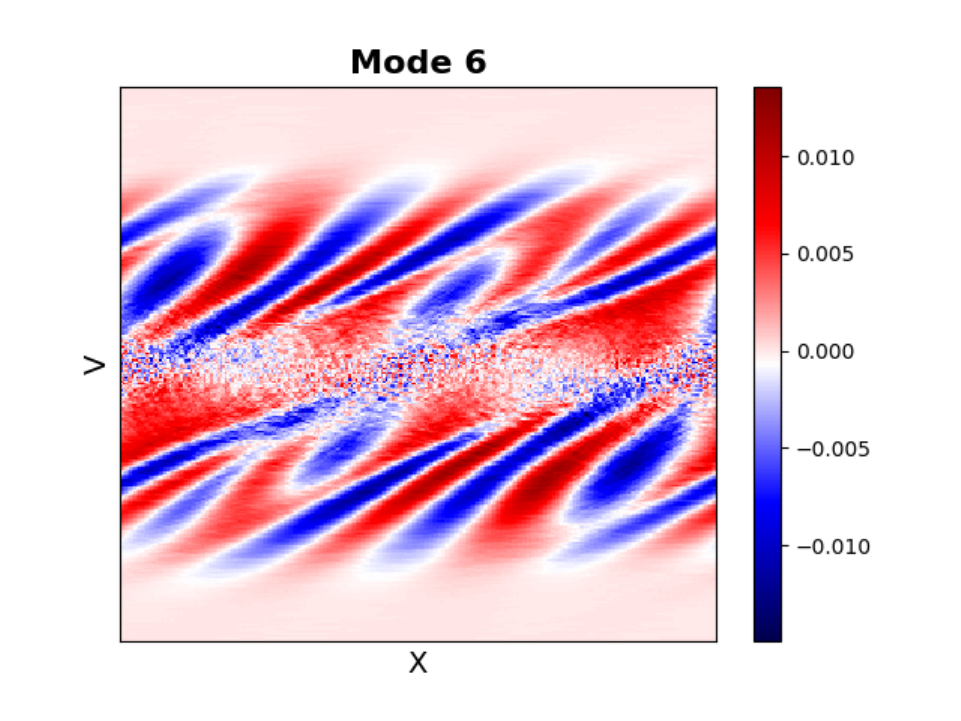} \\

        \includegraphics[width=0.4\textwidth]{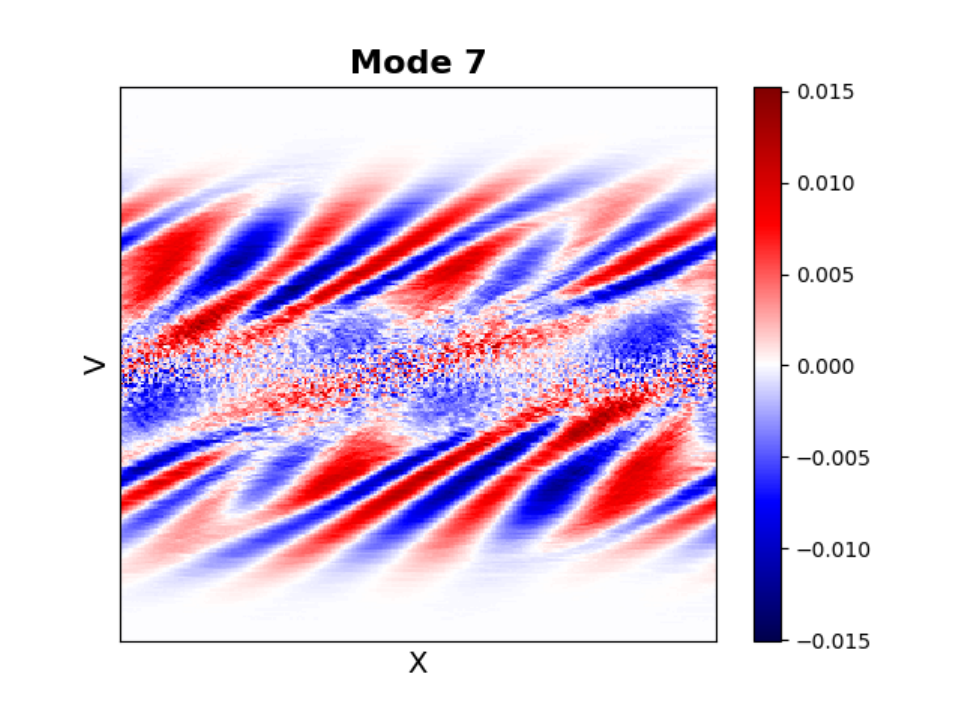} & 
        \includegraphics[width=0.4\textwidth]{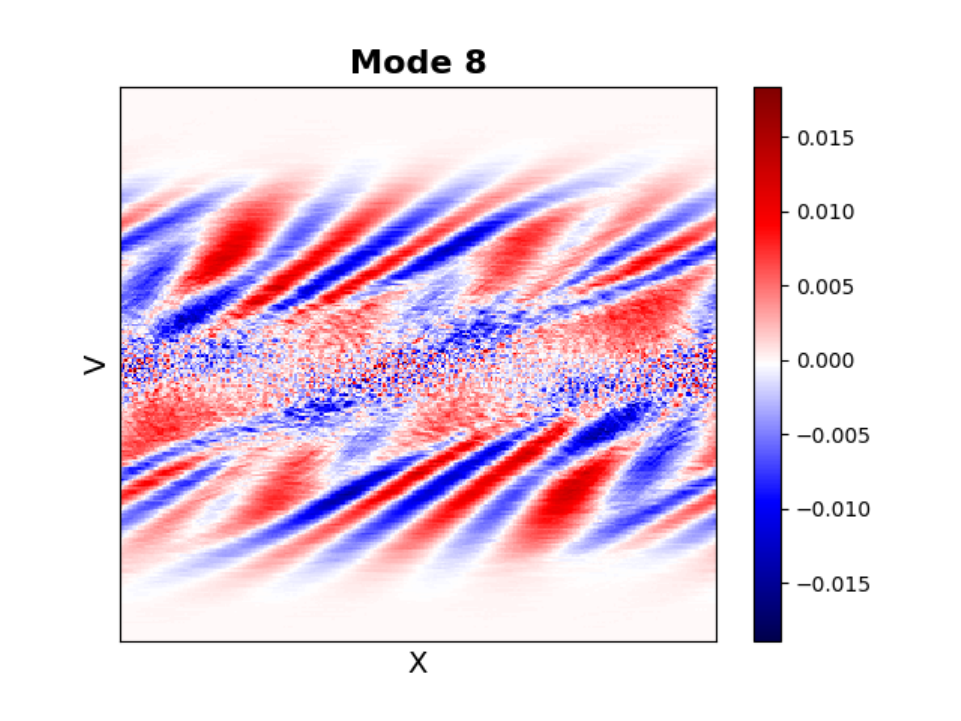} \\

        \includegraphics[width=0.4\textwidth]{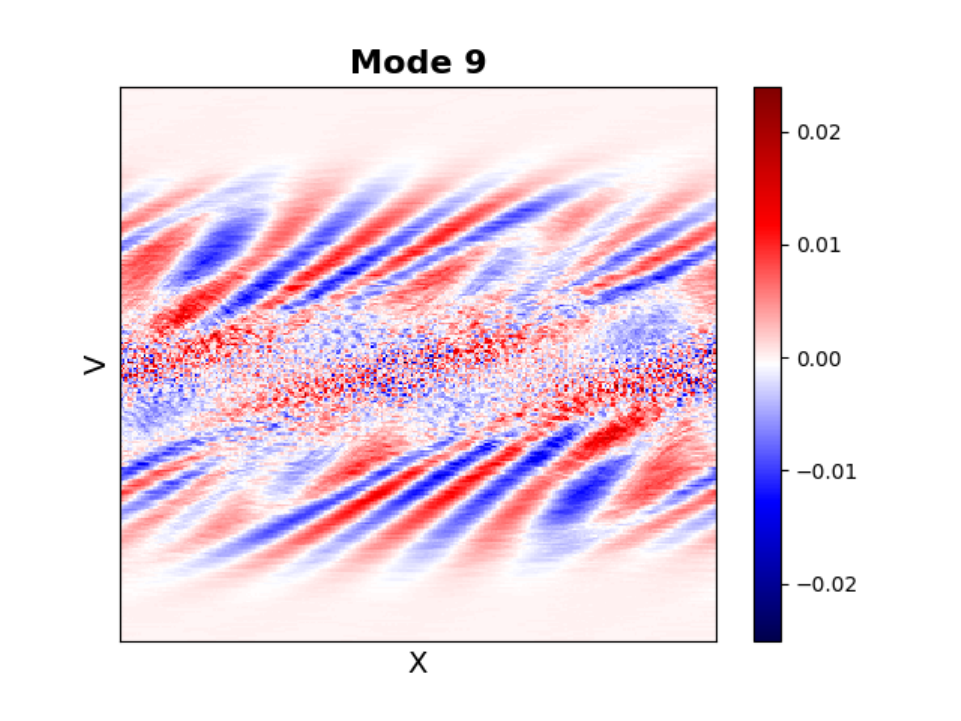} & 
        \includegraphics[width=0.4\textwidth]{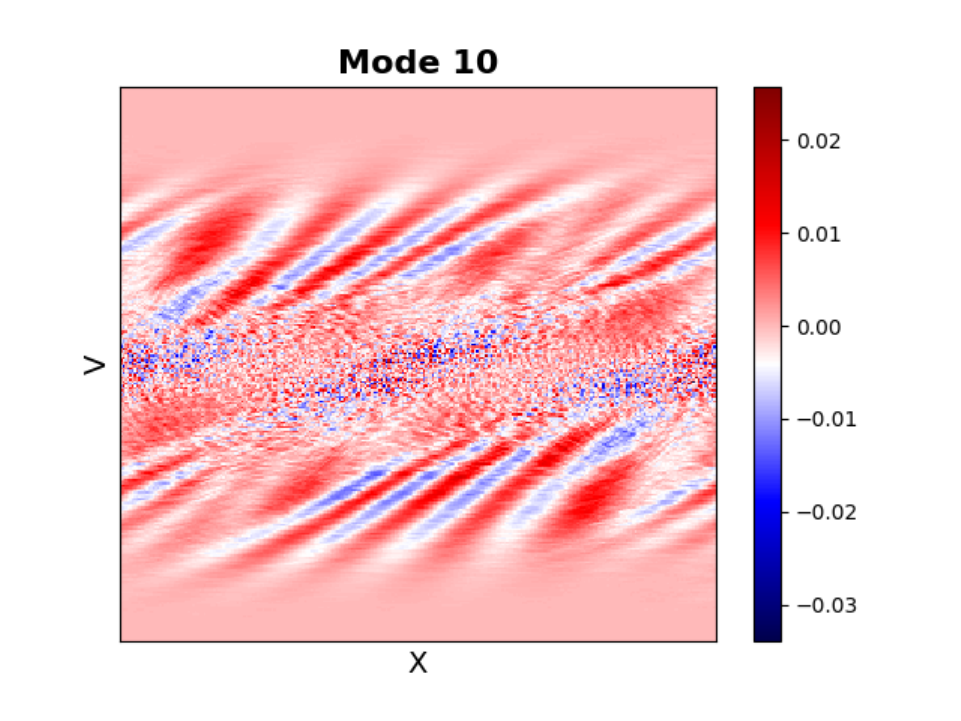} \\
  
    \end{tabular}
    
    \caption{Spatial distribution of first ten modes from POD in the self-consistent electrostatic kinetic simulation 5K dataset.}
    \label{fig:Modes_5kE}  

\end{figure}

\begin{figure}[p] 
    \centering 
    
    \begin{tabular}{ | c | c | c |} 
        
        \hline
        \textbf{Original Data} & \textbf{POD Reconstructed} & \textbf{POD--SINDy Reconstructed} \\
        \hline 
        \rule{0pt}{15pt} 
        
        \includegraphics[width=0.3\textwidth]{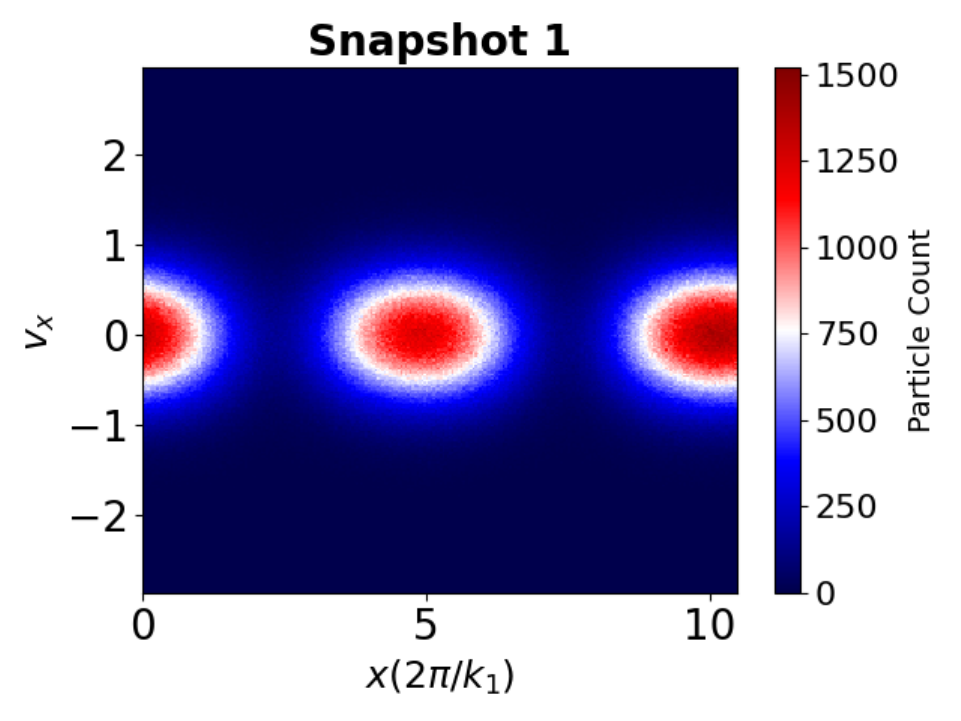} & 
        \includegraphics[width=0.3\textwidth]{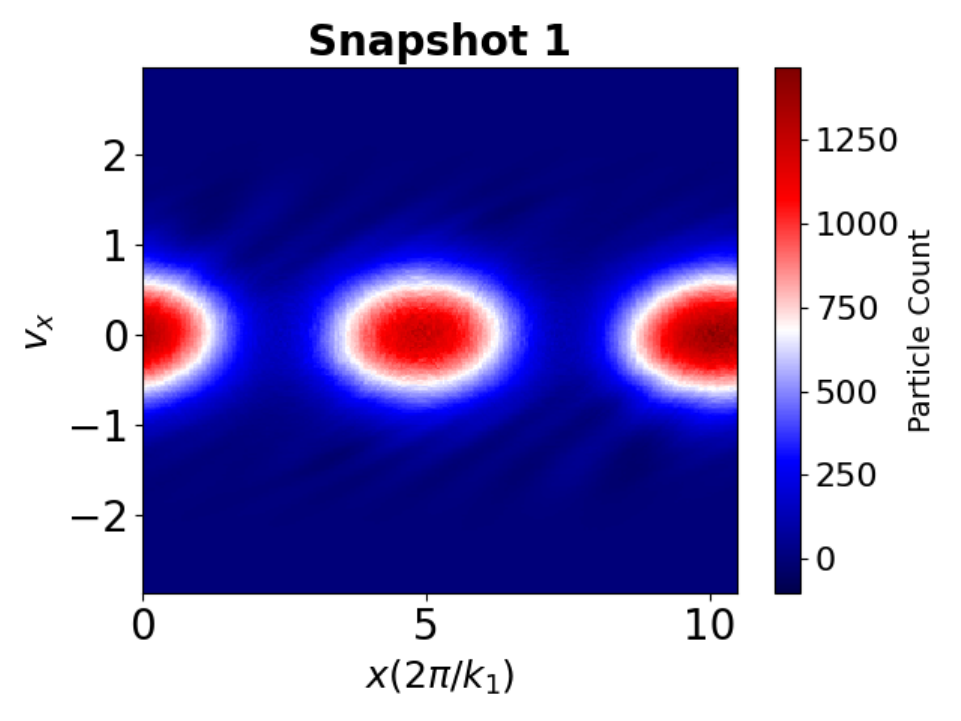} & 
        \includegraphics[width=0.3\textwidth]{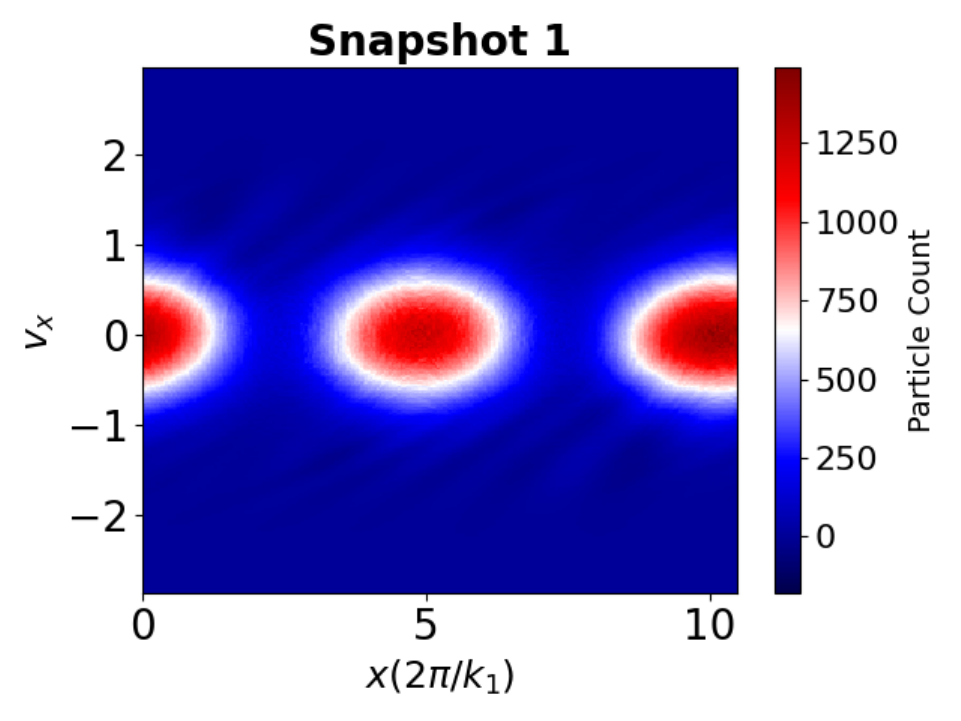} \\
        
        \includegraphics[width=0.3\textwidth]{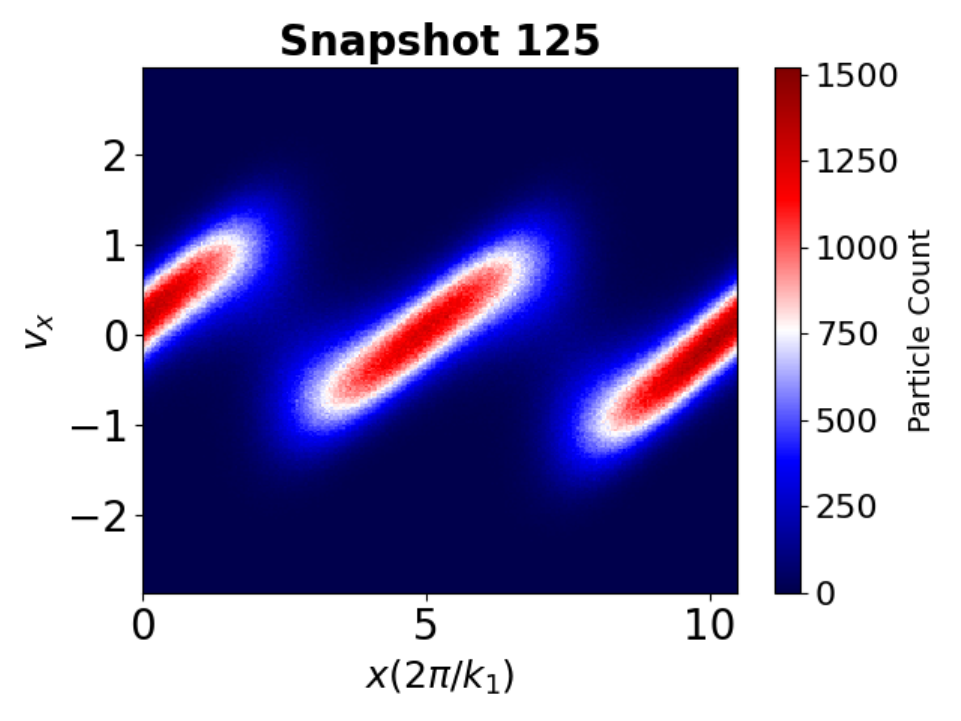} & 
        \includegraphics[width=0.3\textwidth]{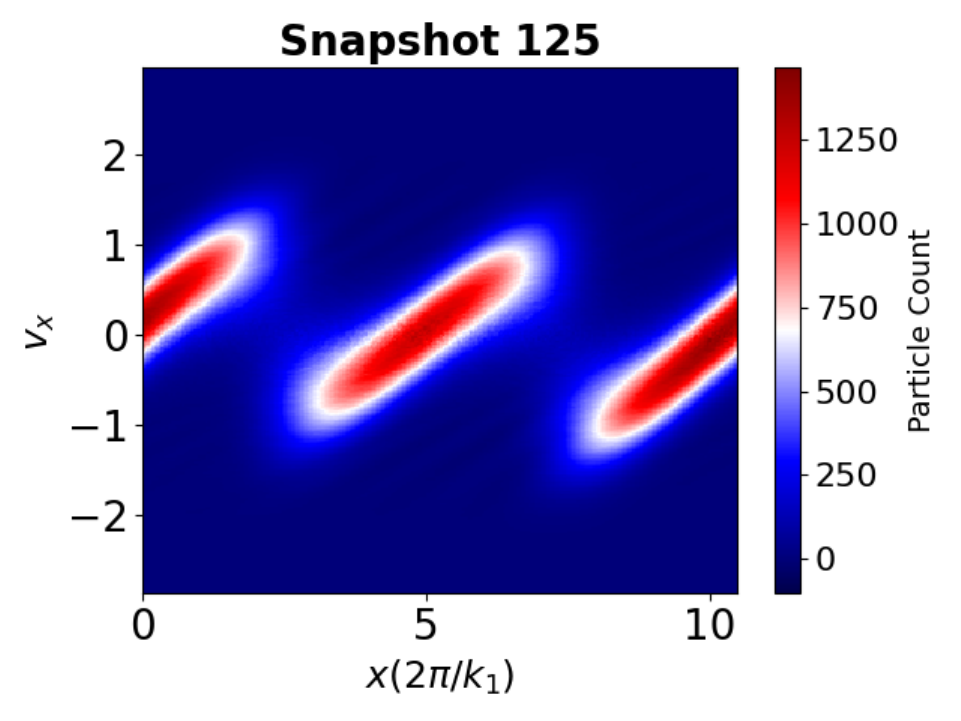} & 
        \includegraphics[width=0.3\textwidth]{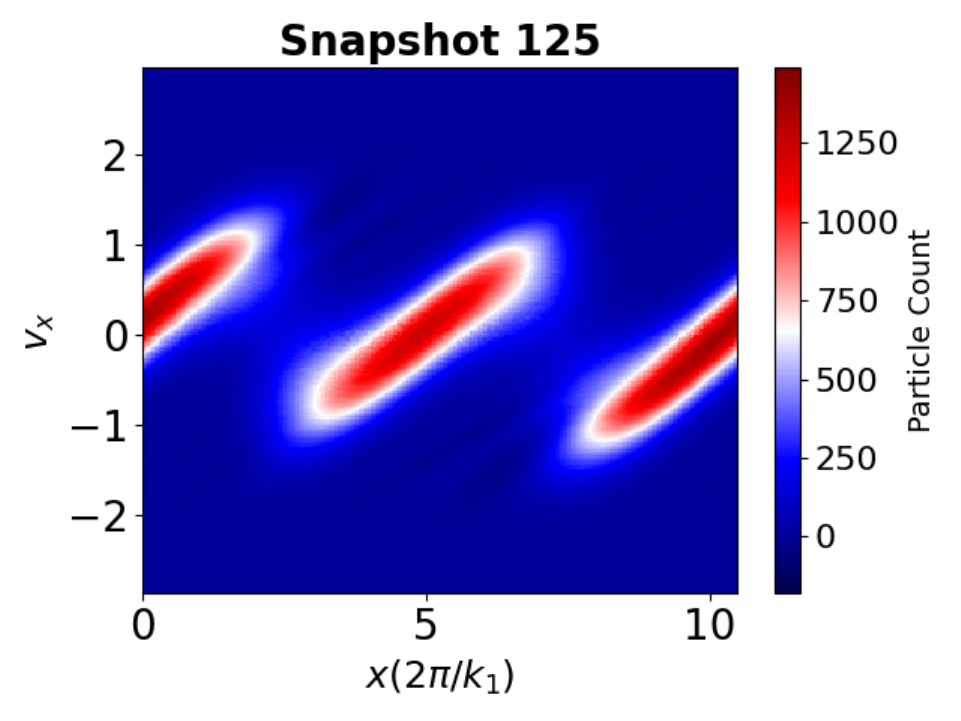} \\

        \includegraphics[width=0.3\textwidth]{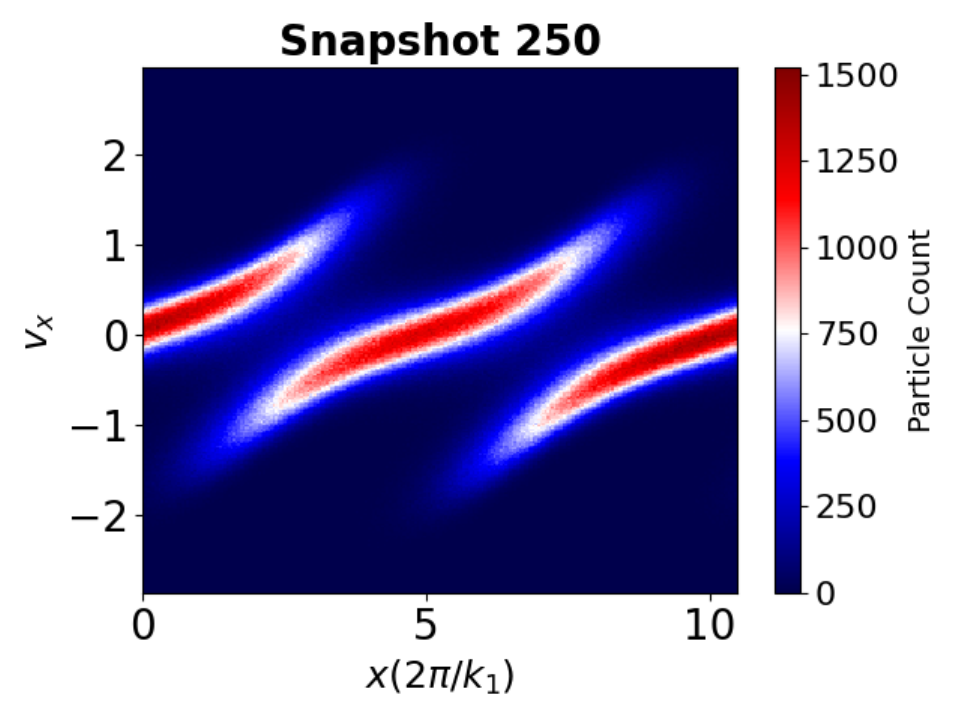} & 
        \includegraphics[width=0.3\textwidth]{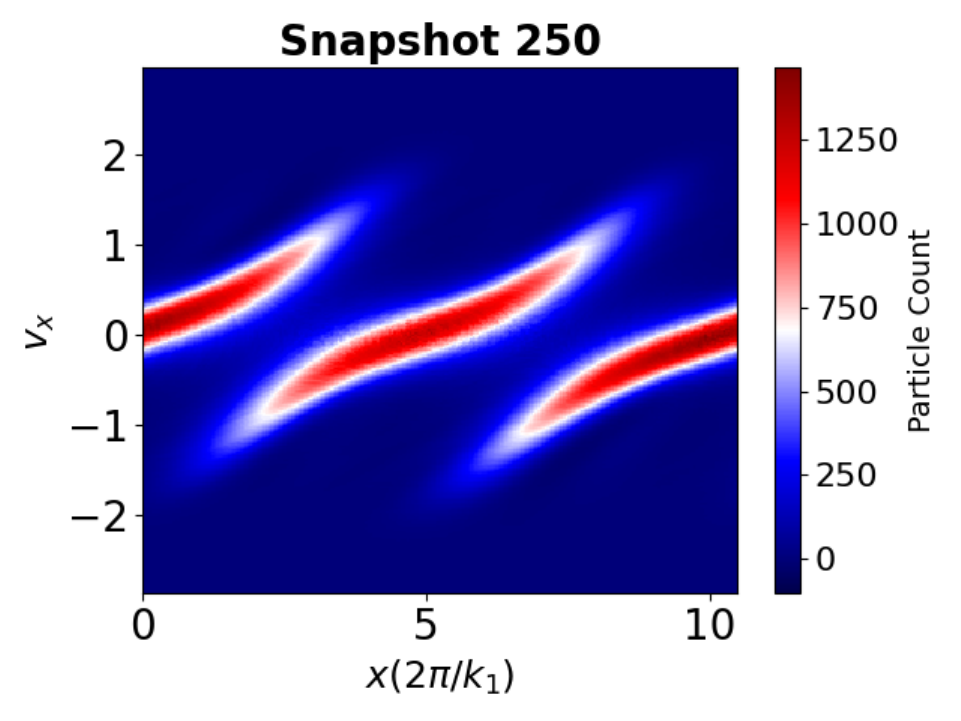} & 
        \includegraphics[width=0.3\textwidth]{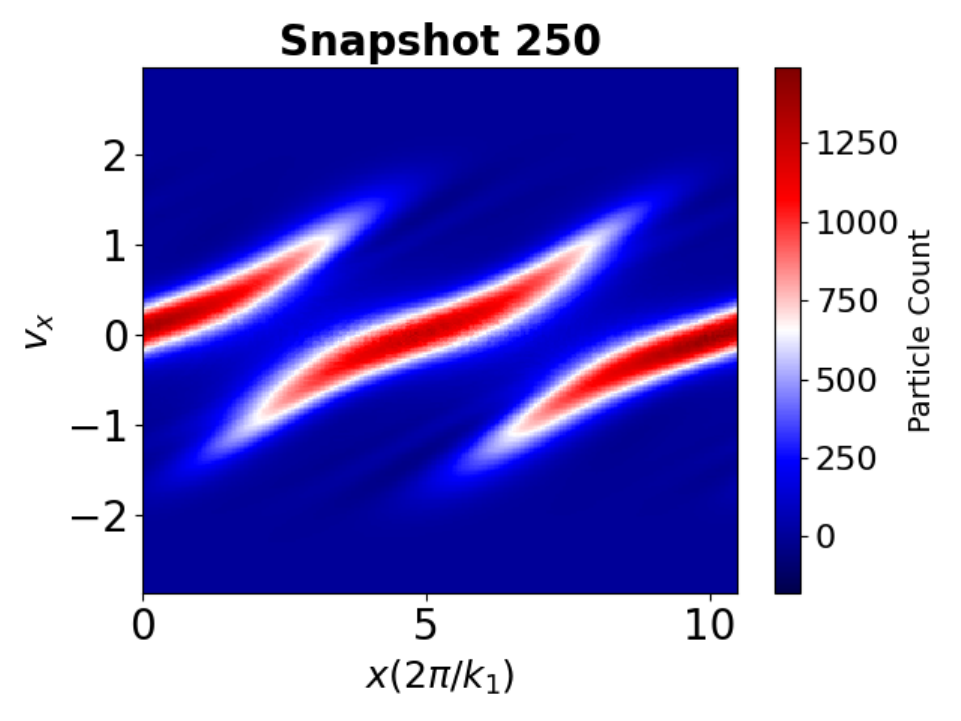} \\

        \includegraphics[width=0.3\textwidth]{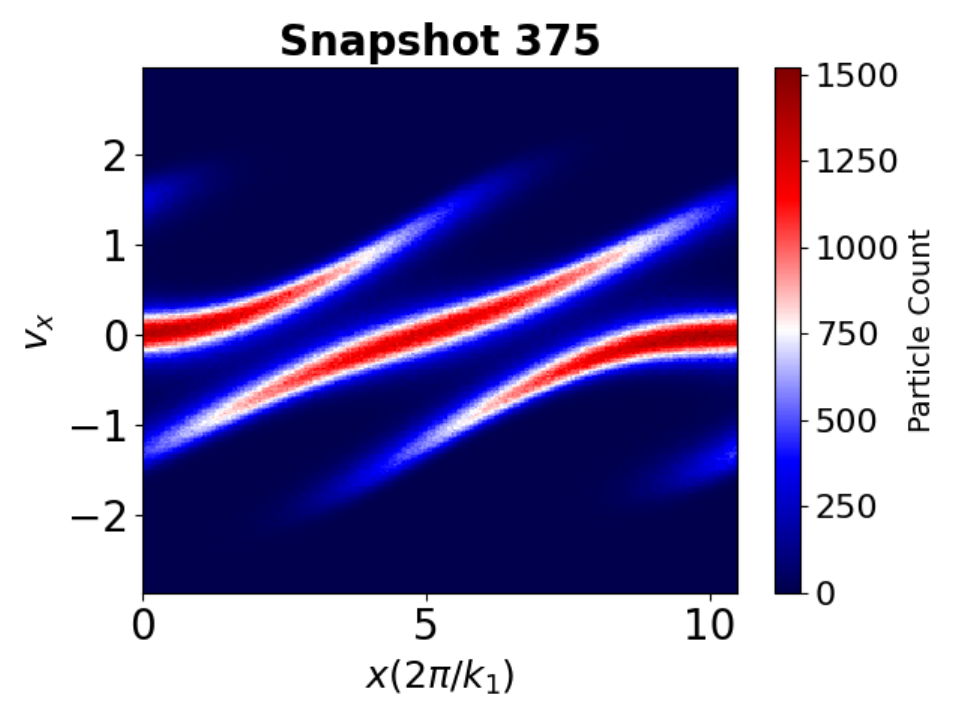} & 
        \includegraphics[width=0.3\textwidth]{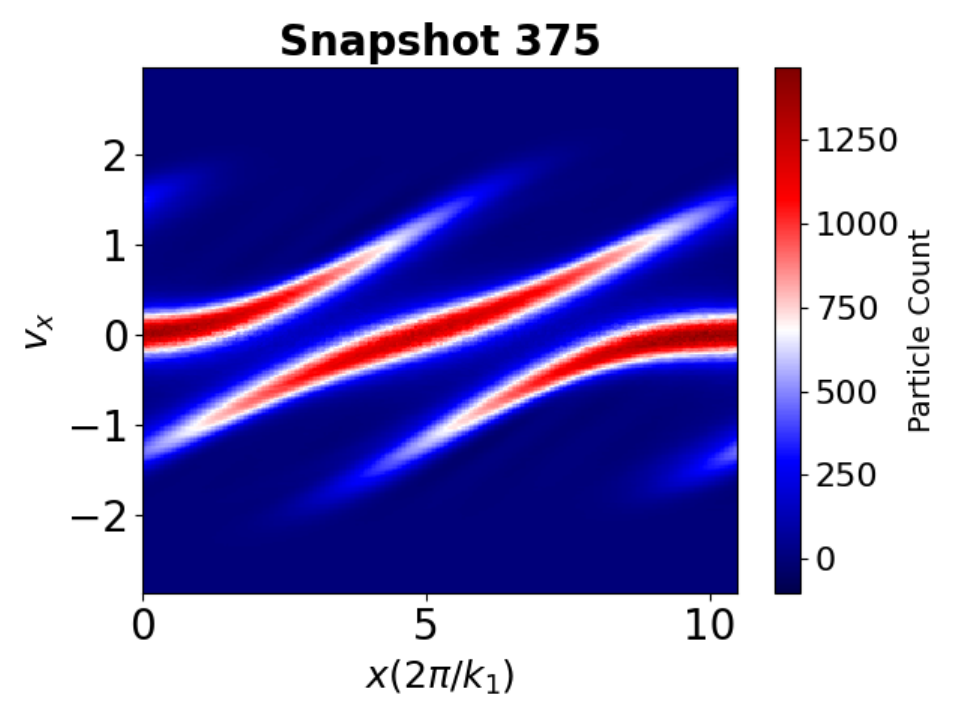} & 
        \includegraphics[width=0.3\textwidth]{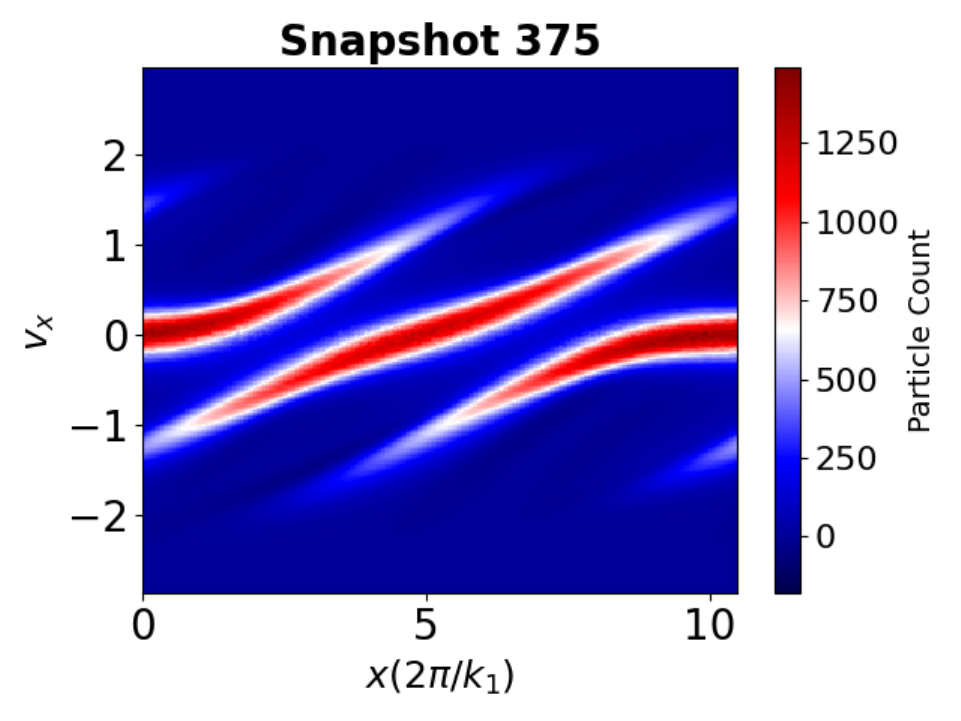} \\

        \includegraphics[width=0.3\textwidth]{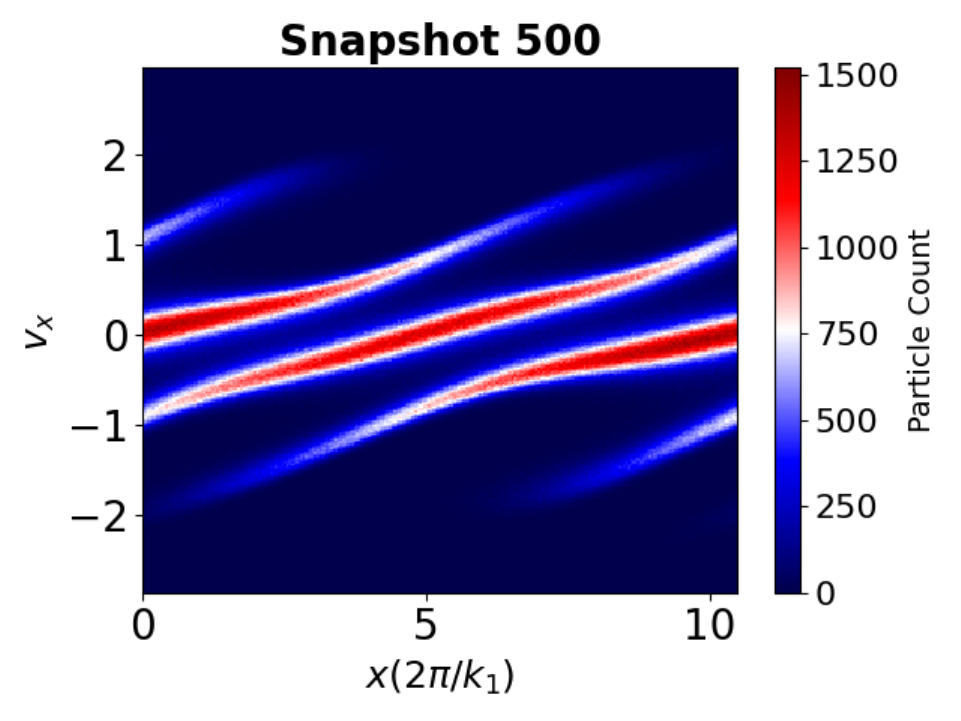} & 
        \includegraphics[width=0.3\textwidth]{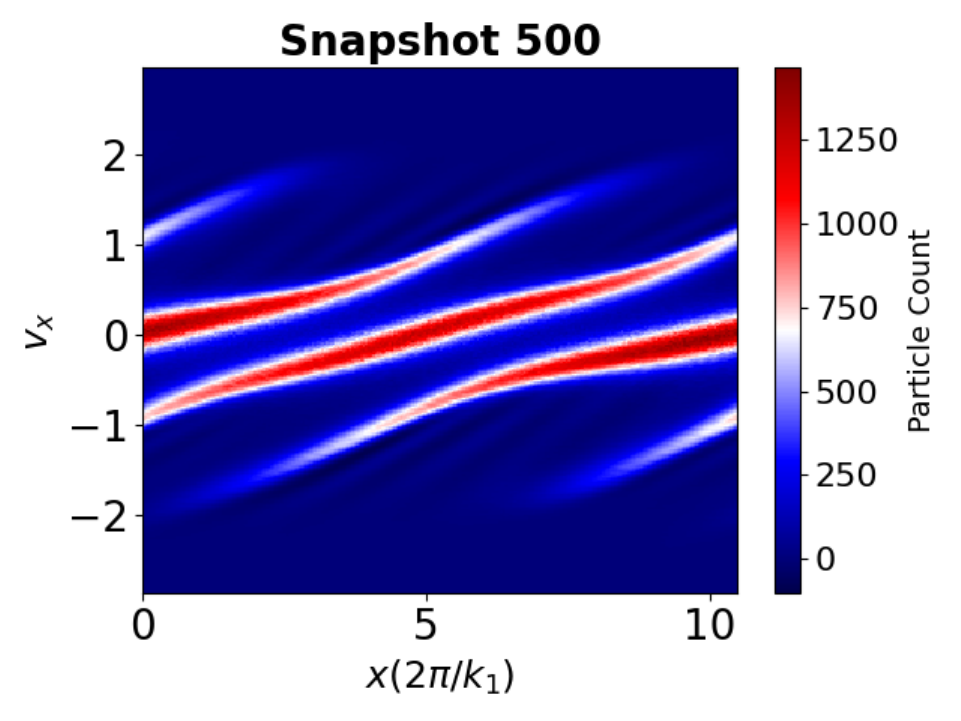} & 
        \includegraphics[width=0.3\textwidth]{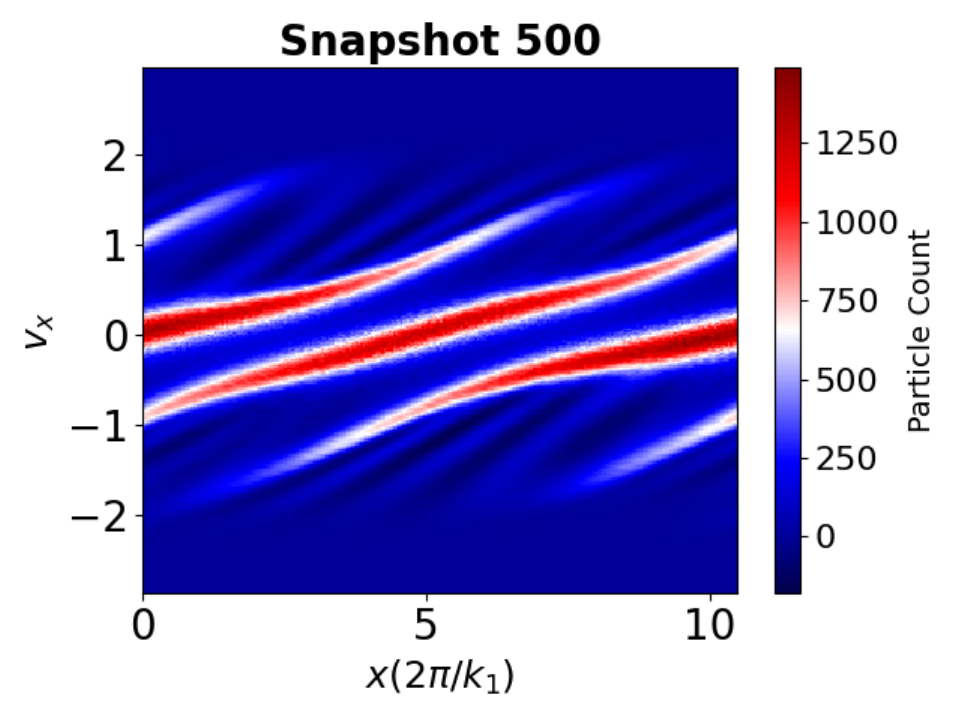} \\
    \hline
    \end{tabular}
    
    \caption{Comparison of phase mixing process with original data (left), POD reconstructed data (center) and POD--SINDy reconstructed data (right) for self-consistent electrostatic kinetic simulation 40K dataset.}
    \label{fig:Snapshots_40kE}  

\end{figure}

\begin{figure}[p] 
    \centering 
    
    \begin{tabular}{ | c | c | c |} 
        
        \hline
        \textbf{Original Data} & \textbf{POD Reconstructed} & \textbf{POD--SINDy Reconstructed} \\
        \hline 
        \rule{0pt}{15pt} 
        
        \includegraphics[width=0.3\textwidth]{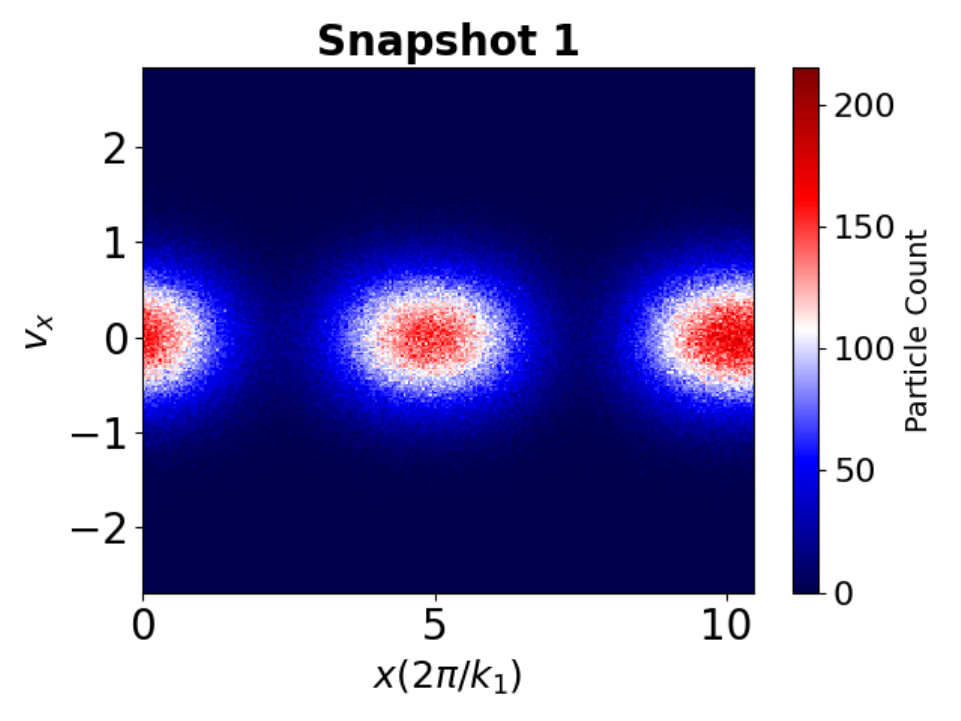} & 
        \includegraphics[width=0.3\textwidth]{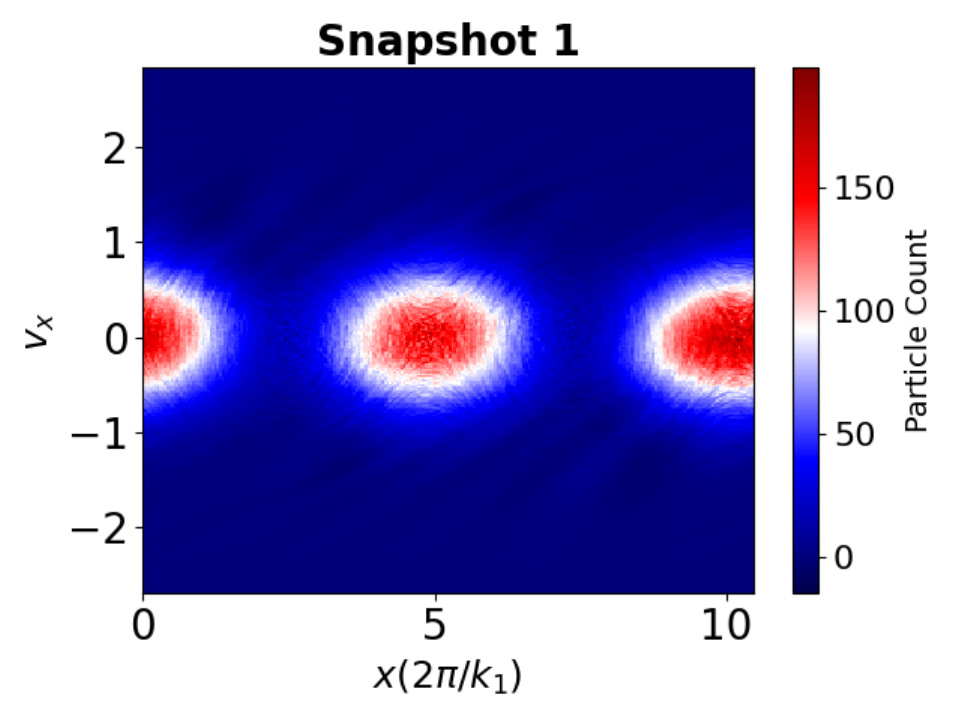} & 
        \includegraphics[width=0.3\textwidth]{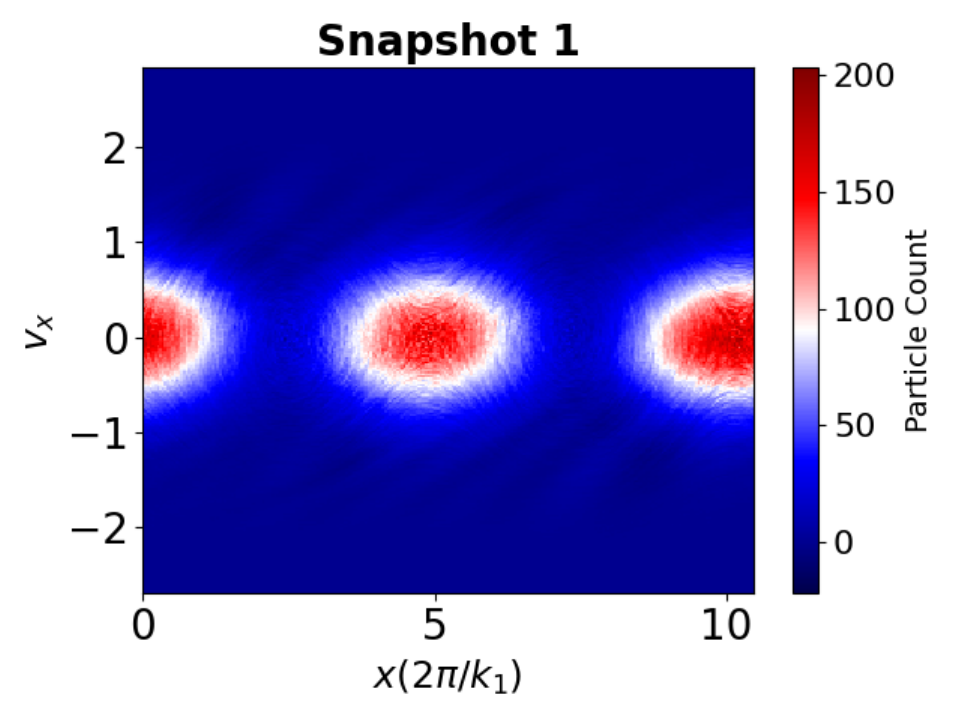} \\
        
        \includegraphics[width=0.3\textwidth]{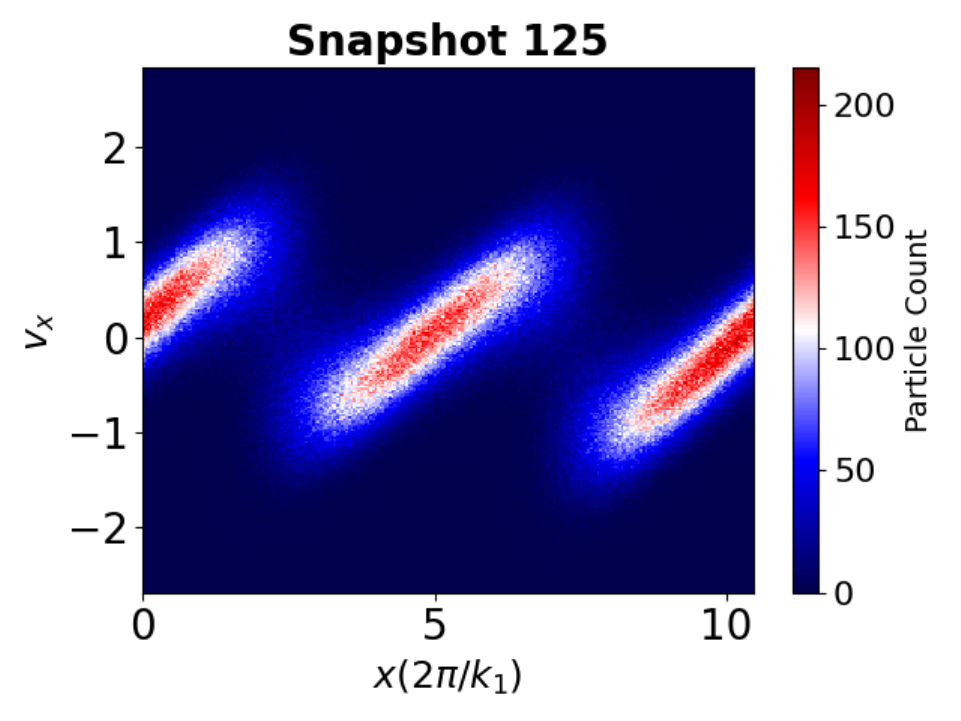} & 
        \includegraphics[width=0.3\textwidth]{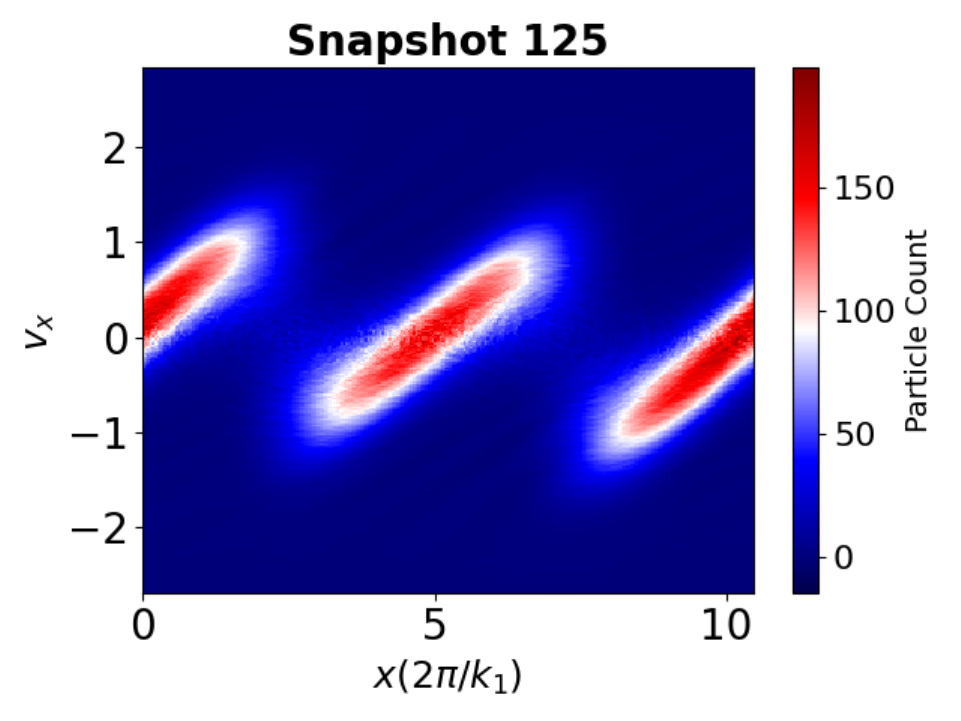} & 
        \includegraphics[width=0.3\textwidth]{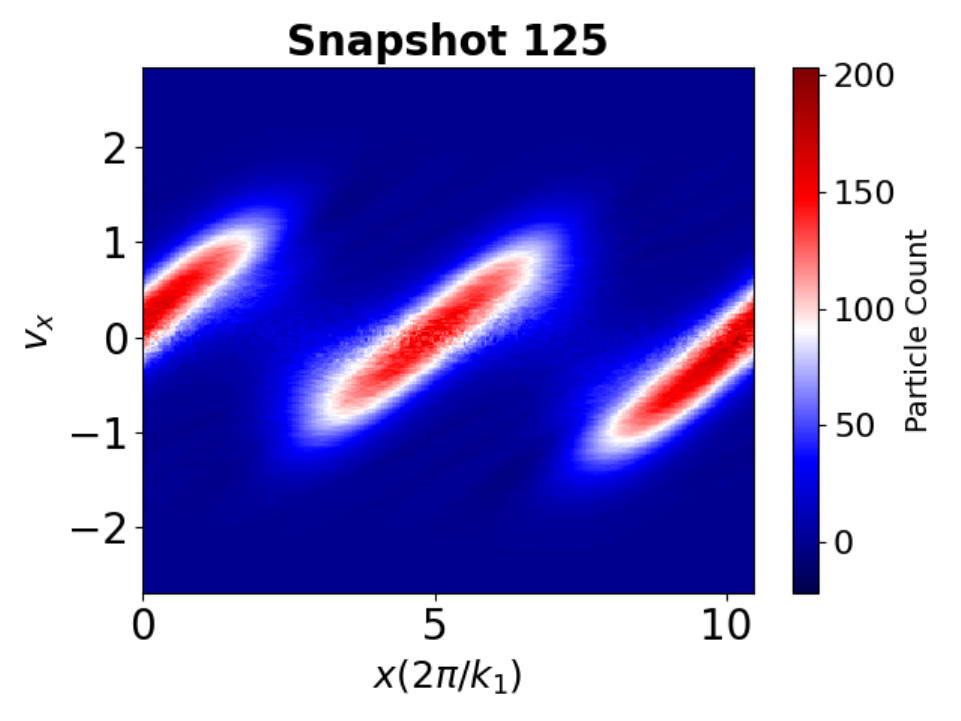} \\

        \includegraphics[width=0.3\textwidth]{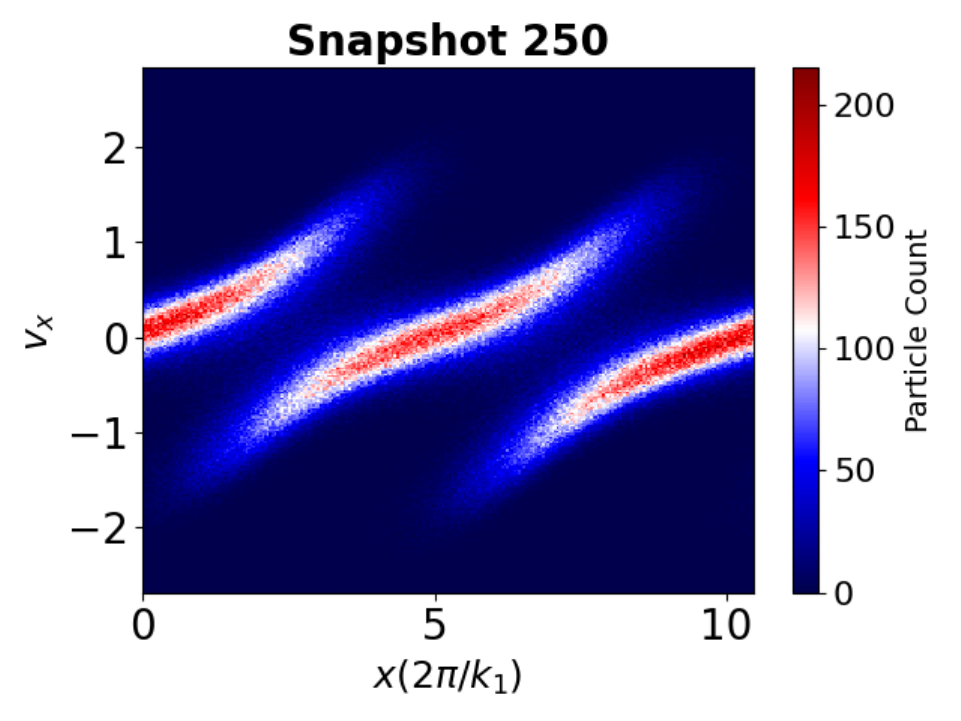} & 
        \includegraphics[width=0.3\textwidth]{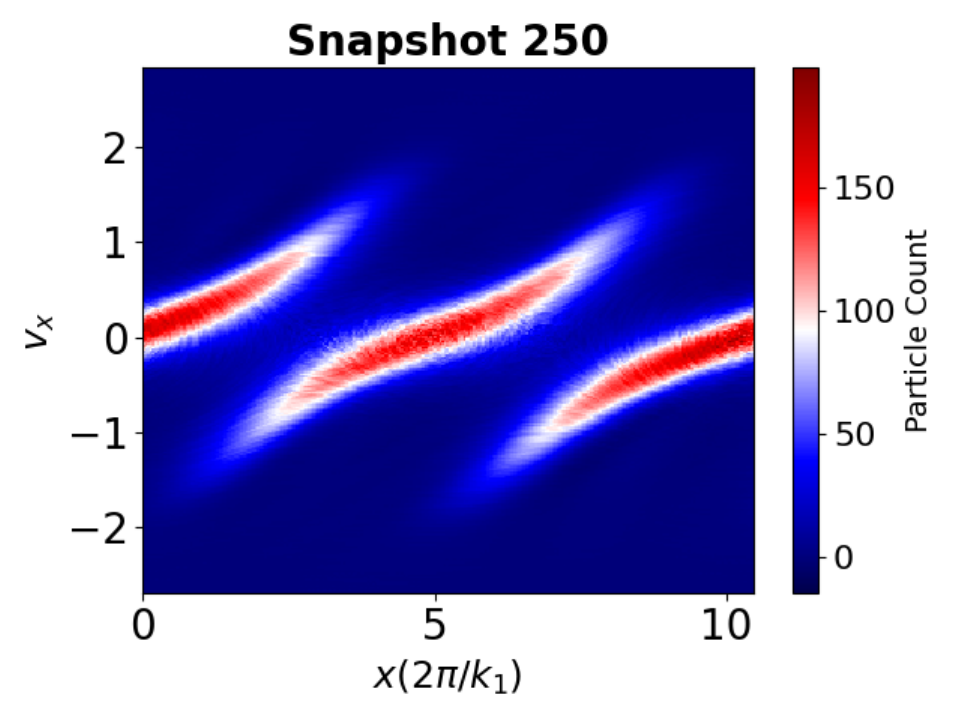} & 
        \includegraphics[width=0.3\textwidth]{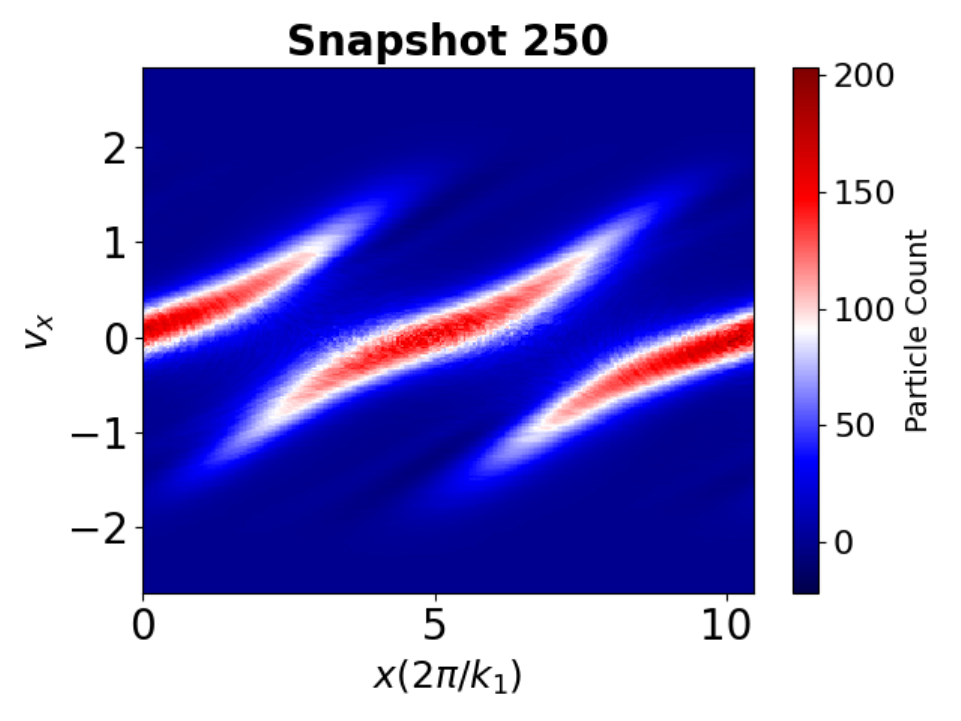} \\

        \includegraphics[width=0.3\textwidth]{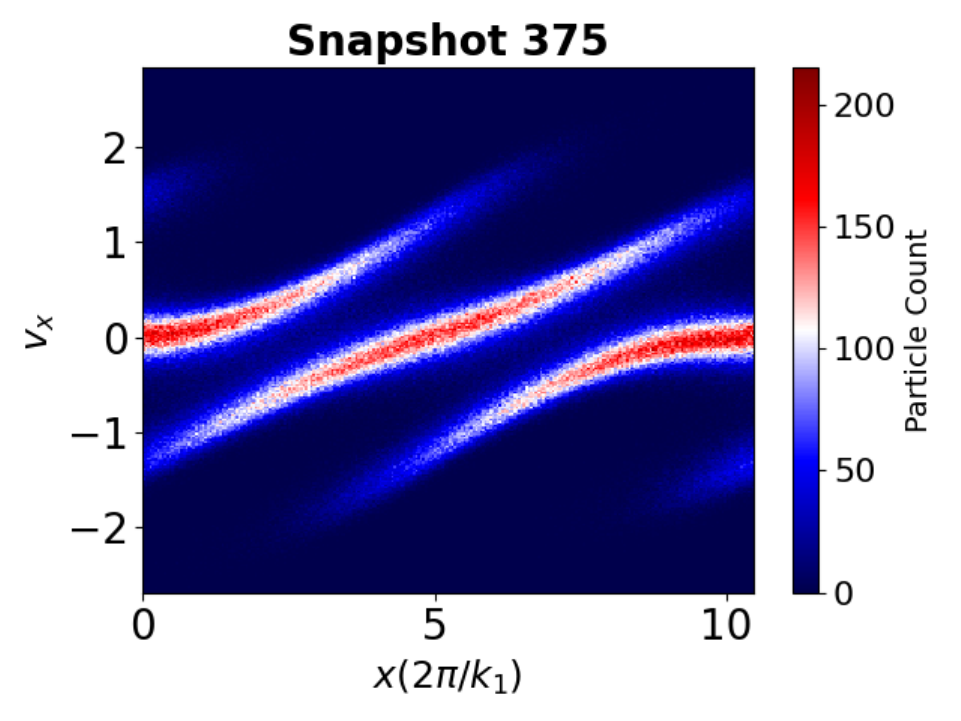} & 
        \includegraphics[width=0.3\textwidth]{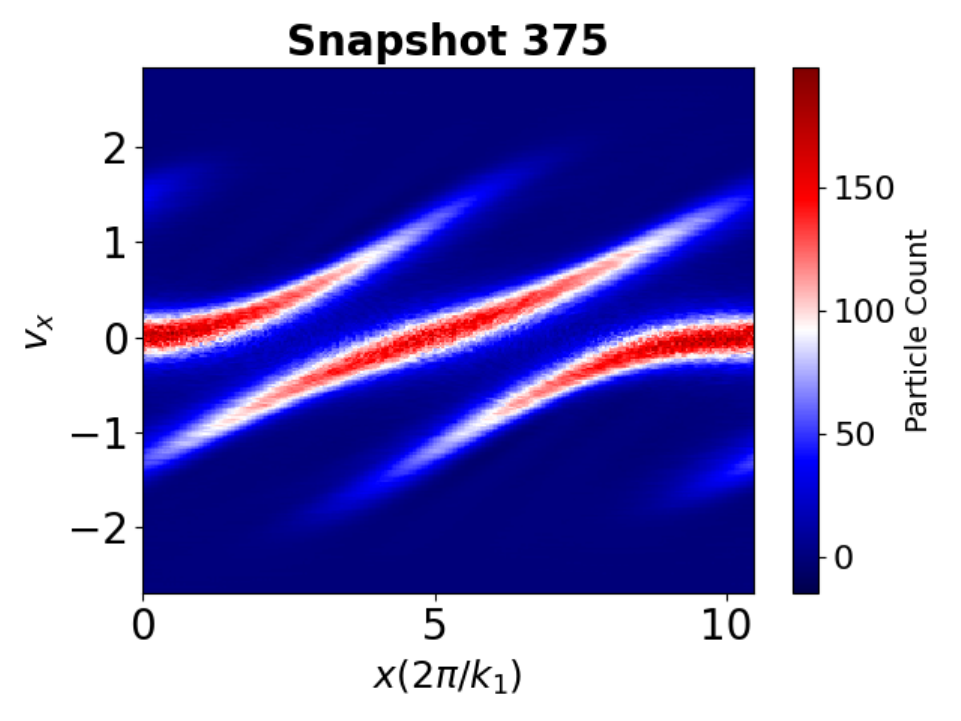} & 
        \includegraphics[width=0.3\textwidth]{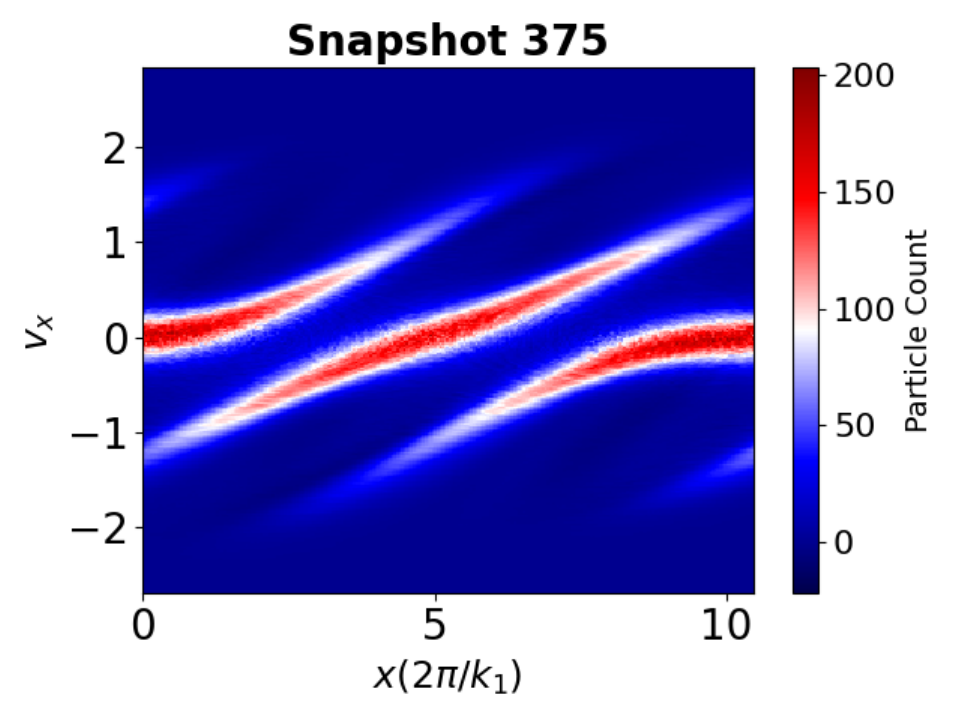} \\

        \includegraphics[width=0.3\textwidth]{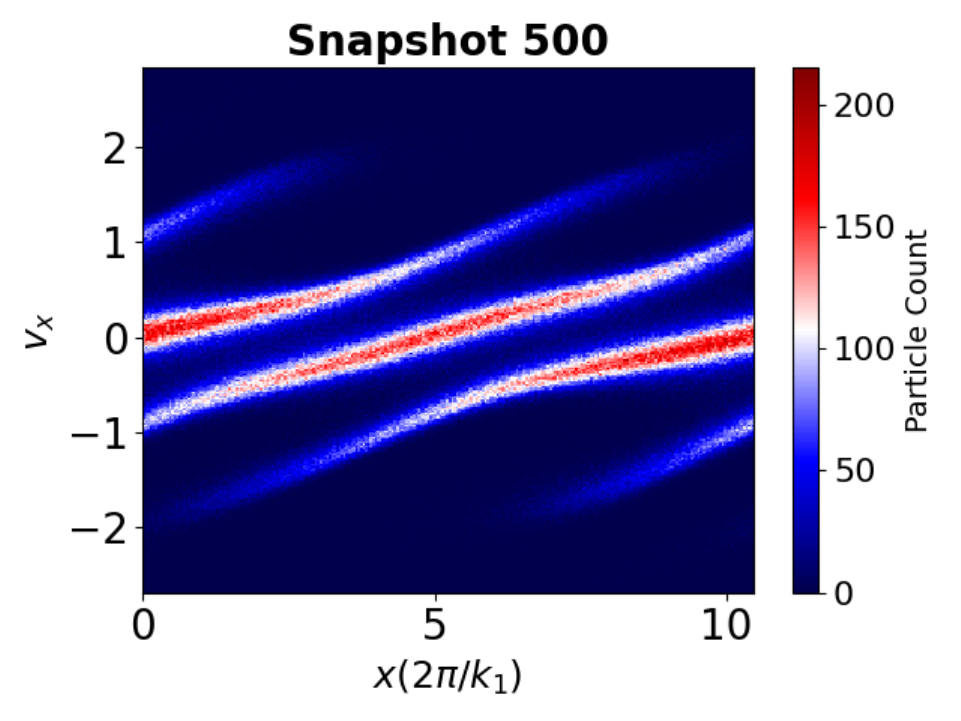} & 
        \includegraphics[width=0.3\textwidth]{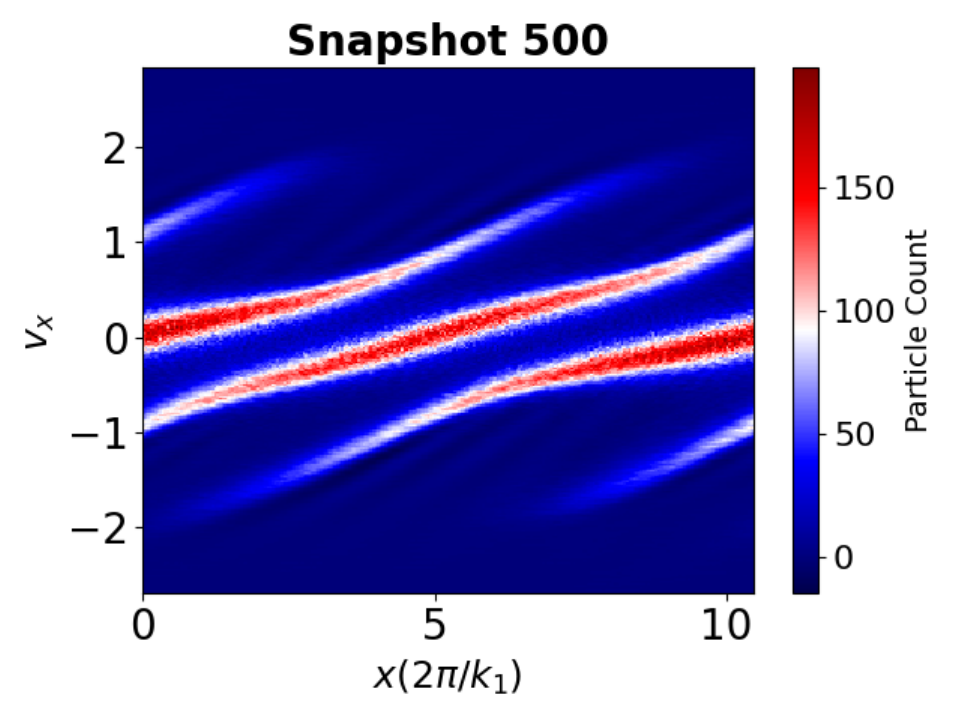} & 
        \includegraphics[width=0.3\textwidth]{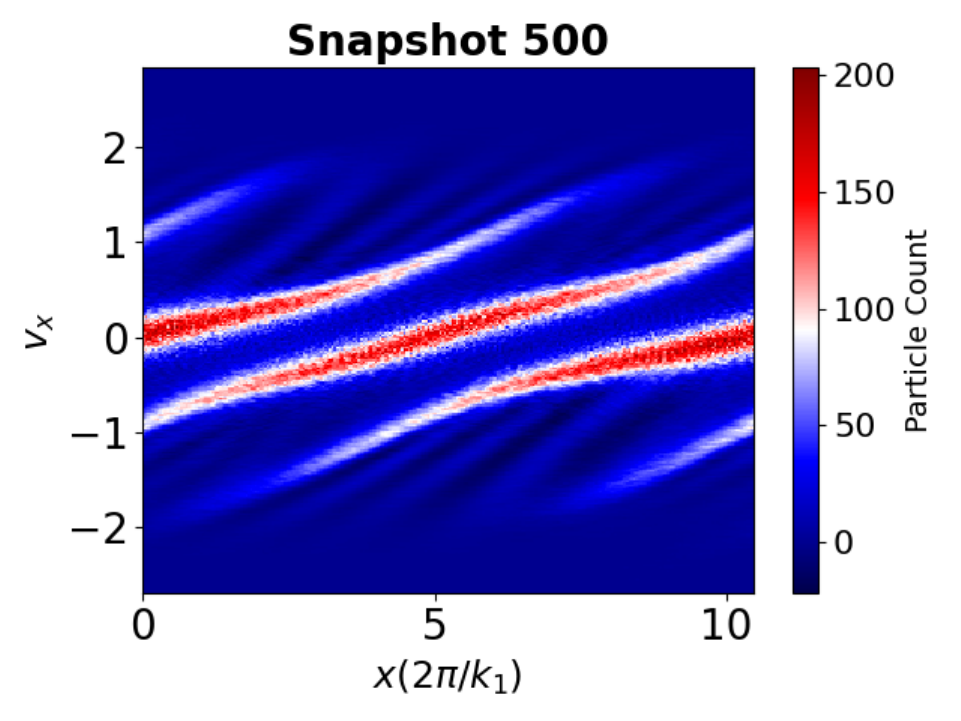} \\
    \hline
    \end{tabular}
    
    \caption{Comparison of phase mixing process with original data (left), POD reconstructed data (center) and POD--SINDy reconstructed data (right) for self-consistent electrostatic kinetic simulation 5K dataset.}
    \label{fig:Snapshots_5kE}  

\end{figure}

Figure \ref{fig:Modes_40kE} provides a clear insight of the above mentioned complex spatial structures introduced in the presence of self-consistent electric field. It can be appreciated how the first mode still shows relatively coherent features, but with increased filamentation compared to the passive kinetic simulation case. As we move to higher modes, modes clearly exhibit how the spatial structures become progressively finer and more fragmented. By modes five through eight, the clean sinusoidal patterns from the passive kinetic  simulation case have already degraded into dense filaments, capturing the strong mixing introduced by the field-particle interaction. Once more, this progressive degradation tells us that the variance of the system is distributed across many more modes, indicating that the dynamics of the ROM system cannot be represented by a few dominant sinusoidal structures anymore. Compared to the passive kinetic simulation case, where just a handful of modes captured almost everything of the energy, here the electric field drives nonlinear spatial interactions that forces us to rely on higher-order modes to describe the full filamentation process.

Similar effect is present in the high-noise 5K-E dataset. Figure \ref{fig:Modes_5kE} shows the spatial structure of the first ten obtained modes for the 5K-E dataset. Looking at the spatial distribution of these modes we can clearly observe the same overall filamentation patterns as in the 40K-E case. Nevertheless, the increment of particle noise has a noticeable effect on the integrity of the filamented structures as it is shown how the coherent structures lose definition much faster compared to the 40K-E. When examining closely the spatial patterns from the fifth mode it can be seen how noise becomes more dominant inducing the modes to appear more fragmented compared to the cleaner and smoother structures in the 40K-E dataset. This evaluation highlights the impact of reduced particle resolution on the POD. It is appreciated how the underlying physics are still captured in relatively low-rank modes, but the modal clarity degrades faster due to the increased statistical noise. While the essential filamentation dynamics are preserved, the higher noise level makes it harder for POD to isolate clean coherent structures in the later high-order modes, which are essentially related to small and noisy fluctuations of the data, i.e., noise in data propagates into POD basis.

Above analysis clarifies how increased reconstruction error stems not from POD--SINDy modeling of the amplitudes, but rather from the higher variance in the small-scale and noisy high-order modes which are predominant in these nonlinear and noisy datasets. Nevertheless it is important to remember that although the average reconstruction error of the POD--SINDy framework is relatively higher compared to the benchmark passive kinetic simulation dataset, we can still achieve a close reconstruction of the major filamented structures in the original phase space, as illustrated in the following section \ref{final_result_section}.

\subsubsection{POD--SINDy framework snapshot reconstruction}
\label{final_result_section}

Figure \ref{fig:Snapshots_40kE} shows the comparison of the evolution of the phase mixing process for the original 40K-E dataset, POD reconstructed data, and POD--SINDy reconstructed snapshots, all employing truncation of ten modes. It can be noticed how both reduced models are able to track the overall evolution of the phase mixing dynamics with remarkable accuracy. In this sense, the large-scale structures, the filamentation process, and the dynamics of the phase mixing are all retained not only in the POD reconstruction but also in the one achieved by the POD--SINDy framework.

Small differences compared to the original dataset can be appreciated, especially at fine-scale background structures at the beginning and end of the evaluated simulation interval. Nevertheless the evolution of the phase mixing process in the general interval shows a remarkable accuracy for the reconstructed data. It is important to notice that although the number of modes selected from truncation in these self-consistent electrostatic datasets are doubled from the benchmark passive kinetic simulation dataset, with only ten modes out of a total of five hundred, both POD and POD--SINDy framework are capable to capture the vast majority of the dynamics in a compact and efficient way, highlighting the strength of the reduced--order modeling technique: high compression with only minimal loss of fidelity. Even in the more complex self-consistent electrostatic case, POD--SINDy framework achieve compact, low-rank reconstructions that preserve the essential dynamics of phase mixing with minimal loss of fidelity.

Considering the performance with increased level of noise, figure \ref{fig:Snapshots_5kE} shows the comparison of the evolution of the phase mixing process for the original 5K-E dataset, POD reconstructed data, and POD--SINDy reconstructed snapshots, all employing again truncation of ten modes. These snapshots exhibit similar characteristic with the low-noise dataset such as clear differences at the edges of the simulation interval while achieving consistent description of the phase mixing process in the simulation interval. It can be shown then that major phase mixing are captured with the truncation of ten modes, while the accurate description of finer back-ground structures are dominated by high-order modes which must then be included to capture those dynamics in the overall reconstruction.

Analysis of the snapshots evolution in both reduced--order modeling techniques implies that despite the increased noise level, both POD and POD--SINDy framework are capable of recovering the same phase mixing dynamics while efficiently filtering out much of the particle noise present in the original dataset. This is a particularly important outcome of this study; the reconstruction of the phase space data by the POD--SINDy framework not only capture the underlying physics but also presents it in a cleaner, more interpretable way. Hence, while the reconstruction errors are relatively higher compared to the benchmark passive kinetic simulation dataset and low-noise case, these results showcase that the POD--SINDy methodology remains a robust and reliable tool for reduced--order modeling of phase mixing process, even when applied to more challenging datasets. 

\section{Conclusions}

This paper introduced the first exploratory assessment of a joint POD--SINDy framework for reduced-order modeling of early stages of phase mixing in plasma kinematic simulations. Employing kinematic data from one dimensional 1D1V electrostatic PIC simulations, we inquired on whether compact low-rank models, coupled with sparse identification of modal dynamics, can effectively capture and reproduce the essential filamentation physics while achieving substantial reductions in dimensionality and storage. By progressing from a passive kinetic simulation benchmark to self-consistent electrostatic cases with increased particle noise, we evaluated robustness across regimes of rising complexity.

Several clear findings emerge during evaluation of the obtained results. In the passive kinetic simulation regime 40K-NoE dataset, POD–SINDy framework attains near-optimal behavior as with only five modes ($\approx$1\% of the 500-mode basis), reconstructions capture almost all variance and reproduce filamentation with average errors below 4\%. With the inclusion of self-consistent fields, variance spreads across more modes due to nonlinear coupling and the formation of fine velocity-space structures. This implies that singular values decay more slowly and strict low-rank truncations become more challenging. Nevertheless, expanding the truncation to ten modes, which accounts for only 2\% of the total modes, stabilizes the evolution of POD--SINDy reconstructions per snapshot, yielding average errors around 7\% for the low-noise 40K-E dataset and approximately 13\% for the noisy 5K-E dataset. Notably, when evaluating the reconstruction snapshots of the phase space data considered in this study, the first ten modes still retain the dominant phase-mixing structures even in the most demanding, noisy case; the remaining discrepancies arise primarily in subtle background features near the beginning and end of the simulation window, pointing to the role of additional higher-order modes in capturing localized corrections rather than core dynamics of the phase mixing process.

Beyond these challenges, the joint POD--SINDy framework offers distinct advantages.

\begin{enumerate}
    \item Symbolic interpretability: SINDy yields sparse, coupled ODEs for modal amplitudes, providing compact linear descriptions of the reduced dynamics even when derived from nonlinear data.

    \item Noise attenuation: when applying POD to noisy datasets, particle noise tends to concentrate in the high-order modes. As a result, truncation naturally filters out this noise, allowing reconstructions to preserve coherent filamentation while suppressing spurious fluctuations. This effect is particularly evident in the 5K-E case.

    \item Strong data compression: the achieved ROMs cut storage requirements by orders of magnitude while keeping essential physics, more specifically:

    \begin{itemize}
        \item 40K-NoE: 76.4 GB $\rightarrow$ 1.27 MB ($\approx$ 60,000 times smaller).
        \item 40K-E: 76.4 GB $\rightarrow$ 2.53 MB ($\approx$ 30,000 times smaller).
        \item 5K-E: 9.58 GB $\rightarrow$ 2.53 MB ($\approx$ 3,800 times smaller).
    \end{itemize}
\end{enumerate}

Together, these results show that the increased reconstruction error observed in nonlinear and noisy settings primarily reflect the broader variance distribution that POD must capture, which in practice requires the inclusion of additional modes. This does not indicate a breakdown of SINDy’s ability to model modal dynamics. With modest increases in the number of retained modes, the POD–SINDy framework remains capable of capturing the dominant phase-mixing physics in a compact and sparse manner.

Looking ahead, several extensions are promising. This methodology offers potential use in the discovery of SINDy equations to construct reduced-order fluid closures that systematically embed phase-mixing physics, bridging kinetic simulations and fluid models. Additionally, extensions of the present framework to higher-dimensional kinetic problems and turbulence is viable, in which multiscale coupling and echo-like phenomena may test the limits of both compression and sparse identification. Furthermore, since the present study has focused on the early stages of phase mixing, methodological improvements are required to ensure applicability across arbitrary stages and extended durations of the phase-mixing process.

In summary, the POD–SINDy framework stands as a promising methodology for reduced-order modeling of phase mixing. It provides a robust, interpretable pathway that preserves dominant filamentation structures with very compact representations. Within this framework, the governing amplitude dynamics are uncovered in sparse form, while noise attenuation is simultaneously achieved. Furthermore, the results clearly show how complexity and noise shift variance toward higher modes, thereby offering principled guidance for the selection of an appropriate ROM rank.

\printbibliography
\end{document}